\documentclass[%
 aip,
 amsmath,amssymb,
reprint,%
]{revtex4-1}

\usepackage{graphicx}
\usepackage{dcolumn}
\usepackage{bm}
\usepackage{subfigure}
\usepackage[utf8]{inputenc}
\usepackage[T1]{fontenc}
\usepackage{mathptmx}
\usepackage{xcolor} 
\usepackage{hyperref}
\hypersetup{pdftex,colorlinks=true,allcolors=blue,
            pdfpagemode=UseOutlines }

\usepackage[normalem]{ulem} 

\newcommand\siesta{\textsc{Siesta}}
\newcommand\tsiesta{\textsc{TranSiesta}}
\newcommand\tbtrans{\textsc{TBtrans}}
\newcommand\phtrans{\textsc{PHtrans}}

\newcommand\blacs{\textsc{BLACS}}
\newcommand\scalapack{\textsc{ScaLAPACK}}
\newcommand\sisl{\textsc{sisl}}

\let\Im\relax
\DeclareMathOperator\Im{Im}

\DeclareUnicodeCharacter{2009}{\,}

\begin{document}

\title[\siesta\ ]{\siesta: recent developments and applications}

\author{Alberto Garc\'{\i}a}
 \email{albertog@icmab.es}
  \affiliation{Institut de Ci\`encia de Materials de Barcelona (ICMAB-CSIC), Bellaterra E-08193, Spain}

\author{Nick Papior} \email{nicpa@dtu.dk} \affiliation{DTU Computing
  Center, Technical University of Denmark, 2800 Kgs. Lyngby, Denmark
}%

\author{Arsalan Akhtar} \email{arsalan.akhtar@icn2.cat}
\affiliation{Catalan Institute of
  Nanoscience and Nanotechnology - ICN2, CSIC and BIST, Campus UAB,
  08193 Bellaterra, Spain
}%

\author{Emilio Artacho}\email{ea245@cam.ac.uk}
\affiliation{CIC Nanogune BRTA, Tolosa Hiribidea 76, 
             20018 San Sebasti\'an, Spain}
\affiliation{Donostia International Physics Center (DIPC), Paseo Manuel de Lardizabal 4, 20018 Donostia-San Sebastian, Spain}
\affiliation{Ikerbasque, Basque Foundation for Science, 48011 Bilbao, Spain}
\affiliation{Theory of Condensed Matter,
             Cavendish Laboratory, University of Cambridge,
             Cambridge CB3 0HE, United Kingdom}

\author{Volker Blum}\email{volker.blum@duke.edu}
\affiliation{Department of Mechanical Engineering and Materials Science, Duke University, Durham, NC 27708, USA}
\affiliation{Department of Chemistry, Duke University, Durham, NC 27708, USA}

\author{Emanuele Bosoni} \email{ebosoni@icmab.es}
\affiliation{Institut de Ci\`encia de Materials de Barcelona
  (ICMAB-CSIC), Bellaterra E-08193, Spain}

\author{Pedro Brandimarte}\email{pedro_brandimarte001@ehu.eus}
\affiliation{Donostia International Physics Center (DIPC), Paseo Manuel de Lardizabal 4, 20018 Donostia-San Sebastian, Spain}

\author{Mads Brandbyge}
\email{mabr@dtu.dk}
\affiliation{DTU Physics, Center for Nanostructured Graphene (CNG),
  Technical University of Denmark, Kgs. Lyngby, DK-2800, Denmark}
  
\author{J. I. Cerd\'a}%
 \email{jcerda@icmm.csic.es}
\affiliation{%
Instituto de Ciencia de Materiales de Madrid ICMM-CSIC, Cantoblanco, 28049 Madrid, Spain
}%

\author{Fabiano Corsetti}\email{fabiano.corsetti@gmail.com}
\affiliation{CIC Nanogune BRTA, Tolosa Hiribidea 76, 
             20018 San Sebasti\'an, Spain}

\author{Ram\'on Cuadrado} \email{ramon.cuadrado@gmail.com}
\affiliation{Catalan Institute of Nanoscience and Nanotechnology -
  ICN2, CSIC and BIST, Campus UAB, 08193 Bellaterra, Spain
}%

\author{Vladimir Dikan} \email{vdikan@icmab.es}
\affiliation{Institut de Ci\`encia de Materials de Barcelona
  (ICMAB-CSIC), Bellaterra E-08193, Spain}

\author{Jaime Ferrer} \email{ferrer@uniovi.es}
\affiliation{Department of Physics, University of Oviedo, Oviedo, 33007, Spain}
\affiliation{Nanomaterials and Nanotechnology Research Center,
             CSIC - Universidad de Oviedo, Oviedo, 33007, Spain}

\author{Julian Gale}\email{J.Gale@curtin.edu.au}
\affiliation{Curtin Institute for Computation, Institute for
  Geoscience Research (TIGeR), School of Molecular and Life Sciences,
  Curtin University, PO Box U1987, Perth, WA 6845, Australia}

\author{Pablo Garc\'{i}a-Fern\'andez} \email{garciapa@unican.es}
\affiliation{ Departamento de Ciencias de la Tierra y F\'{\i}sica de
  la Materia Condensada, Universidad de Cantabria, Cantabria Campus
  Internacional, Avenida de los Castros s/n, 39005 Santander, Spain
}

\author{V. M. Garc\'{i}a-Su\'arez} \email{vm.garciasuarez@gmail.com}
\affiliation{Department of Physics, University of Oviedo, Oviedo, 33007, Spain}
\affiliation{Nanomaterials and Nanotechnology Research Center,
             CSIC - Universidad de Oviedo, Oviedo, 33007, Spain}

\author{Sandra Garc\'{i}a} \email{sandragil@gmail.com}
\affiliation{Catalan Institute of
  Nanoscience and Nanotechnology - ICN2, CSIC and BIST, Campus UAB,
  08193 Bellaterra, Spain
}%

\author{Georg Huhs}\email{ghuhs@physik.hu-berlin.de}
\affiliation{Barcelona Supercomputing Center, c/ Jordi Girona, 29,
  08034 Barcelona, Spain}

\author{Sergio Illera}
 \email{sergiollera22@gmail.com}
\affiliation{%
Catalan Institute of Nanoscience and Nanotechnology - ICN2, CSIC and BIST, Campus UAB, 08193 Bellaterra, Spain
}%

\author{Richard Koryt\'ar} \email{korytar@karlov.mff.cuni.cz}
\affiliation{Department of Condensed Matter Physics, Faculty of
  Mathematics and Physics, Charles University, Ke Karlovu 5, 121 16
  Praha 2, Czech Republic}
 
\author{Peter Koval}\email{koval.peter@gmail.com}
\affiliation{Simune Atomistics S.L., Tolosa Hiribidea, 76, 20018, Donostia-San Sebastian, Spain}

\author{Irina Lebedeva}\email{i.lebedeva@nanogune.eu}
\affiliation{CIC Nanogune BRTA, Tolosa Hiribidea 76, 
             20018 San Sebasti\'an, Spain}

\author{Lin Lin}\email{linlin@math.berkeley.edu}
\affiliation{Department of Mathematics, University of California, Berkeley, CA 94720, USA}
\affiliation{Computational Research Division, Lawrence Berkeley
  National Laboratory, Berkeley, CA 94720, USA}

\author{Pablo L\'opez-Tarifa}\email{pablolopeztarifa@gmail.com}
\affiliation{Centro de F\'{\i}sica de Materiales,
  Centro Mixto CSIC-UPV/EHU,
  Paseo Manuel de Lardizabal 5, 20018 Donostia-San Sebastian, Spain}

\author{Sara G. Mayo}
\email{sara.garciamayo@uam.es}
\affiliation{ 
Departamento de F\'{\i}sica de la Materia Condensada, Universidad Aut\'onoma de Madrid, 28049 Madrid, Spain
}

\author{Stephan Mohr}\email{stephan.mohr@bsc.es}
\affiliation{Barcelona Supercomputing Center, c/ Jordi Girona, 29,
  08034 Barcelona, Spain}

\author{Pablo Ordej\'on} \email{pablo.ordejon@icn2.cat} 
\affiliation{Catalan Institute of
  Nanoscience and Nanotechnology - ICN2, CSIC and BIST, Campus UAB,
  08193 Bellaterra, Spain
}%
\author{Andrei Postnikov}
\email{andrei.postnikov@univ-lorraine.fr}
\affiliation{LCP-A2MC, Université de Lorraine, 1 Bd Arago, F-57078 Metz, France}

\author{Yann Pouillon} \email{yann.pouillon@unican.es} 
\affiliation{ Departamento de Ciencias de la Tierra y F\'{\i}sica de
  la Materia Condensada, Universidad de Cantabria, Cantabria Campus
  Internacional, Avenida de los Castros s/n, 39005 Santander, Spain
}%

\author{Miguel Pruneda} \email{miguel.pruneda@icn2.cat} 
\affiliation{Catalan Institute of
  Nanoscience and Nanotechnology - ICN2, CSIC and BIST, Campus UAB,
  08193 Bellaterra, Spain
}%

\author{Roberto Robles}\email{roberto.robles@ehu.eus}
\affiliation{Centro de F\'{\i}sica de Materiales,
  Centro Mixto CSIC-UPV/EHU,
  Paseo Manuel de Lardizabal 5, 20018 Donostia-San Sebastian, Spain}

\author{Daniel S\'anchez-Portal}\email{daniel.sanchez@ehu.eus}
\affiliation{Centro de F\'{\i}sica de Materiales,
  Centro Mixto CSIC-UPV/EHU,
  Paseo Manuel de Lardizabal 5, 20018 Donostia-San Sebastian, Spain}
\affiliation{Donostia International Physics Center (DIPC), Paseo Manuel de Lardizabal 4, 20018 Donostia-San Sebastian, Spain}

\author{Jose M. Soler} \email{jose.soler@uam.es}
\affiliation{ 
Departamento de F\'{\i}sica de la Materia Condensada, Universidad Aut\'onoma de Madrid, 28049 Madrid, Spain
}
\affiliation{Instituto de F\'{\i}sica de la Materia Condensada (IFIMAC),
  Universidad Aut\'onoma de Madrid, 28049 Madrid, Spain }

\author{Rafi Ullah} \email{ullah1@llnl.gov}
\affiliation{CIC Nanogune BRTA, Tolosa Hiribidea 76, 
             20018 San Sebasti\'an, Spain}
\affiliation{Departamento de F\'{\i}sica de Materiales, UPV/EHU, Paseo Manuel de Lardizabal 3, 20018 Donostia-San Sebastián, Spain}

\author{Victor Wen-zhe Yu}\email{wenzhe.yu@duke.edu}
\affiliation{Department of Mechanical Engineering and Materials Science, Duke University, Durham, NC 27708, USA}

\author{Javier Junquera}\email{javier.junquera@unican.es}
\affiliation{ Departamento de Ciencias de la Tierra y F\'{\i}sica de
  la Materia Condensada, Universidad de Cantabria, Cantabria Campus
  Internacional, Avenida de los Castros s/n, 39005 Santander, Spain
}%


\date{ April 20, 2020. Accepted by Jour. of Chem. Phys. After
  publication it can be found at https://doi.org/10.1063/5.0005077}

\begin{abstract}
A review of the present status, recent enhancements, and applicability
of the \siesta\ program is presented. Since its debut in the
mid-nineties, \siesta's flexibility, efficiency and free distribution
has given advanced materials simulation capabilities to many
groups worldwide. The core methodological scheme of
\siesta\ combines finite-support pseudo-atomic orbitals as basis sets,
norm-conserving pseudopotentials, and a real-space grid for the
representation of charge density and potentials and the computation of
their associated matrix elements. Here we describe the more recent
implementations on top of that core scheme, which include: full
spin-orbit interaction, non-repeated and multiple-contact ballistic
electron transport, DFT$+U$ and hybrid functionals, time-dependent
DFT, novel reduced-scaling solvers,
density-functional perturbation theory,
efficient Van der Waals non-local density functionals,
and enhanced molecular-dynamics options.  In addition, a
substantial effort has been made in enhancing interoperability and
interfacing with other codes and utilities, such as {\sc wannier90}
and the second-principles modelling it can be used for, an AiiDA
plugin for workflow automatization, interface to Lua for
steering \siesta\ runs, and various
postprocessing utilities.  \siesta\ has also been engaged in the
Electronic Structure Library effort from its inception, which has
allowed the sharing of various low level libraries, as well as data
standards and support for them, in particular the PSML definition and
library for transferable pseudopotentials, and the interface to the
ELSI library of solvers. Code sharing is made easier by the new
open-source licensing model of the program.  This review also presents
examples of application of the capabilities of the code,
as well as a view of on-going and future developments.
\end{abstract}

\maketitle

\section{\label{sec:introduction} Introduction.}

The possibility of treating large systems with first-principles
electronic-structure methods has opened up new research avenues in
many disciplines. The \siesta\ method and its implementation have been
key in this development, offering an efficient and flexible simulation
paradigm based on the use of strictly localized basis sets. This
approach enables the implementation of reduced scaling algorithms, and
its accuracy and cost can be tuned in a wide range, from quick
exploratory calculations to highly accurate simulations matching the
quality of other approaches, such as plane-wave methods.

The \siesta\ method has been described in detail in
Ref.~\onlinecite{Soler-02}, with an update in
Ref.~\onlinecite{Artacho-08}.  In this paper we shall describe its
present status, highlighting its strengths and documenting the steps
that have recently been taken to improve its capabilities,
performance, ease of use, and visibility in the electronic-structure
community.

As we shall see, the improvements touch many areas. We can underline
the implementation of new core electronic-structure features (DFT+U,
spin-orbit interaction, hybrid functionals), modes of operation
(improved time-dependent density functional theory (TD-DFT), density
functional perturbation theory (DFPT), and analysis methods and
procedures to access new properties. A major effort has been spent in
enhancing the interoperability of the code at various levels (sharing
of pseudopotentials, a new wannierization interface opening the way to
sophisticated post-processing, and an interface to multiscale
methods). Very significant performance enhancements have been made,
notably to the \tsiesta\ module through improved algorithms, and to
the core electronic structure problem through the development of
interfaces to new solvers.  These advances have put \siesta\ in a
prominent place in the high-performance electronic-structure
simulation scene, a role reinforced by its participation in important
international initiatives and by its new open-source licensing model.

The manuscript is organized as follows. We provide an overview of the
underlying methodology and the capabilities of \siesta\ in
section~\ref{sec:siesta-main}, which serves to place the code in the
wider ecosystem of electronic-structure materials simulation.
Section~\ref{sec:developments} presents the recent developments in and
around the code, which are covered in sub-sections.
To showcase \siesta's utility in the context of electronic-structure
calculations, we present briefly some relevant applications
and survey a few areas in which \siesta\ is being profitably used in
section~\ref{sec:applications}. Plans for the future evolution of
\siesta\ are outlined in section~\ref{sec:future}.

\section{\label{sec:siesta-main} Key concepts of \siesta\ }

\subsection{\label{sec:background} Theory background and context}

\siesta\ appeared as a consequence of the push for linear-scaling
electronic structure methods of the mid nineties, which has been
reviewed, for example, in Refs.~\onlinecite{Galli1996} and
~\onlinecite{Goedecker1999}. \siesta\ was the first linear-scaling
\textsl{self-consistent} implementation of density functional theory
(DFT).\cite{Ordejon1996,SanchezPortal1997}

The \siesta\ method relies on atomic-like functions of 
finite support as basis sets~\cite{Sankey-89,Artacho-99} 
-- of arbitrary number, angular momentum, 
radial shape, and centers --
combined with a discretization of space
for the computation of the Kohn-Sham Hamiltonian terms that involve more than 
two centers. The electron-ion interaction is represented by norm-conserving pseudopotentials.
  These key ingredients, through the optimized handling of sparse matrices, 
are used  to compute the self-consistent Hamiltonian and overlap matrices
with a computational expense that scales linearly with system size. 
  The method is completed with a choice of solvers for that Hamiltonian, from
optimized (but cube-scaling) diagonalization methods, to reduced-scaling
solvers of different flavors. 

The orbitals in the \siesta\ basis set are
made of the product of a real spherical harmonic and a radial
function, which is numerically tabulated in a grid.
The shape of the radial part is in principle totally arbitrary, but
the experience accumulated has proven that the numerical solution
of the Schrodinger equation for a (confined) isolated atom with the corresponding pseudopotential is a 
very good choice in terms of accuracy versus computational cost.
Fuller descriptions of the mechanisms to generate and optimize these
pseudo-atomic orbitals (PAOs) are given in Refs.~\onlinecite{Artacho-99,Junquera-01,Anglada-02}.

The auxiliary real-space grid is an essential ingredient of the
method, as it allows the efficient representation of charge densities
and potentials, as well as the computation of the matrix elements of
the Hamiltonian that cannot be handled as two-center integrals. This
grid can be seen as the reciprocal space of a set of plane waves, and
its fineness is most conveniently parametrized by an energy cutoff
(the ``density'' cutoff of plane-wave methods). There are limits to
the softness of the functions that can be described with such a grid,
so core electrons are not considered (although semi-core electrons
usually are), and their effect is incorporated into pseudopotentials.
The real-space grid is also used to solve the Poisson equation
involved in the computation of the electrostatic potential from the
charge density, through the use of a fast-fourier-transform
method. This means that \siesta\ uses periodic boundary
conditions. Non periodic systems, such as molecules, tubes, or slabs,
are treated using appropriate supercells.

\siesta\ is now a mature code with more than 20 years of existence.
In this period, the most important algorithms behind our implementation
have been already fully described and documented in a series of papers.
Readers interested in the details of
how the basic elements defining the method are combined, as well as
other relevant implementation details that make the method practical,
can find them in the main \siesta\ reference~\cite{Soler-02},
and in the update with the new capabilities of the code~\cite{Artacho-08}.

We note that the term \siesta\ is regularly used to describe both the method
(as outlined in the earliest
papers\cite{Ordejon1996,SanchezPortal1997}) and its implementation in
a computer program.  The
\siesta\ method is at the basis of later independent implementations, such as
OpenMX,\cite{OpenMX} and QuantumATK.\cite{QuantumATK} Other subsequent
codes built on the method, revising some of the fundamental
ingredients. This is the case of FHI-aims,\cite{fhiaims_blum_2009} which uses a more sophisticated
real-space grid (atom-centered), thus extending the core scheme to
all-electron calculations.

In this paper we describe new additions to the \siesta\ code,
based on independent methodological advances, either pre-existent
or specifically developed for \siesta, as specified and cited
in each section below.

\subsection{\label{sec:capabilities} Overview of \siesta\ capabilities}

As a general purpose implementation, \siesta\ can provide the standard
functionality available in mainstream DFT codes: energies, forces,
molecular-dynamics simulations, band structures, densities of states,
etc., and shares with those codes the basic current limitations of DFT
(notably the description of strongly-correlated systems).

What makes \siesta\ different from most other codes, and is at the
root of its key strengths, is the atomic-like, and strictly localized,
character of its basis set.  The use of a ``good first approximation''
to the full problem implies, first, that a much smaller number of
basis functions is needed. Second, the finite-support of the orbitals
leads to sparsity and the possibility to use reduced-scaling
methods. Thus high performance emerges almost by default.

Take first the basis cardinality: the number of basis orbital per atom
in a typical \siesta\ calculation is of the order of 10-20. This is to
be compared with a few hundred in the typical plane-wave (PW)
calculation. Furthermore, for systems whose description needs a vacuum
region (e.g., slabs for surface calculations, 2D monolayers, etc),
empty space is essentially ``free'' for \siesta, whereas PW codes
still need a basis set determined by the total size of the simulation
cell. \siesta\ is then quite capable of dealing with systems composed of
dozens to hundreds of atoms on modest hardware, 
even when using cubic-scaling
diagonalization solvers, which are the default as they are universally
applicable. 

Electronic-structure solvers with a more favorable
size-scaling can be applied to suitable systems. For example, one of
\siesta's earlier calculations, in 1996, was a linear-scaling
run for a strand of DNA with 650 atoms, performed on a desktop
workstation of the era.\cite{SanchezPortal1997} Reduced size-scaling is
also a feature of the PEXSI solver described in
section~\ref{sec:stand-alone-solvers} below, and of the NTPoly solver
mentioned in section~\ref{sec:elsi}. In addition to time-to-solution
efficiency, these solvers have a smaller memory footprint than
diagonalization, as the relevant matrices are kept in sparse form
rather than converted to a dense format.

Crucially, \siesta's baseline efficiency can be scaled up to
ever-larger systems by parallelization. Both distributed (MPI) and
shared-memory (OpenMP) parallelization options are implemented in the
code. As some of the examples in section~\ref{sec:applications} show, 
non-trivial calculations with thousands of atoms are
used in applications in different contexts, from molecular
biology to electronic transport.

\begin{figure}[htbp]
  \centering
  \includegraphics[width=0.99\linewidth]{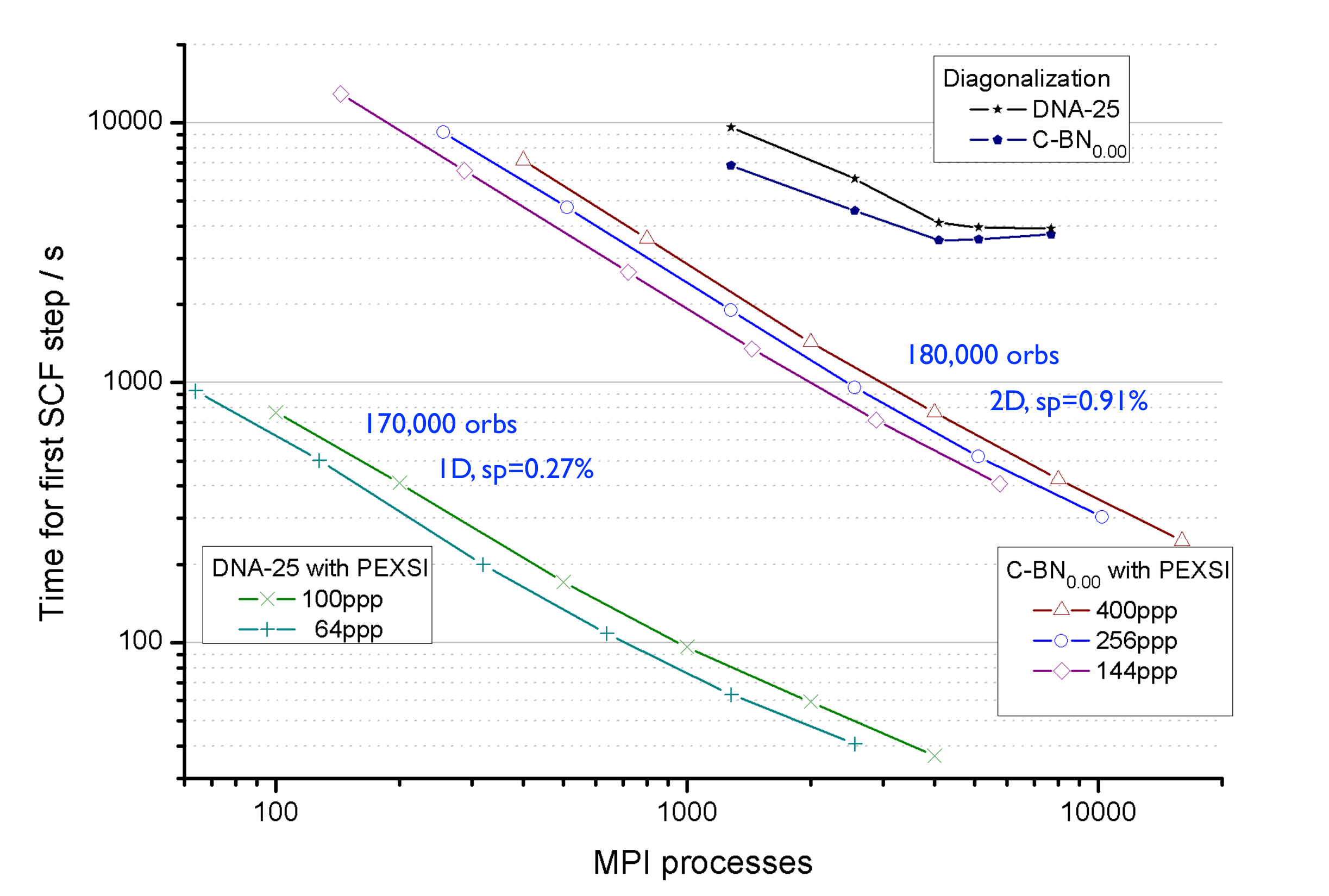}
  \caption{ Parallel strong scaling of SIESTA-PEXSI and the
    (Scalapack) diagonalization approach for a
    DNA chain and a Graphene-Boron Nitride stack, prototypes of large
    (hundreds of thousands of orbitals) quasi
    one-dimensional and two-dimensional systems. “ppp” stands for the
    number of MPI processes used in each pole computation, and ``sp''
    the sparsity of the Hamiltonian. (For more details, see section~\ref{sec:stand-alone-solvers})
    \label{fig:pexsi-scaling}
  }
\end{figure}

Work on the performance aspects of the code is continuous, mostly on
the solvers, which usually take most of the computer time due to the
very high efficiency of the Hamiltonian setup module in \siesta. This
task is facilitated (see section~\ref{sec:software-eng}) by leveraging
external libraries and developments generated by a number of
international initiatives in which \siesta\ participates. The code can
still run efficiently in modest hardware, while also being able to exploit
massive levels of parallelism in large supercomputers (see
Fig.~\ref{fig:pexsi-scaling}).

It is worth noting also that the atomic character of the
basis set enables the use of a very intuitive suite of analysis tools,
sice most of the concepts relating to chemical bonding use the
language of atomic orbitals. Hence \siesta\ has a natural advantage in
this area. Partial densities of states and atomic and crystal
populations (COOP/COHP) are routinely used to gain insights into the stability and
other properties of materials. For a recent example, see Ref.~\onlinecite{Carreras2019}.
Similarly, an atomic basis provides a very natural and
adequate language for the first-principles simulation of 
electronic ballistic transport in nanosized systems, via 
the Green's-function based Keldysh formalism implemented 
in \tsiesta,\cite{Brandbyge2002} a part of the \siesta\ package.

The very high number of citations of
the \siesta\ papers testify to the successful application of the code to widely
different systems. With regard to specific capabilities and the levels of accuracy achievable, 
we can distinguish several levels. First, \siesta\ implements DFT, one of the most versatile
materials simulation frameworks.  DFT has its shortcomings, notably in regard to the
description of strongly-correlated systems, but these are being
addressed  (see sections on DFT+U and hybrid
functionals below).  Second, \siesta\ uses pseudopotentials to
represent the electron-ion interaction. The pseudopotential approach
is firmly rooted in a sound physical approximation (that bonding
effects depend mostly on the valence electrons); however, it is at a
disadvantage when core-electrons effects are important (but see
section~\ref{sec:core-levels} below).
Third, \siesta\ employs periodic boundary conditions (PBC) for the
solution of the Poisson problem,
sharing with plane-wave codes the need to resort to
repeated supercells for the study of low-dimensional systems,
and to special techniques for the treatment of charged systems.
  It is important to note, however, that, unlike plane-wave codes,
\siesta\ is only bound to PBC because of the present treatment of
the Hartree term of the single-particle Hamiltonian.
  This limitation is lifted by the incorporation of alternative
Poisson solvers, as described in Sec.~\ref{sec:software-eng}, which
allow for open boundary conditions, as for isolated nano-systems, 
and hybrid open/periodic boundary conditions in different dimensions, 
as for isolated wires and slabs.
  It should be remembered that the three approximations mentioned 
in this paragraph are very widely used in the community, shared by 
some of the most popular electronic structure codes.

Fourth, with regard to \siesta\ specific approximations, particularly
the basis set, it should be stressed that \siesta\ is limited to
basis sets composed of functions that are product of a radial part and
spherical harmonics, but it does not constrain on how many, where such
functions are centered, and the size of their finite-support
region. Calculations can flexibly range from quick exploration to very
high-quality simulations (one may recall that accuracy gold standards
in electronic structure are provided by quantum-chemistry methods,
based on LCAO).

\begin{figure}[htb]
\centering
\includegraphics[width=1.0\linewidth]{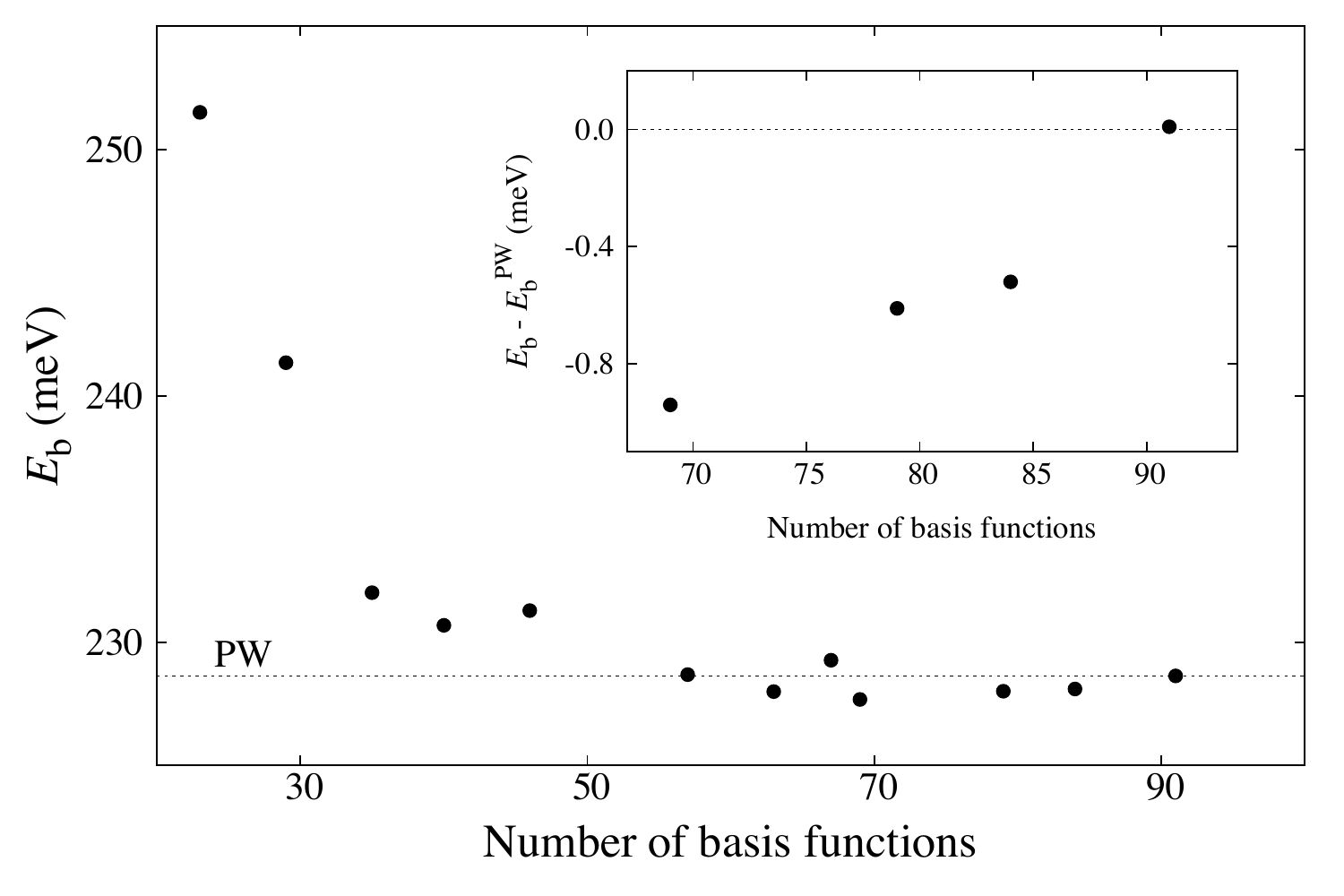} 
\caption{\label{fig:basesH2O} 
Basis set convergence for the binding energy ($E_b$) of 
a water dimer. Details can be found in 
Ref.~\onlinecite{Corsetti2013}. The horizontal dotted line
represents the converged plane-wave (PW) calculation (1300 eV cutoff)
for the same system (dimer geometry and box), pseudopotentials 
and density functional, using the ABINIT code.\cite{gonze_abinit_2009} Inset: deviation
of $E_b$ versus the PW reference. The deviation for the last
point is of 10 $\mu$eV.}
\end{figure}

The use of an atomic-orbital basis set implies however the limitation of
non-uniformity of convergence. As opposed to plane-wave methods, in
which a single energy cutoff parameter monotonically determines the
quality of the calculation, there is no univocal procedure for the
choice of an appropriate basis set. 
It is a well-known problem, shared by the whole quantum-chemistry 
community, on which there is widely used and tested know-how.
As
Fig.~\ref{fig:basesH2O} shows, it is possible to attain in practice an
accuracy comparable to that of well-converged plane-wave
calculations. The reader is also referred to
sections~\ref{sec:topoferro} and~\ref{sec:cdw} for
showcase examples of the accuracy of the code, among many others in the literature.

To close this section, we stress that it has been a traditional and
deliberate attitude by the \siesta\ team that, although proposing
sensible starting points to users as defaults, the choice of
fundamental approximations and inputs to the program (not only basis
sets, but also density functionals, and pseudopotentials) is a
responsibility of the users, who retain full control and the flexibility
to adapt the code to their specific needs. 
Nevertheless, tools for basis optimization
are provided with the program, new curated databases of
pseudopotentials are coming online, and new ways to ameliorate the
correlation problem are being implemented. Some of these developments
are described in the following sections.

\section{\label{sec:developments} Recent developments in \siesta\ }

\subsection{\label{sec:code-development}New distribution model and
  development infrastructure}

A few years ago, in 2016, a decision was made to change the licensing
model for \siesta: traditionally it had always been free of charge
to academics, but non-academic use required a special license and
redistribution was not permitted. Now \siesta\ is formally an
open-source program, distributed according to the terms of the GPL
license~\cite{GPL}. At the same time, the development infrastructure
was made more transparent and scalable, using first the Launchpad
platform~\cite{launchpad} and now the Gitlab service~\cite{gitlab}.
The net effect of the changes has been a more fertile and dynamic
development, with more contributors who can have direct access to the
various branches of development, and a better experience for users,
who can download code and raise issues in an integrated platform.

These changes have been substantial for the core developers, and the
transitory period is still being felt. The main code base is gradually
absorbing new developments, both those that were planned long in
advance, and new ones made possible by the greater openness and
fluidity of the development model. Most of the new features described
below are already part of public releases, but a few are undergoing
the last stages of testing before release. The work-flow is also
moving from long-lived releases, hard to maintain with bug-fixes, to
more frequent releases that will be maintained for a shorter time.

\subsection{\label{sec:psml}New pseudopotential format for interoperability}

PSML (for PSeudopotential Markup
Language)~\cite{psml_Garcia_2018,psml_info} is a file format for
norm-conserving pseudopotential data which is designed to encapsulate
as much as possible the abstract concepts involved in the domain,
and to provide appropriate metadata and provenance information.  This
extra level of formalization aims at removing the interoperability
problems associated to bespoke pseudopotential formats, which usually were
designed to serve the needs of specific generators and client codes,
and thus contain implicit assumptions about the meaning of the data or
lack information not considered relevant.

PSML files can be produced by the \textsc{ONCVPSP}\cite{Hamann2013}
and \textsc{ATOM}\cite{Froyen} pseudopotential generator programs,
and are a download-format option in the Pseudo-Dojo database of
curated pseudopotentials~\cite{dojo_vanSetten_2018,pseudo-dojo-site}.

The software library libPSML~\cite{psml_Garcia_2018,psml_info} can be
used by electronic structure codes to transparently extract the
information in a PSML file and incorporate it into their own data structures,
or to create converters for other formats. It is currently used by
\siesta\ and {\sc Abinit},\cite{gonze_abinit_2009,Gonze2016_106} making possible a full pseudopotential
interoperability and facilitating comparisons of calculation results.

The use of this new format opens the door to benefit from the
availability of a periodic table of reliable and accurate
norm-conserving pseudopotentials, easing in most cases the task of
pseudopotential quality control.

\subsection{\label{sec:DFT+U}DFT+U for correlated systems}

The LDA+U method, initially developed by Anisimov and coworkers~\cite{Anisimov-91}
with the objective to improve
the treatment of the electron-electron interaction for localized electrons within the bare LDA 
description, has been implemented in \siesta.
The idea behind the LDA+U consists in describing the ``strongly correlated'' 
electronic states of a system (typically, localized $d$ or $f$ orbitals) 
using the Hubbard model, whereas the rest of valence electrons are treated
at the level of ``standard'' approximate DFT functionals.~\cite{Himmetoglu-14}
In the current version of \siesta\ the implementation is based on the 
simplified rotationally invariant functional proposed by Dudarev and coworkers.~\cite{Dudarev-98}
Here, the corrections are made invariant under rotation of the atomic orbitals used to define
the occupation number of the correlated subspace, at the cost of retaining only the lowest order
Slater integrals in the factorization of the integrals of the Coulomb kernel of the electron-electron
interaction, and neglecting the higher order ones (i.e. taking the exchange
interaction J as 0).
The expression of the corrective term as a functional of the 
occupation number $n^{I\sigma}_{\ell m}$ of the localized correlated orbital
$\ell m$ with spin $\sigma$ within the atom $I$ is given by

\begin{equation}
    E_{\mathrm U} = \sum_{I \sigma \ell} \frac{U^{I\ell}}{2} 
    \left[\sum_{m}  n^{I\sigma}_{\ell m} \left( 1 - n^{I\sigma}_{\ell m} \right) \right],
\end{equation}
\noindent where only one interaction parameter $U^{I\ell}$ is needed
to specify the interaction per atom and $\ell$-shell.
In the practical \siesta\ implementation, the populations on the
correlated orbitals are computed using non-overlapping (i. e. orthogonal)
localized projectors.
They can be generated using either $(i)$ the same algorithm used to produce the first-$\zeta$ orbitals
of the basis set, but with a larger energy shift, or $(ii)$
cutting the exact solution of the pseudoatom with a Fermi function.

The results of the LDA+U method are sensitively dependent on the numerical
value of the effective on-site electronic interaction, the Hubbard $U$.
Although in principle the value of $U$ can be computed from 
first principles using linear response methods,~\cite{Cococcioni-05}
a common practice is to tune it semiempirically, 
seeking agreement of certain properties (for instance band gaps or
magnetic moments) with available experimental measurements.
Then, the fitted $U$ is used in subsequent calculations to predict 
other properties.

\begin{figure}[htb]
\centering
  \includegraphics[width=0.7\linewidth]{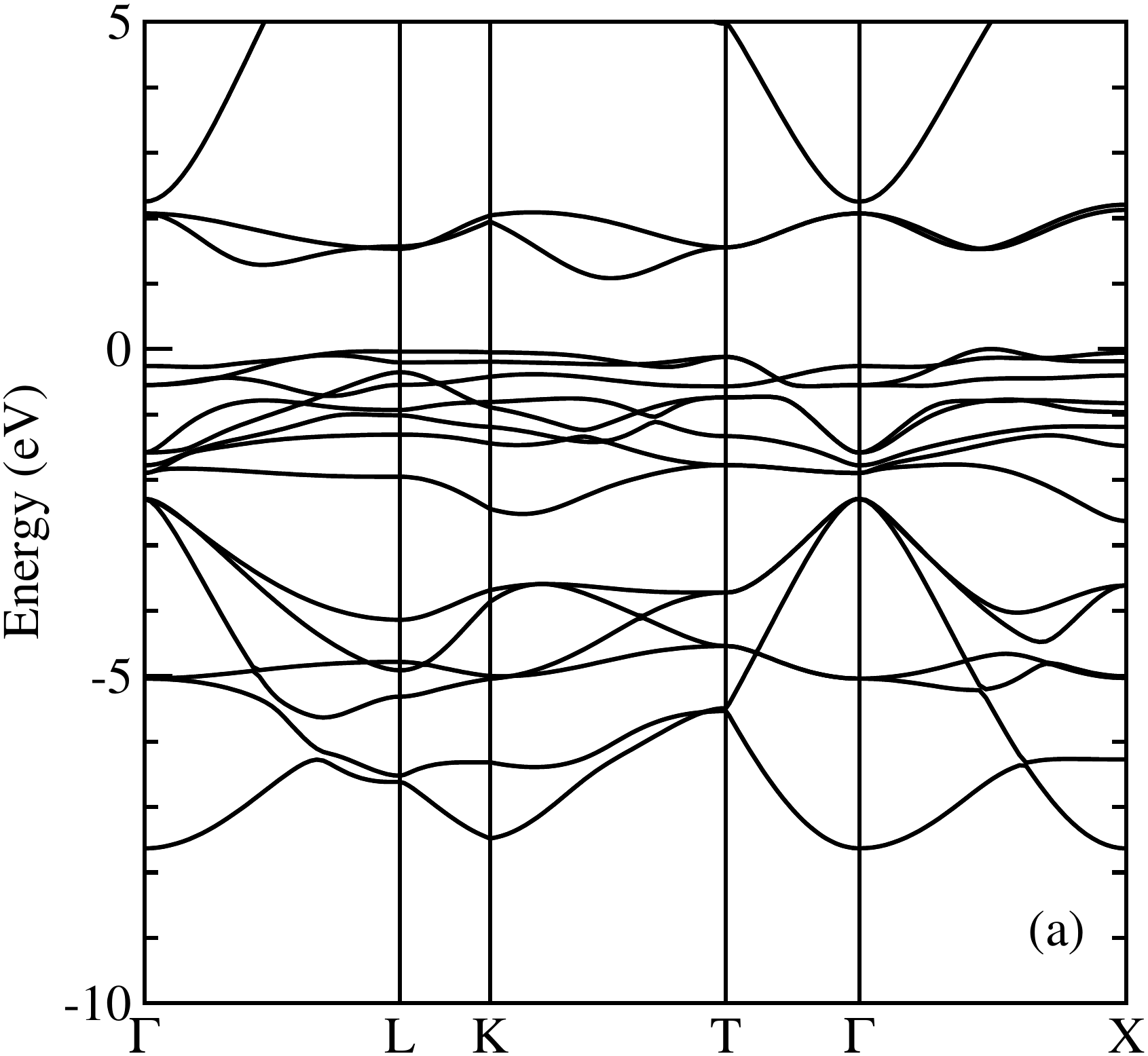} \\
  \includegraphics[width=0.7\linewidth]{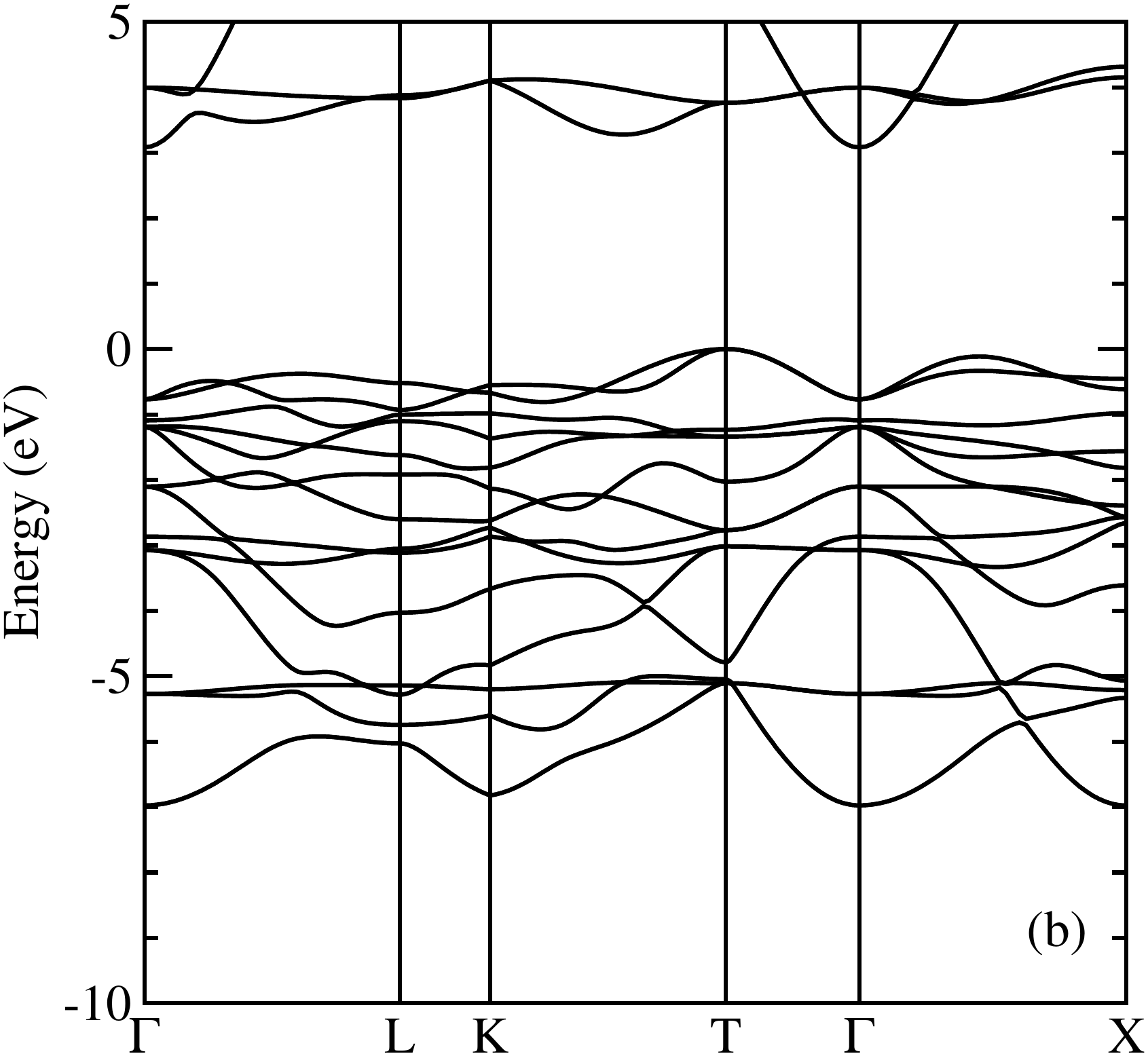}
\caption{\label{fig:ldau} Band structure of NiO in the undistorted 
         rock-salt type structure with rhombohedral symmetry
         introduced by a type-II antiferromagnetic order. 
         The experimental lattice spacing is used.
         The bands obtained within GGA-Perdew-Burke-Ernzerhof functional
         (panel a), and with a Hubbard U correction of
         4.6 eV applied on the $d$-orbitals of Ni (panel b), 
         as in Ref.~\onlinecite{Cococcioni-05}, are shown.
         The zero of the energy is set at the top of the valence band.
        }
\end{figure}

The LDA+U corrects localized states, for which the 
self-interaction correction is expected to be stronger, and is an 
effective method to improve the description of the 
(underestimated) band gap of insulators, as shown in 
Fig.~\ref{fig:ldau} for the case of NiO. 
Once the Hubbard correction is switched on, the optical band gap
increases up to 3.08 eV (from the bare GGA-PBE value of 1.08 eV),
very close to the experimental value for the onset of optical
absorption in NiO~\cite{Powell-70} (3.10 eV). 
The magnetic moment on the Ni atom is also properly described,
with a value of 1.67 $\mu_{\mathrm B}$ which lies well within the experimental
range of values (between 1.64 $\mu_{\mathrm B}$~\cite{Alperin-62} 
and 1.9 $\mu_{\mathrm B}$~\cite{Cheetham-83}),
and improves on the result of 1.39 $\mu_{\rm B}$ obtained with 
a bare GGA-PBE functional.

\subsection{\label{sec:vdw}Van der Waals functionals}

An efficient calculation of van der Waals (vdW) functionals
\cite{Dion2004,Berland2014} was developed and first implemented in
\siesta\ using a polynomial expansion in the local variables
$(q_1,q_2)$ of the nonlocal interaction kernel
$\Phi(q_1,q_2,r_{12})$ and a Fourier expansion in the relative
position $r_{12}$ \cite{RomanPerez2009}.  As a result, the scaling
of the vdW computation decreases from $O(N^2)$ to $O(N \log N)$ and it
becomes marginal within the overall cost.  This scheme was later
extended~\cite{Corsetti2013} to a more complex
kernel~\cite{Vydrov2010} of the form $\Phi(n_1,|\nabla
n_1|,n_2,|\nabla n_2|,r_{12})$, and it has been applied to a large
variety of systems, like carbon nanotubes~\cite{RomanPerez2009},
hydrogen adsorption~\cite{Kong2009,GonzalezHerrero2016}, or liquid
water~\cite{Wang2011}.

\subsection{\label{sec:hybrids}Hybrid functionals}

\begin{figure}[htb]
\centering
  \includegraphics[width=0.9\linewidth]{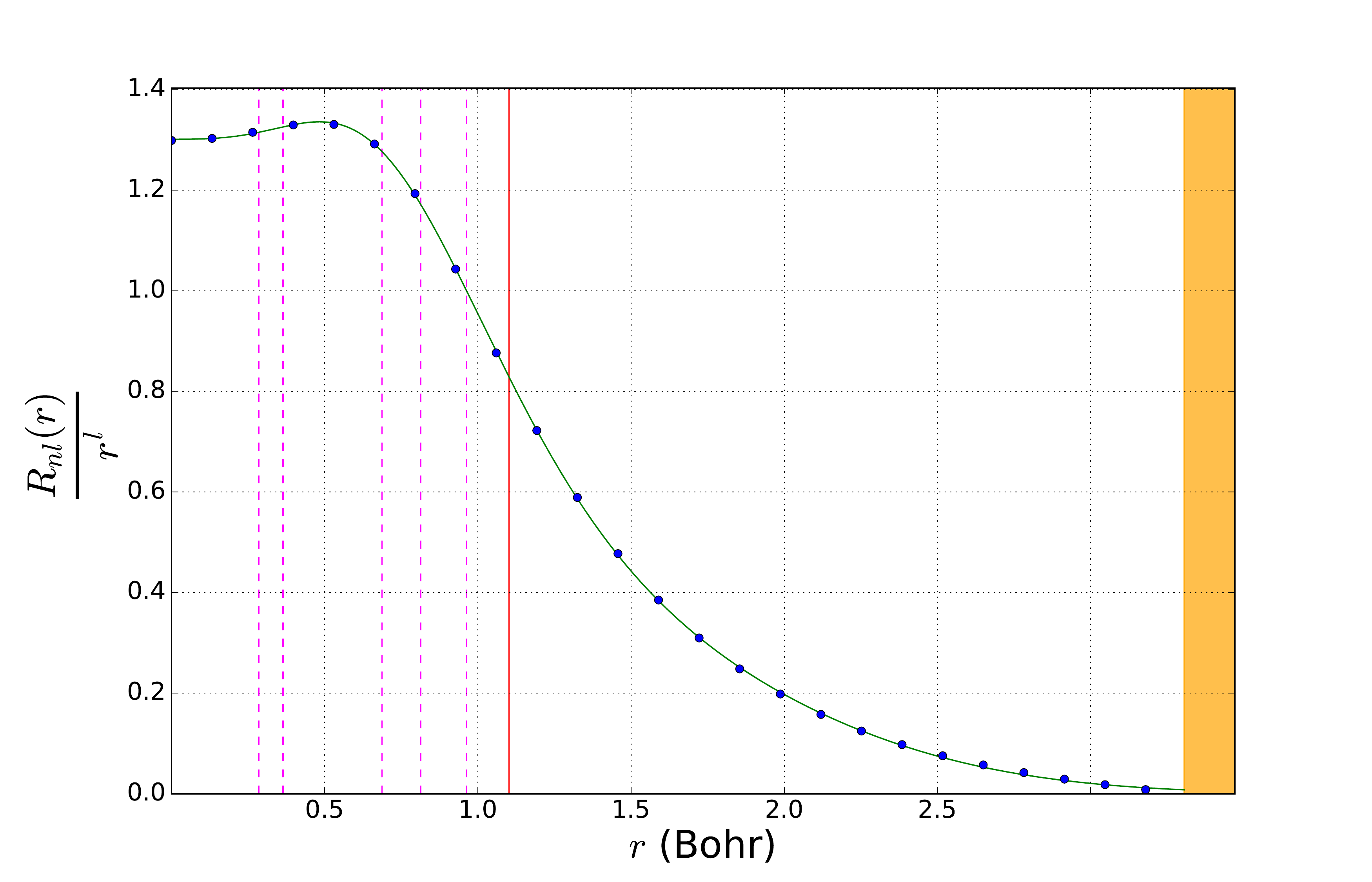} \\
  \includegraphics[width=0.9\linewidth]{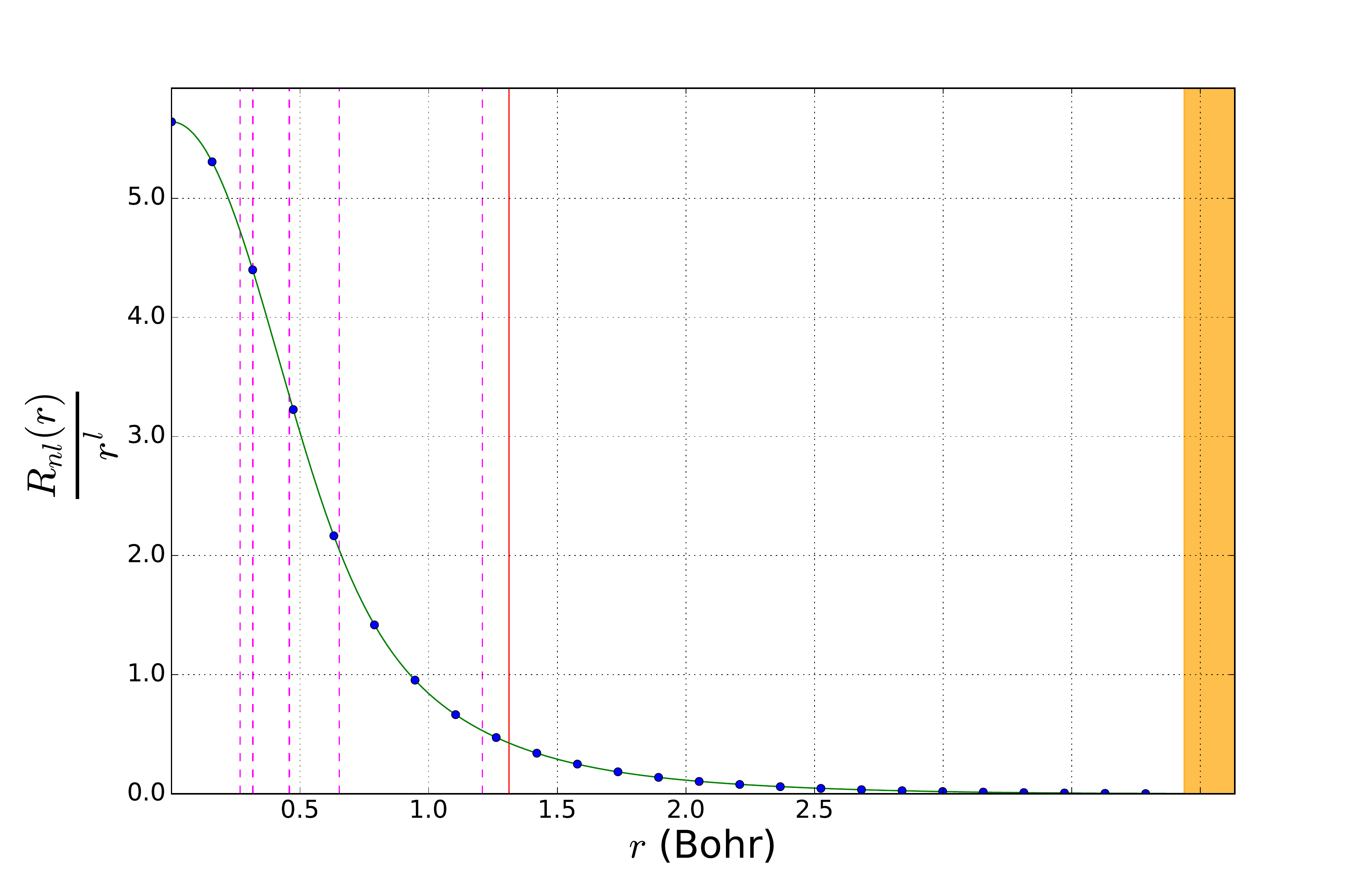}
\caption{\label{fig:fitting-orb} Gaussian fits of the radial part of oxygen 2$s$ (a) and 2$p$ (b) orbitals using
6 Gaussian functions. The orbitals to fit are represented by blue dots and the
corresponding Gaussian expansions by green continuous lines. Dashed vertical
lines represent the standard deviations of individual Gaussians and a red
continuous line marks their upper limit. The orbitals are set to zero in the
yellow area, marking their cutoff radii.}
\end{figure}

The screened hybrid functional HSE06~\cite{Heyd-03,Heyd-06,HSE06} has been implemented in 
\siesta\ building on the work of Ref.~\onlinecite{Shang-11}.
This functional is the result of adding nonlocal Hartree-Fock type exact exchange (HFX) into 
semilocal density functionals. 
The Coulomb potential that appears in the exchange interaction is screened, so it has a shorter range than $1/r$.
Here, to reduce the big prefactor involved in the computation of the HFX potential matrix elements,
we fit the NAO of the basis set with Gaussian-type orbitals, specially suited to computing the
four center electron repulsion integrals (ERIs) in a straightforward and efficient analytical way.
An example of this fitting for the $2s$ and $2p$ atomic orbitals basis set of the oxygen is
shown in Fig.~\ref{fig:fitting-orb}.
The \textsc{libint} package~\cite{libint-site} is required to calculate primitive ERIs,
where recursive schemes of the Obara-Saika~\cite{Obara-84}
method and the Head-Gordon and Pople's variation~\cite{Head-Gordon-88} thereof are implemented.
ERIs are calculated in the first SCF cycle and then stored in disk.
Only the ERIs with non-negligible contributions are calculated, keeping the HFX Hamiltonian also sparse.

\begin{figure}[htb]
\centering
  \includegraphics[width=0.9\linewidth]{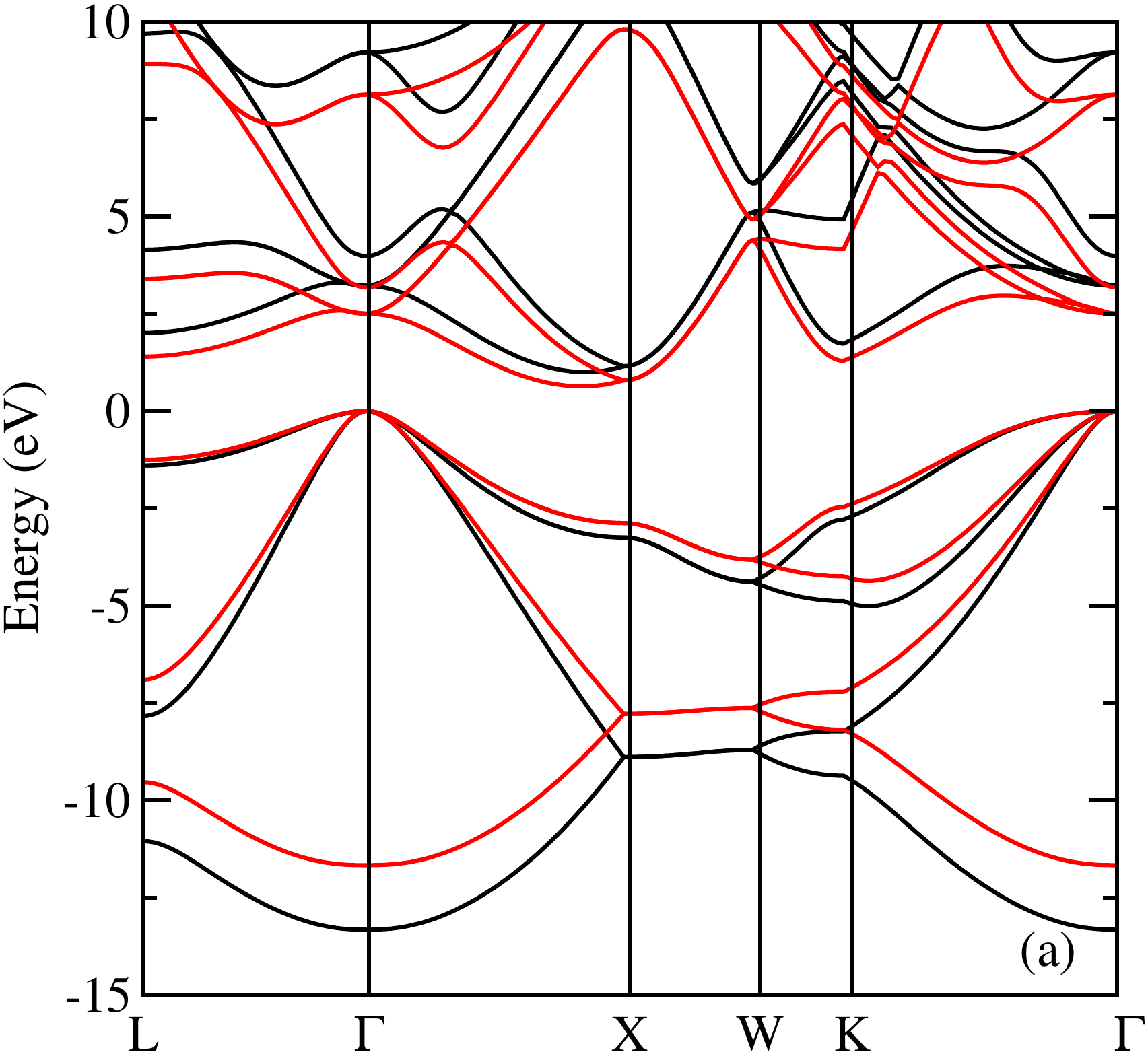} \\
  \includegraphics[width=0.9\linewidth]{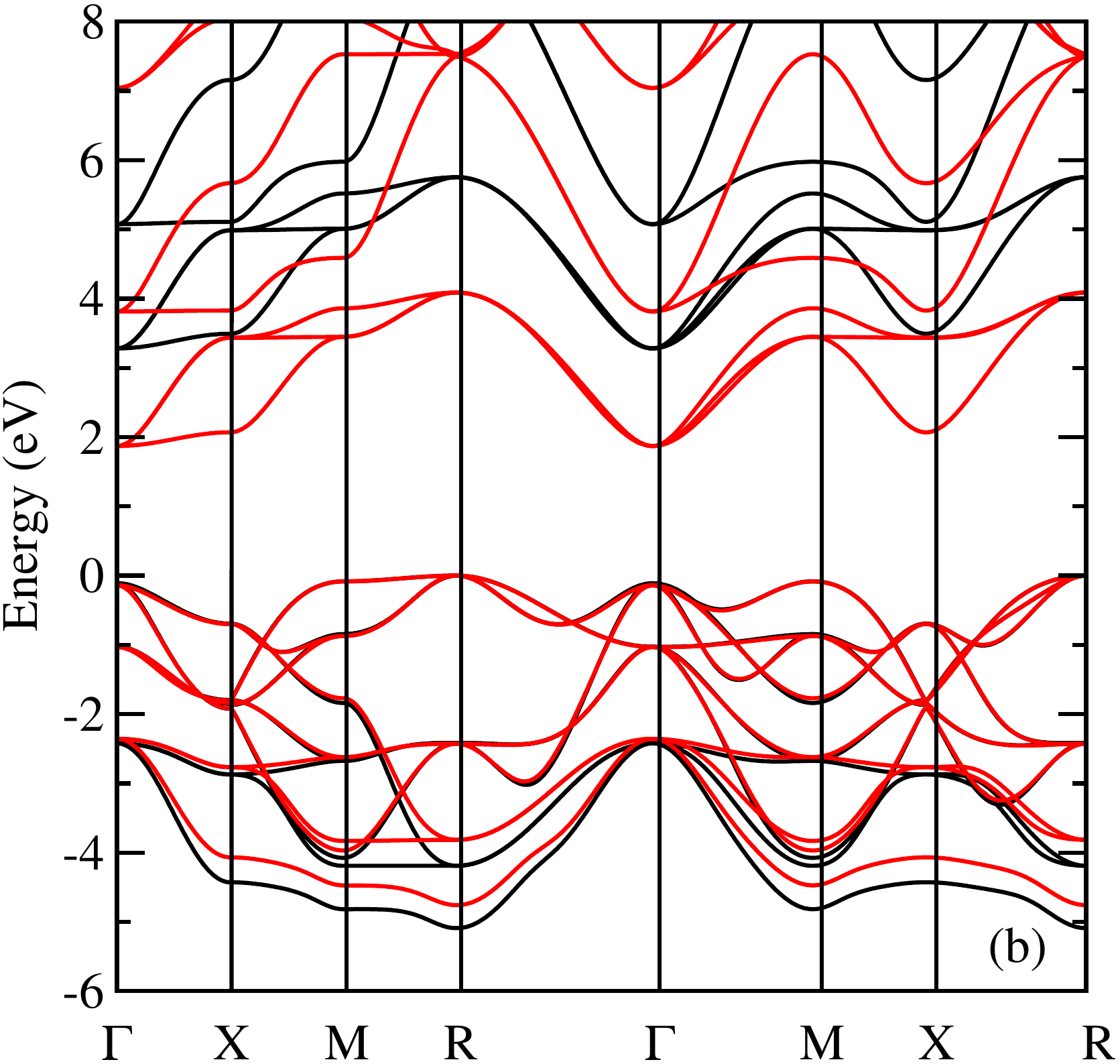}
\caption{\label{fig:bands-hybrids} Band structure of (a) bulk Si in the diamond structure,
 and (b) bulk BaTiO$_{3}$ in the cubic structure obtained with the 
 Perdew-Burke-Ernzerhof functional (red lines) and with the HSE06 hybrid functional 
 (black lines). The zero of energies have been set to the valence band maximum.}
\end{figure}

This HSE06 functional has been used to compute the band structure of bulk Si [diamond structure; 
Fig.~\ref{fig:bands-hybrids}(a)] and BaTiO$_3$ [cubic structure; Fig.~\ref{fig:bands-hybrids}(b)] 
with a double-zeta polarized basis set at the equilibrium lattice constant of the
Perdew-Burke-Ernzerhof functional~\cite{Perdew-96} within the Generalized Gradient Approximation
(5.499 \AA\ for Si and 4.033 \AA\ for BaTiO$_{3}$). 
In both cases, the gap is opened with respect to the value obtained with the semilocal functional.
In bulk Si the band gap is indirect: the top of the valence band is located at $\Gamma$
and the bottom of the conduction band at a point along the $\Gamma\rightarrow X$ high-symmetry line.
It increases from 0.64 eV within GGA to 1.00 eV with the hybrid functional, 
in good agreement with the experimental value of 1.17 eV~\cite{Kittel}.
For the case of the perovskite oxide BaTiO$_{3}$, the band gap is also indirect,
from $R$ to $\Gamma$, and its value increases from 1.87 eV with GGA to 3.28 eV with the HSE06 
functional, almost matching the experimental value of 3.2 eV estimated by Wemple in the cubic
phase~\cite{Wemple-70}.

\subsection{\label{sec:soc}Spin-orbit coupling}

The capability to include the spin--orbit~(SO) interaction 
in \siesta\ and in the analysis tools is seen as a strategic asset 
for the project in view of the recent interest in topological insulators and 
quasi--two--dimensional systems with important spin--orbit effects, like some 
of the transition metal dichalcogenides. Also, it brings the possibility 
to obtain  the magnetic crystalline anisotropy~(MCA)~(change in the total energy
of the system upon changing the spin quantization axis). 

In a standard collinear-spin DFT calculation, the total KS Hamiltonian 
is represented by two independent spin--blocks, $\hat H^{\sigma\sigma}_{\mu\nu}$
[$\sigma$=$\uparrow,\downarrow$]. However, when the SO coupling is included,  
off--diagonal spin blocks arise 
(i.e., there are non--zero couplings between the two spin components). Therefore, 
and similar to the non-collinear spin case, the 
Hamiltonian becomes a full 2$\times$2 matrix in spin space
\begin{equation}
\hat H^{KS}_{\mu\nu}=\left(\hspace{-0.1cm}\begin{array}{cc}
\vspace{0.1cm}
\hat H^{\uparrow\uparrow}_{\mu\nu}&\hat H^{\uparrow\downarrow}_{\mu\nu} \\
\hat H^{\downarrow\uparrow}_{\mu\nu}&\hat H^{\downarrow\downarrow}_{\mu\nu}\end{array}\hspace{-0.1cm}\right)
\end{equation}
where $\mu\nu$ subindexes refer to the \siesta\ basis orbitals. The fully 
relativistic Hamiltonian $\hat H^{KS}$ is expressed as a sum of the
kinetic energy $\hat T$, the scalar-relativistic pseudo-potential part in
the form of Kleinman--Bylander projectors $\hat V^{KB}$, the
spin--orbit $\hat V^{SO}$ term and the
Hartree $\hat V^H$ and exchange--correlation $\hat V^{XC}$ potentials:
\begin{equation}
\hat H^{KS} = \hat T + \hat V^{KB} + \hat V^{SO} + \hat V^H + \hat V^{XC}
\end{equation}
The first three terms of the right hand side do not depend on the charge density, 
$\rho({\bf r})$, and therefore do not change in the self--consistent cycle,
while
$\hat V^{SO}$ and $\hat V^{XC}$ are the only spin--dependent terms
that couple both spin components. 

In order to compute the MCAs, different orientations of the spin
quantization axis need to be considered.  This may be done by rotating
either $\hat V^{SO}$ (as done by \citet{Cuadrado_2012}) or the density
matrix, which is the approach currently followed by \siesta\ for
compatibility with the non--collinear case.

In the current implementation the SO term is included non-perturbatively, 
so that the fully relativistic Hamiltonian is solved self-consistently
after extending the Kohn--Sham wave--functions to full spinors. 
Two different approaches have been implemented in \siesta\ to account for the SO term, $\hat V^{SO}$:
\begin{itemize}
\item [-] {\it on--site} approximation: \\
Based on the work of \citet{Fernandez_Seivane_2006,Fernandez_Seivane_2007}, only the intra--atomic SO contributions
within each $l$--shell of each atom are considered. In this approach the SO
terms are obtained from analytical
simple expressions for the angular integrals while the
radial integrals are computed numerically.
\item [-] {\it off--site} approach: \\
Here, $\hat V^{SO}$ is built following the Hemstreet formalism\cite{Hemstreet_1993,Cuadrado_2012}
whereby a fully-relativistic pseudo-potential (FR-PP) operator is constructed in 
a fully separable form, i.e., non--local in the radial part as well as in the angular variables, in order to substantially reduce the computational cost. 
The necessary $lj$ Kleinman--Bylander projectors may be either 
constructed by \siesta\ itself from relativistic semilocal PPs, or 
directly read from appropriately generated PSML files, as provided by the Pseudo--Dojo
project\cite{dojo_vanSetten_2018,pseudo-dojo-site}. Moreover, we note that the
FR-PP formalism (as well as the original one implemented in Ref.~\citenum{Cuadrado_2012}) uses the correct
normalization constants $C_{l\pm1/2}$, in contrast with what was erroneously
stated in Ref.~\citenum{Zirkelbach_2015}.
\end{itemize}

Although we consider the {\it off--site} approach more accurate, as it includes
inter-shell and inter-atomic SO couplings, both approximations yield very 
similar results in most of the tested systems, with relevant qualitative 
differences only found in a few specific cases. Furthermore,
the construction of the $V^{SO}_{\mu\nu}$ matrix is very fast under both
schemes and involves a tiny fraction of the entire self-consistent calculation.

\subsection{\label{sec:solvers}New electronic-structure solvers}

For most problems, \siesta\ spends the largest fraction of cpu-time in
the solver stage (solution of the generalized eigenvalue problem $H
\Phi = \epsilon S \Phi$). The stage devoted to the calculation of the
hamiltonian H and overlap S is typically much lighter weight, as those
matrices are intrinsically sparse due to the use of a finite-support
basis set. Accordingly, \siesta's performance is almost completely linked to
the use of appropriate external solver libraries.

Over the past few years we have expanded the choices available to
users and refined the relevant interfaces. Initially, we added support
for new individual solvers as detailed below, but recently we have
consolidated some of the most important functionality under a new
common interface to the ELSI library of solvers\cite{elsi_yu_2018,yu2019elsi}.

\subsubsection{\label{sec:stand-alone-solvers} Solvers with a native interface}

Diagonalization (solution of the generalized eigenproblem appropriate
for non-orthogonal orbitals) is the default method for obtaining the
density-matrix in \siesta. A number of standard routines are contained
in the \scalapack\ library~\cite{choi1996scalapack}, but more
efficient alternatives are possible. In particular, the ELPA
library~\cite{elpa_auckenthaler_2011,elpa_marek_2014,Kus2019} uses an
extra intermediate step in the tridiagonal conversion of the matrices
to obtain better scalability and significant speedups over
\scalapack. An interface to ELPA is offered in \siesta, so this solver
can be used as a drop-in replacement for \scalapack\ throughout the
code.

In addition, \siesta\ has implemented interfaces to several methods not
based on diagonalization. In most cases, the use of a finite-support
basis set, leading to the appearance of sparse matrices, is a
significant factor to achieve good performance:

\begin{itemize}
\item The Fermi Operator Expansion method (FOE)~\cite{Goedecker1993}
  uses the formal relationship between Hamiltonian and density-matrix,
  $\hat\rho=f_{FD}(\hat H-\mu)$, where $f_{FD}$ is the Fermi-Dirac
  function. A simple polynomial expansion of $f_{FD}$ can then be used to obtain
  $\hat\rho$ without diagonalization. This method is implemented in
  the CheSS library~\cite{Mohr2017}, developed within the BigDFT
  project~\cite{BigDFT}.

\item The PEXSI method~\cite{pexsi_lin_2013,Lin-14} uses a pole expansion of $f_{FD}$
  to get $\hat\rho$ in the form:
  
\begin{equation}
  \hat\rho = \Im\left(
    \sum_{l=1}^{P}\frac{\omega^{\rho}_l}{H - (z_l+\mu) S}\right)
\end{equation}
where $\omega^{\rho}_l$ and $z_l$ are the weights and poles for the
corresponding expansion of the Fermi-Dirac function. The number of poles needed is
significantly smaller than for the polynomial version of the FOE, as
its dependence on the spectrum size is only logarithmic.

It would appear that having to invert matrices would still render this
approach cubic-scaling, but in fact only \textit{selected} elements of
$\hat\rho$ have to be actually computed. This ``pole expansion and
selected inversion'' method offers a reduced complexity (at most
$\mathcal O(N^2)$ for dense systems, and $\mathcal O(N)$ for
quasi-one-dimensional systems), and trivial parallelization over
poles, so it is well-suited for very large problems on large
machines. For example, \cite{Hu-2014} computed the electronic
structure of large (up to 11,700 atoms) graphene nanoflakes using
\siesta-PEXSI.

\item The electronic structure problem can also be cast as a
  minimization problem (of an extended functional) without orthogonalization.
  When additional localization constraints are put in place, the
  original linear-scaling method in \siesta\ results.  Without the extra
localization constraints, the cubic-scaling Orbital Minimization
Method (OMM)\cite{Corsetti2_2014} can be competitive with respect to
diagonalization, as data can be reused across scf-cycle steps.

\end{itemize}

\subsubsection{\label{sec:elsi} The ELSI interface}

We have considerably extended the range of solver choices and the
performance enhancement possibilities of the code with the integration
of the open-source ELSI library (\url{https://elsi-interchange.org}),
that provides a unified software interface that connects electronic
structure codes to various high-performance solver libraries to solve
or circumvent eigenproblems encountered in electronic structure theory
\cite{elsi_yu_2018}. ELSI also ships with its own tested versions of
the individual solver libraries, but additionally, linking against
already compiled upstream versions from each solver library is
supported as much as possible.

\begin{figure}[h!]                                          
\centering 
\includegraphics[width=0.6\linewidth]{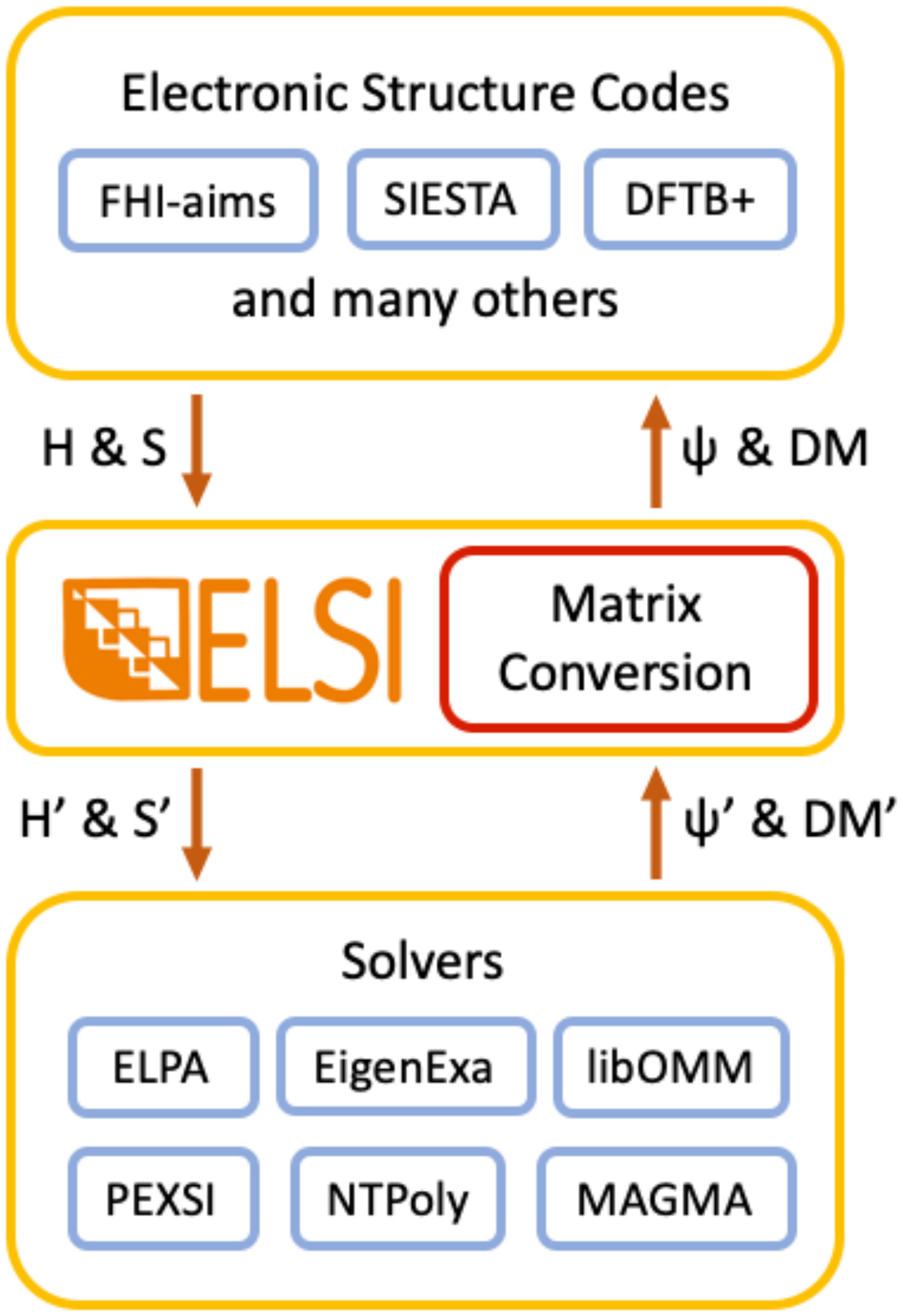}
\caption{Interaction of the ELSI interface with electronic structure
  codes. ELSI serves as a bridge between electronic structure codes
  and solver libraries. An electronic structure code has access to
  various eigensolvers and density matrix solvers via the ELSI
  API. Whenever necessary, ELSI handles the conversion between
  different units, conventions, matrix formats, and programming
  languages.}
\label{fig:elsi}
\end{figure}

The ELPA, OMM, and PEXSI solvers, which had their own ad-hoc
interfaces as described in the previous section, are now available
through ELSI, which also supports other conventional dense
eigensolvers (EigenExa \cite{eigenexa_imamura_2011}, MAGMA
\cite{magma_dongarra_2014}), sparse iterative eigensolvers (SLEPc
\cite{slepc_hernandez_2005}), and linear scaling density matrix
purification methods (NTPoly~\cite{ntpoly_dawson_2018}). As sketched in
Fig.~\ref{fig:elsi}, an electronic structure code interfacing to ELSI
automatically has access to all the eigensolvers and density matrix
solvers supported in ELSI. In addition, the ELSI interface is able to
convert arbitrarily distributed dense and sparse matrices to the
specification expected by the solvers, taking this burden away from
the electronic structure code. A comprehensive review of the
capabilities in the latest version of ELSI, including parallel
solution of problems found in spin-polarized systems (two spin
channels) and periodic systems (multiple \textbf{\textit{k}}-points),
scalable matrix I/O, density matrix extrapolation, iterative
eigensolvers in a reverse communication interface (RCI) framework, has
recently been completed~\cite{yu2019elsi}.

\begin{figure}[htb]
\centering
  \includegraphics[width=0.98\linewidth]{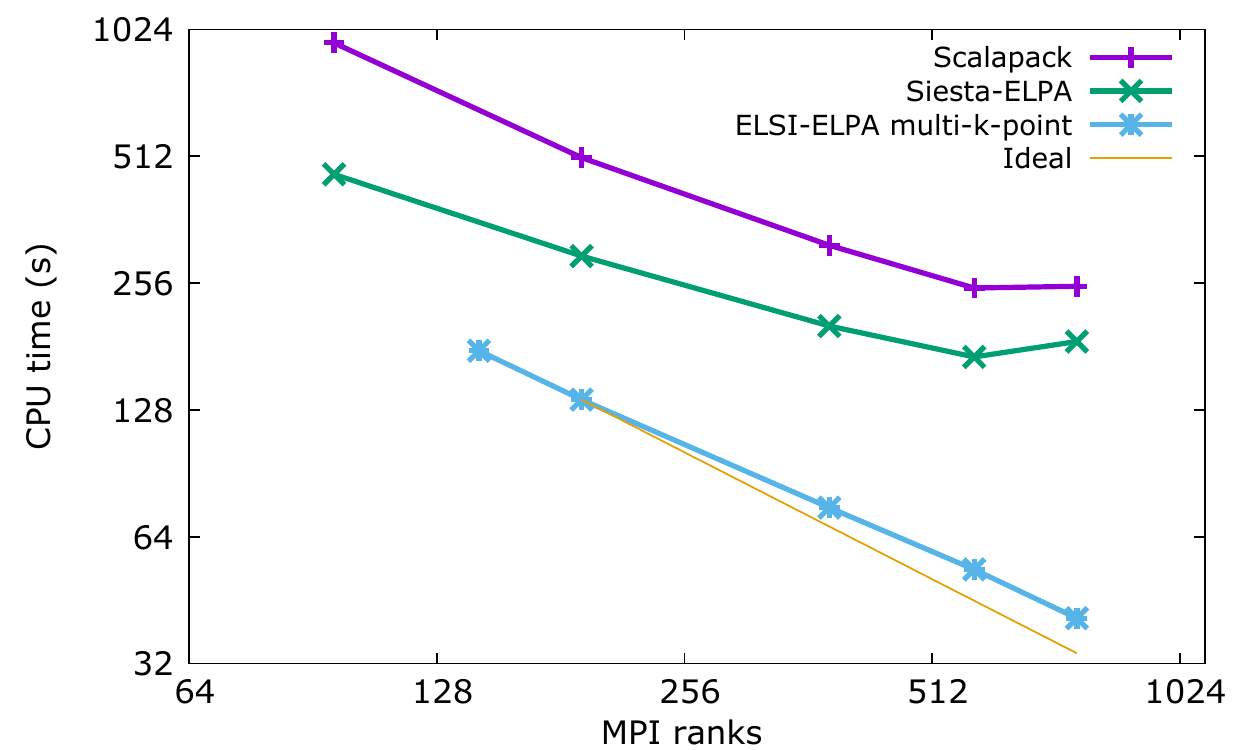}
\caption{\label{fig:elsi-parallel-k} Performance improvement from the
  use of the extra level of parallelization over k-points in \siesta\
  using the ELSI interface with the ELPA solver, compared to the
  previous diagonalization scheme (using both the standard \scalapack\
  solver and the existing ELPA interface in \siesta). The system is
  bulk Si with H impurities, with 1040 atoms, 13328 orbitals, and a
  sampling of 8 k-points. The multi-k scheme is able to stay closer to
  ideal scalability for larger numbers of MPI processes.}
\end{figure}

With the common interface in place, any additions and enhancements to
the supported solvers can be used in \siesta\ with almost no code
changes. This is particularly relevant for performance
enhancements. For example:
\begin{itemize}
\item Further levels of parallelization: A feature common in principle
  to all solvers is that the \siesta-ELSI interface can exploit the
  full parallelization over k-points and spins mentioned above. This
  means that these calculations can use two extra levels of
  parallelization in the solver step beyond the standard one of
  parallelization over orbitals (see Fig.~\ref{fig:elsi-parallel-k}).
\item The new version of the PEXSI solver integrated in ELSI can
  achieve the same level of precision with fewer poles, and offers an
  extra level of parallelization over trial points for the
  determination of the chemical-potential.
\item Mixed-precision support: The ELPA solver can be invoked in
  single-precision mode, which can speed up the initial steps of the
  electronic self-consistent-field (scf) cycle.
\item Accelerator offloading: The ELPA library offers GPU support
  in some kernels~\cite{Kus2019}, and there is scope for extending it
  to more kernels. ELSI also offers an interface to the
  accelerator-enabled MAGMA library. Finally, the PEXSI developers are
  working on adding GPU support to the solver.

\end{itemize}

\subsection{\label{sec:tddft} Time dependent DFT}

  Time-dependent density-functional theory (TD-DFT) was first implemented 
into \siesta\ in its real-time propagating form. 
  It was first described in Ref.~\onlinecite{Tsolakidis2002}, and then
briefly in Ref.~\onlinecite{Artacho-08}.
  It was based on the Crank-Nicolson algorithm, by which, the effect of
the evolution operator for an infinitesimal time step
\begin{equation}
\hat{U}(t_0+\Delta t, t_0)=\exp \left[ -i \hat{H}(t)\Delta t \right ]
\end{equation}
on the wave-function coefficients matrix at a given time $t_0$, $c(t_0)$ is 
approximated by 
\begin{equation}\label{Eq_default0}
c(t_0+\Delta t)= \left [S+iH(t_0 + \Delta t) \frac{\Delta t}{2} \right ]^{-1} 
			   \left [S-iH(t_0) \frac{\Delta t}{2} \right ]
c(t_0)
\end{equation}
where $\Delta t$ represents the finite time-step resulting from time
discretization, and $S$ and $H$ represent the overlap and Hamiltonian
matrices, respectively, in the representation given by a non-orthogonal 
basis set, as used by \siesta.
  That expression is obtained from equating the first-order evolution of the
coefficients forward, from $t_0$ to $t_0 +\Delta t /2$, to the backward evolution
from $t_0 + \Delta t$ to the same intermediate step.

  It can be further simplified to 
\begin{equation}\label{Eq_default}
c(t_0+\Delta t)= \left [S+iH(t_0) \frac{\Delta t}{2} \right ]^{-1} 
			   \left [S-iH(t_0) \frac{\Delta t}{2} \right ]
c(t_0)
\end{equation}
for a smooth-enough variation of the Hamiltonian itself and a small
enough $\Delta t$, thereby avoiding the self-consistency implied in 
propagation using Eq.~\ref{Eq_default0}. In Section \ref{sec:P-TDDFT} below, 
recent developments on efficient treatments of Eq.~\ref{Eq_default0}
beyond Eq.~\ref{Eq_default} are presented. 
  Here we describe the parallelization and related features in the 
TD-DFT implementation now found in standard \siesta\ releases.


  The implemented propagation is based on Eq.~\eqref{Eq_default} (with the 
improvement possibilities described in Section~\ref{sec:P-TDDFT}), but proper 
consideration must be taken of the fact that not only the coefficients change
in time, but also the basis set and the Hilbert space spanned by it when the 
atoms move.
  An analysis of the geometrical implications of this fact is presented in
Ref.~\onlinecite{Artacho-17}.
  The time-dependent Kohn-Sham equation 
\begin{equation}
    H |\psi \rangle = i \partial_t |\psi \rangle
\end{equation}
becomes 
\begin{equation}
    H c = i \, S \, (\partial_t + D) c 
\end{equation}
where $H$, $S$, and $c$ are the Hamiltonian, overlap and coefficients matrices,
respectively, as before, and the $D$ matrix is the connection in the manifold given by 
the evolving Hilbert space\cite{Artacho-17}, $D_{\mu\nu} = \langle \phi_{\mu} | 
\partial_t \phi_{\nu}\rangle$, for $\phi_{\mu}$ and $\phi_{\nu}$ basis functions.

  A way of taking such evolution into account in the discretized implementation 
was proposed by Tomfohr and Sankey \cite{Tomfohr_2001}, and is based on a 
L\"{o}wdin orthonormalization. 
  The scheme consists of two steps. First the wavefunctions are propagated 
using both $S$ and $H$ at time $t_0$ using Eq.~\eqref{Eq_default}, but
to an auxiliary coefficient matrix $\tilde c$,
\begin{equation} \label{eq:ted-step1}
{\tilde c}(t_0+\Delta t)= \left [S+iH(t_0) \frac{\Delta t}{2} \right ]^{-1} 
			   \left [S-iH(t_0) \frac{\Delta t}{2} \right ] c(t_0) \; .
\end{equation}
  Then, the propagation is followed by a change of basis operation 
(only needed if the ionic positions have changed),
\begin{equation}
c(t_0+\Delta t)=  S^{-\frac{1}{2}}(t_0+\Delta t)\, S^{\frac{1}{2}}(t_0) {\tilde c}(t_0+\Delta t).
\label{eq:chb}
\end{equation}

  This algorithm is unitary by construction, and so the preservation of 
orthonormality is guaranteed, regardless of the size of $\Delta t$.
  As discussed in detail in Ref.~\onlinecite{Artacho-17}, this algorithm 
can be shown not to be entirely consistent with the connection represented 
by the $D$ matrix defined above. 
  Nevertheless, the discrepancies due to the mentioned inconsistency have 
been shown to be small in a series of studies using this 
formalism~\cite{Correa_2012,Zeb_2012,Ullah_2015,Halliday_2019}, at least 
for low atomic velocities. 
  The practical benefit of separating the two procedures is to perform the 
change of basis only when necessary, allowing for many electronic steps 
per atomic motion step, if the nuclei are still significantly slower 
than electrons, for instance.
  The implementation of the Crank-Nicolson part is the same for both the 
fixed and moving basis.

  The square root and inverse square root are calculated by first computing 
its eigenvalues and eigenvectors,
\begin{equation} \label{eq:alt-CN}
S = U\, E \, U^\dagger ,
\end{equation}
where $E$ is a diagonal matrix with the eigenvalues of $S$. 
And $U$ is a square matrix with the eigenvectors of $S$ as its columns.
Then,
\begin{equation*} \label{eq:lowdin}
    S^{1/2} = U \, E^{1/2} U^{\dagger}, \quad \mathrm{ and } \quad
    S^{-1/2} = U \, E^{-1/2} U^{\dagger}
\end{equation*}
where $E^{1/2}$ and $E^{-1/2}$ are obtained by replacing diagonal 
elements of $E$ with their square root and inverse square root 
(in the latter case neglecting those eigenvalues below certain threshold value), 
respectively.

  The two-stage algorithm has been implemented in \siesta\ in parallel, 
allowing for $k$-point sampling and for collinear spin. 
  The initial occupied states to be propagated are read from a file.
  \siesta\ is prepared to run a conventional DFT calculation of whatever
relevant initial state, and write a wave-function continuation file
that acts as initialization of the ulterior \siesta\ run in
real-time TD-DFT mode.
  As it stands, \siesta\ evolves states defined as fully occupied; partial
occupations are not currently supported.

\subsubsection{Parallelization}

  The two-step procedure described above requires matrix-matrix and matrix-scalar 
multiplication, matrix addition, and matrix inversion, plus the diagonalization 
of the overlap matrix for the L\"owdin step.
  Since only the occupied states are propagated, the $c$ matrix is rectangular
$N\times {\cal N}$, that is, number of propagating states $\times$ number of
basis functions, while $S$ and $H$ are square, ${\cal N}\times {\cal N}$.
  The computation of the overlap and Hamiltonian matrices is handled by 
pre-existing \siesta\ routines, which are already parallelized and well-optimized 
for HPC environments \cite{Corsetti_2014,Corsetti2_2014}. 

  The parallelization of the propagation following Eqs.~\eqref{eq:ted-step1} and 
\eqref{eq:chb} is done simply exploiting the MatrixSwitch 
library\cite{MatrixSwitch}, which allows  
for an abstracted manipulation of matrices, the details of parallelization, 
data formats, conversions etc. being taken care of underneath.
  In this case, MatrixSwitch is called to use the \blacs\cite{Anderson_1991} and 
\scalapack\cite{Blackford_1997} libraries, 
meaning that this part of the code is run on dense-matrix infrastructure, as
already done with conventional diagonalization solvers. 
  As for the latter, although the $H$ and $S$ matrices are sparse, the $c$ matrix 
is dense.

  Conversion between storage formats is an important consideration here. 
  The native matrix storage format employed by \siesta\ is a compressed sparse column 
(CSC) scheme with a one-dimensional block-cyclic distribution (1D-BCD) over MPI processes. 
   A block-cyclic distribution is needed by \blacs\ and \scalapack\ package.
   The matrices can therefore be temporarily converted from sparse to dense using the 
same parallel distribution; this is a very efficient operation, since no MPI 
communication is necessary.
   It should be noted that a two-dimensional (2D) BCD is known to be more efficient in 
terms of parallel scaling \cite{Corsetti2_2014}.
   The conversion from 1D to 2D does however carry a heavier cost, as MPI communication 
is inevitable.


  The parallel efficiency of our implementation is therefore chiefly 
determined by that of the underlying \scalapack\ drivers. 
  The matrix inversion in Eq.~\eqref{eq:ted-step1} is performed using $LU$ factorization.
  For the diagonalization of the overlap matrix we have implemented the option of 
using either a standard diagonalization approach (tridiagonal reduction followed 
by the implicit QR algorithm) or a divide-and-conquer algorithm as described in 
Ref.\onlinecite{Tisseur_1999}. 
  The latter is known to scale better with system size.

\begin{figure}[ht]
\includegraphics[width=0.97\linewidth]{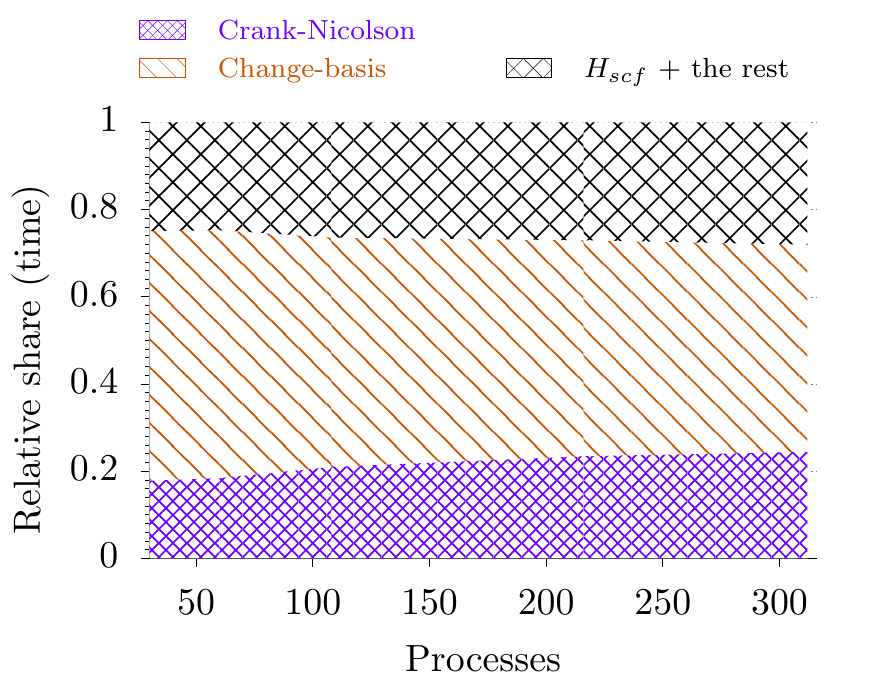}
\caption{The relative share of the total running time for the Crank-Nicolson 
algorithm, the L\"owdin step, and the rest of the program operations 
(including the building of the SCF Hamiltonian) for a system of 5000 Ge 
atoms and one He projectile, using 30-316 processors.}
\label{fig:relative}
\end{figure}

  The scaling with number of processors is very similar to the scaling of a 
conventional DFT \siesta\ run using diagonalization as the solver option, since
both procedures are run on routines of analogous scaling within the same
dense-matrix-algebra library. 
  Figure \ref{fig:relative} shows the relative share in the total running time 
of the three main procedures involved: the Crank-Nicolson algorithm, 
the change of basis, and the calculation of the SCF Hamiltonian plus 
other minor processes in \siesta such as building the density matrix. 
This was performed for a system of 5000 Ge + 1 He atoms described with a single-zeta 
polarized basis set.
  The Crank-Nicolson algorithm takes about 18\% of the total time on 
30 processors, which increases to 25\% on 316 processors.
  Instead, the change of basis procedure takes about 38\% of the total time on 
30 processors, which decreases as the parallelization increases, reflecting 
its better scaling properties.
  The L\"owdin step is the most expensive operation on all numbers of processors, 
which affects TD-DFT simulations (and only those steps) involving atomic motion.

\subsubsection{TD-DFT Beyond the released version}

  There are many possible (and feasible) improvements on what has been described above, some
of them in the pipeline. 
  From a fundamental point of view, the L\"owdin step will be replaced
by another basis-changing step, in the direction of what was proposed in
Ref.~\onlinecite{Artacho-17}.
  It is needed if atoms move at velocities of around 1 atomic unit or more
(1 a.u $\sim c/137$, being $c$ the speed of light).
  In that case the diagonalization step may be avoided (or replaced
by the $N\times N$ diagonalization of the overlap matrix for the evolving
states, instead of the ${\cal N}\times {\cal N}$ for the basis set overlap).

  For fast moving atoms within Ehrenfest dynamics, there is also the need to 
implement correction terms to the forces related to both the change of 
basis and the rotation of the time-dependent Hilbert space (the intrinsic 
curvature of the manifold).
  These terms are well known,\cite{Todorov_2001} and their geometrical
meaning in terms of the relevant curvature\cite{Artacho-17} will
appear in Ref.~\onlinecite{Halliday-20}.
  They are being tested and should be incorporated shortly to a 
visible branch in the open source repository, to be later merged into the
trunk, and further incorporated into \siesta\ releases. 

  For efficiency, iterative inversion options will be explored replacing
the present $LU$ implementation in \scalapack, and quite a few possibilities
exist to incorporate more sophisticated algorithms to the described 
operations. 
  What has been described is robust and quite transparent, but the MatrixSwitch
abstraction should allow easy implementation of other techniques.

\subsubsection{\label{sec:P-TDDFT} Improved real-time propagators}

  Eq.~\eqref{Eq_default} represents a fast approach of electronic propagation 
in real-time TD-DFT, especially suited to study systems where the 
perturbation of the electronic density is relatively small (e.g. optical 
linear response\cite{Tsolakidis2002}).
  If one is interested in simulating systems with heavily perturbed electronic 
densities by external forces (like those exerted by intense laser fields or fast 
atom collisions for instance), one should choose a more elaborate propagation 
scheme that preserves better the time-reversibility of the propagator operator.
  Some of the authors introduced in Ref.\citenum{Ullah_2015} an extrapolation 
algorithm to study the stopping power of prototype semiconductors.
  Briefly, the method uses Eq.~ \eqref{Eq_default} with an extrapolated Hamiltonian:  
\begin{equation}\label{Eq_extrapolation}
c(t_0+\Delta t)= \left [S+iH_{ext} \frac{\Delta t}{2} \right ]^{-1} 
			   \left [S-iH_{ext} \frac{\Delta t}{2} \right ] c(t_0),
\end{equation}
where the extrapolated Hamiltonian $H_{ext}$ reads 
\begin{equation}
H_{ext}=H(t_0)+\frac{1}{2}\Delta H
\end{equation}
and
\begin{equation}
\Delta H=H(t_0)- H(t_0-\Delta t).
\end{equation}
Additionally, the user is given the option to divide each propagation step $\Delta t$ into $n$ sub-steps in an effort to increase the accuracy of the first-order expansion underlying the derivation of Eq.~\eqref{Eq_extrapolation}.  In this case, the final equation reads
\begin{equation}\label{Eq_extrapolation2}
c(t_0+\Delta t)= \prod_{j=1}^n \left [S+iH^j_{ext} \frac{\Delta t}{2n} \right ]^{-1} 
			   \left [S-iH^j_{ext} \frac{\Delta t}{2n} \right ] c(t_0),
\end{equation}
with 
\begin{equation}
H^j_{ext}=H(t_0)+\frac{1}{n}(j-\frac{1}{2})\Delta H
\end{equation}

 Recently we introduced a third algorithm for propagation. Leaving aside in this description the complications associated with the possible subdivision of each time-step, the new algorithm is based on a two-step scheme 
where the electronic wavefunction is first propagated until half of the step, 
$\Delta t/2$, using extrapolation as in Eq.~\eqref{Eq_extrapolation},
\begin{equation}\label{Eq_2step_1}
c(t_0+\frac{\Delta t}{2})= \left [S+iH_{ext} \frac{\Delta t}{4} \right ]^{-1} 
			   \left [S-iH_{ext} \frac{\Delta t}{4} \right ] c(t_0),
\end{equation}
then an explicit calculation of the half-step Hamiltonian, $H(t_0+\Delta t/2)$, is 
performed using the coefficients $c(t_0+\Delta t/2)$ obtained from Eq.~\eqref{Eq_2step_1}.
  In a second step, the coefficients are evolved from the beginning of the step, 
$c(t_{0})$, to the full step, $c(t_{0}+\Delta t)$, using the half-step Hamiltonian:
\begin{equation}\label{Eq_2step_2}
c(t_0+\Delta t)= \left [S+iH(t_0+\frac{\Delta t}{2}) \frac{\Delta t}{2} \right ]^{-1} 
			   \left [S-i H(t_0+\frac{\Delta t}{2}) \frac{\Delta t}{2} \right ] c(t_0)
\end{equation}
  This approach, although increasing the CPU time by around $\sim$35\% as compared to the 
two previous schemes, allows for better energy conservation for highly 
perturbed systems where the Kohn-Sham potential heavily varies in time. 

\begin{figure}[!htb]
\centering
\includegraphics[width=0.98\linewidth]{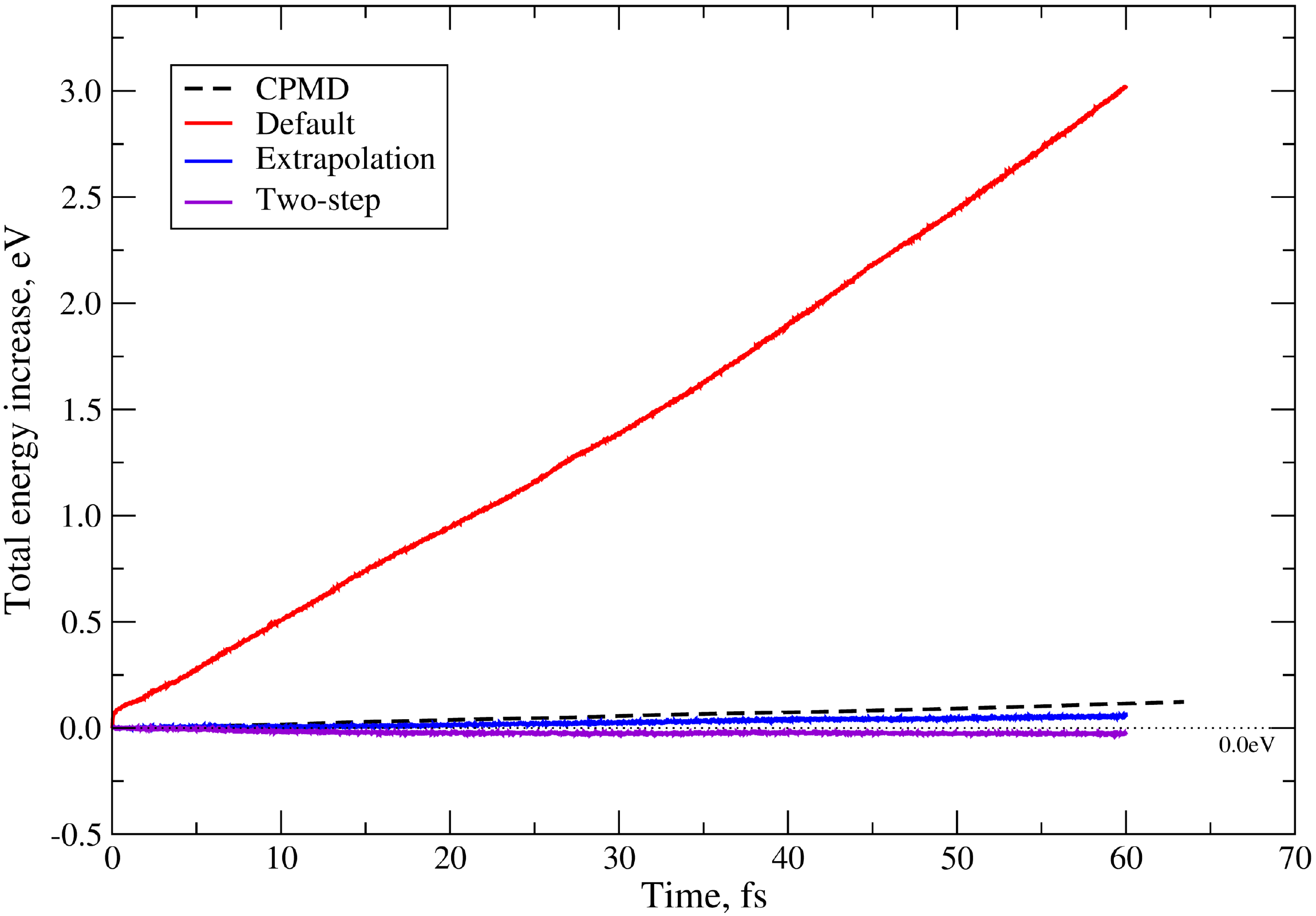}
\caption{Energy drift in an energy conserving TDDFT simulation of ionized-core uracil 
for the three propagation methods described here, as compared with the CPMD 
implementation, for a time step of $\Delta t = 0.24$ attoseconds (0.01 atomic units).}
\label{Scheme2}
\end{figure}

\begin{figure}[htb]
 \centering
\includegraphics[width=0.98\linewidth]{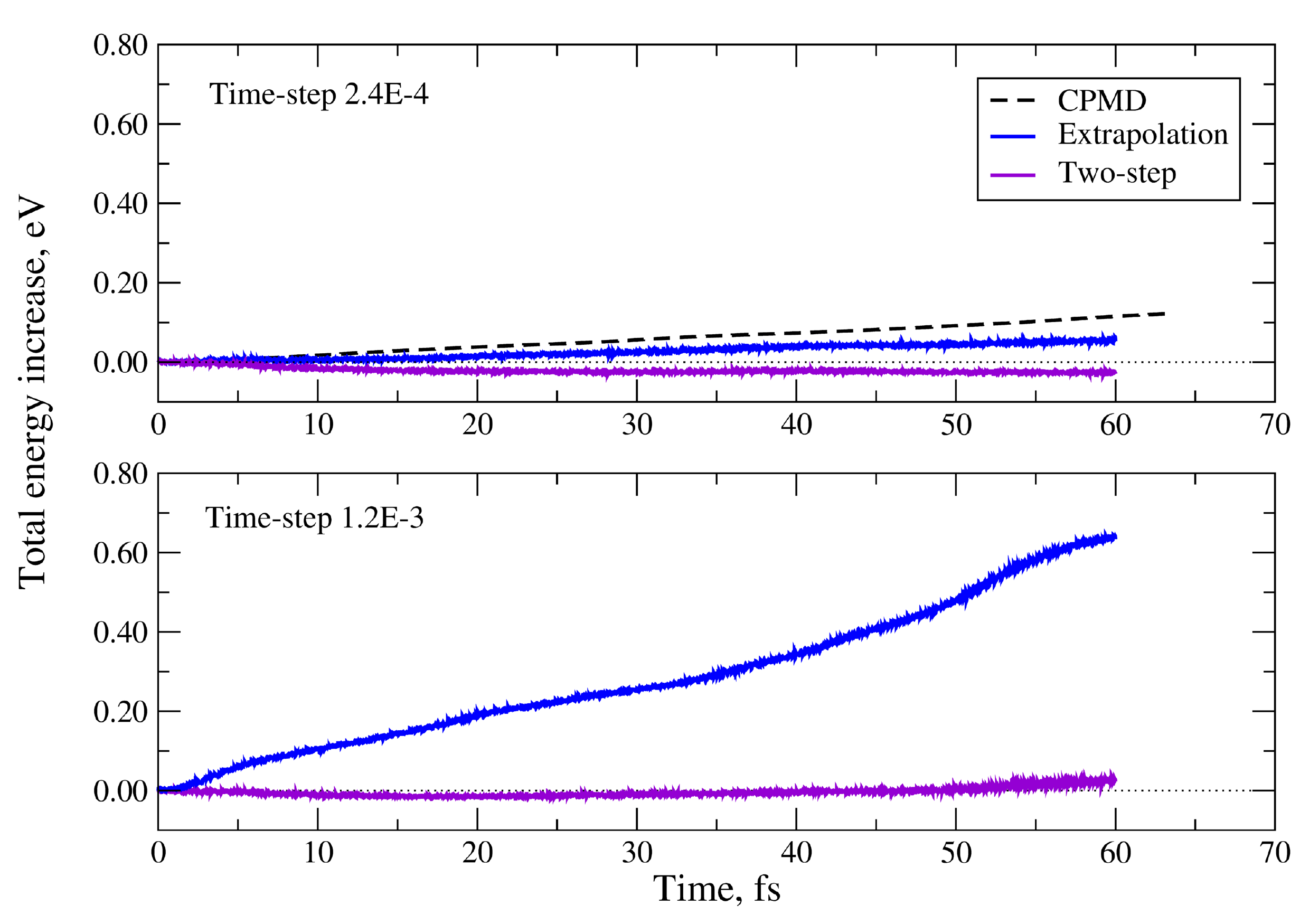}
\caption{Energy drift comparison as in Fig.~\ref{Scheme2}, for time steps 
$\Delta t = 0.24$ and 1.2 attoseconds.}
\label{Scheme1}
\end{figure}

  In order to provide a more quantitative comparison between the three schemes, 
namely, the default propagation of Eq.~\eqref{Eq_default}, the extrapolation 
propagation of Eq.~\eqref{Eq_extrapolation} and the two-step propagation of 
Eqs.~\eqref{Eq_2step_1} and \eqref{Eq_2step_2}, we compare their performance versus 
the P-TDDFT implementation\cite{Tavernelli05} of the CPMD code\cite{CPMD} in the 
case of a double ionization of a uracil molecule in the gas 
phase\cite{LopezTarifa2011, LopezTarifa2014}.
  Simulations of this type provide access to the ultrafast electronic dynamics that 
occurs at the atto and femto time-scales in the ionized genetic material (DNA and RNA) 
as a consequence of collisions with proton or carbon beams\cite{Gaigeot2010}.
  This particular simulation addresses the fragmentation pattern of a doubly-ionized uracil 
molecule (its deepest Kohn-Sham orbital is empty) using the BLYP density 
functional\cite{Becke88,Lee88}.
  The technical details for the CPMD simulation used as a reference can be found in 
Refs.~\onlinecite{LopezTarifa2011, LopezTarifa2014}.
  \siesta\ calculations using the same density functional and the integrators described above use a DZP basis set. As can be seen in Fig.~\ref{Scheme2} the standard \siesta\ implementation cannot properly deal with such a highly excited system. The extrapolation scheme in Eq.~\eqref{Eq_extrapolation} already provides a large improvement and gives an energy conservation similar to the CPMD simulations in Refs.~\onlinecite{LopezTarifa2011, LopezTarifa2014}. Finally, the two-step algorithm further improves the energy conservation. For smaller time steps the improvements given by the two-step scheme are even more clear, as shown in Fig.~\ref{Scheme1}.

\subsubsection{Electronic stopping of atomic projectiles}

  Let us finish the TD-DFT section with a brief mention of its successful application
to the problem of simulating the excitation of the electrons of a condensed matter
system when traversed by a high-energy atomic projectile (so-called electronic
stopping, since the electrons slow down the projectile). 
  This physical problem is very relevant to questions of interest to the nuclear
and aerospace industries, as well as to the treatment of cancer. 
  In spite of its great relevance and of its having been researched since Rutherford's
experiment in 1911, the understanding of electronic stopping processes has been essentially 
limited to either weak effects in the linear-response regime or beyond linear but only for 
target systems close to the homogeneous electron liquid (jellium). 

\begin{figure}[htb]
\includegraphics[width=0.95\linewidth]{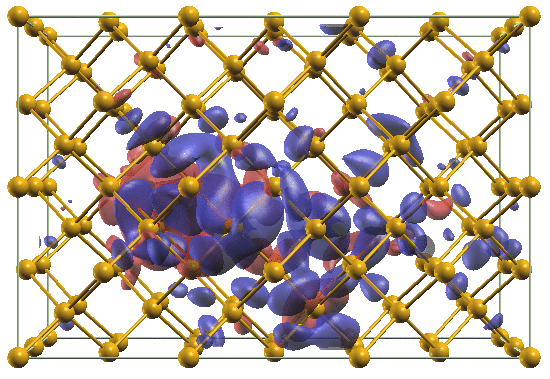}
\caption{Electron deformation density isosurfaces (blue positive, red negative)
for a proton displacing leftwards, at a velocity of 1 a.u. in the bulk of a Ge crystal.}
\label{HinGe}
\end{figure}

  An earlier version of the TD-DFT implementation in \siesta\cite{Tsolakidis2002},
allowed the first explicit first-principles simulation of electronic stopping, for
protons and antiprotons in LiF, a wide-band-gap insulator, which was quite 
successful.\cite{Pruneda-07}
  The difference of sign between protons and antiprotons produces a significant difference
in the stopping power (rate of energy excitation) beyond the linear-response paradigm
(the Barkas effect), and the insulating character of the target makes it inaccessible
to the jellium paradigm. 
  The success stimulated further studies along this line\cite{Correa_2012,Zeb_2012,
Ullah_2015, Halliday_2019} using improved versions of TD-DFT in \siesta, as described here.
  Fig.~\ref{HinGe} displays the electron deformation density around a proton displacing
in a bulk Ge target.\cite{Ullah_2015}
  They were also followed by analogous simulations using plane-wave codes by an increasing
number of groups (for a review see Ref.~\onlinecite{CorreaReview_2018}), 
although the latter calculations do demand considerably larger computational resources. 

\subsection{\label{sec:dfpt}Density Functional Perturbation Theory}

\begin{figure}[htb]
\centering
  \includegraphics[width=0.98\linewidth]{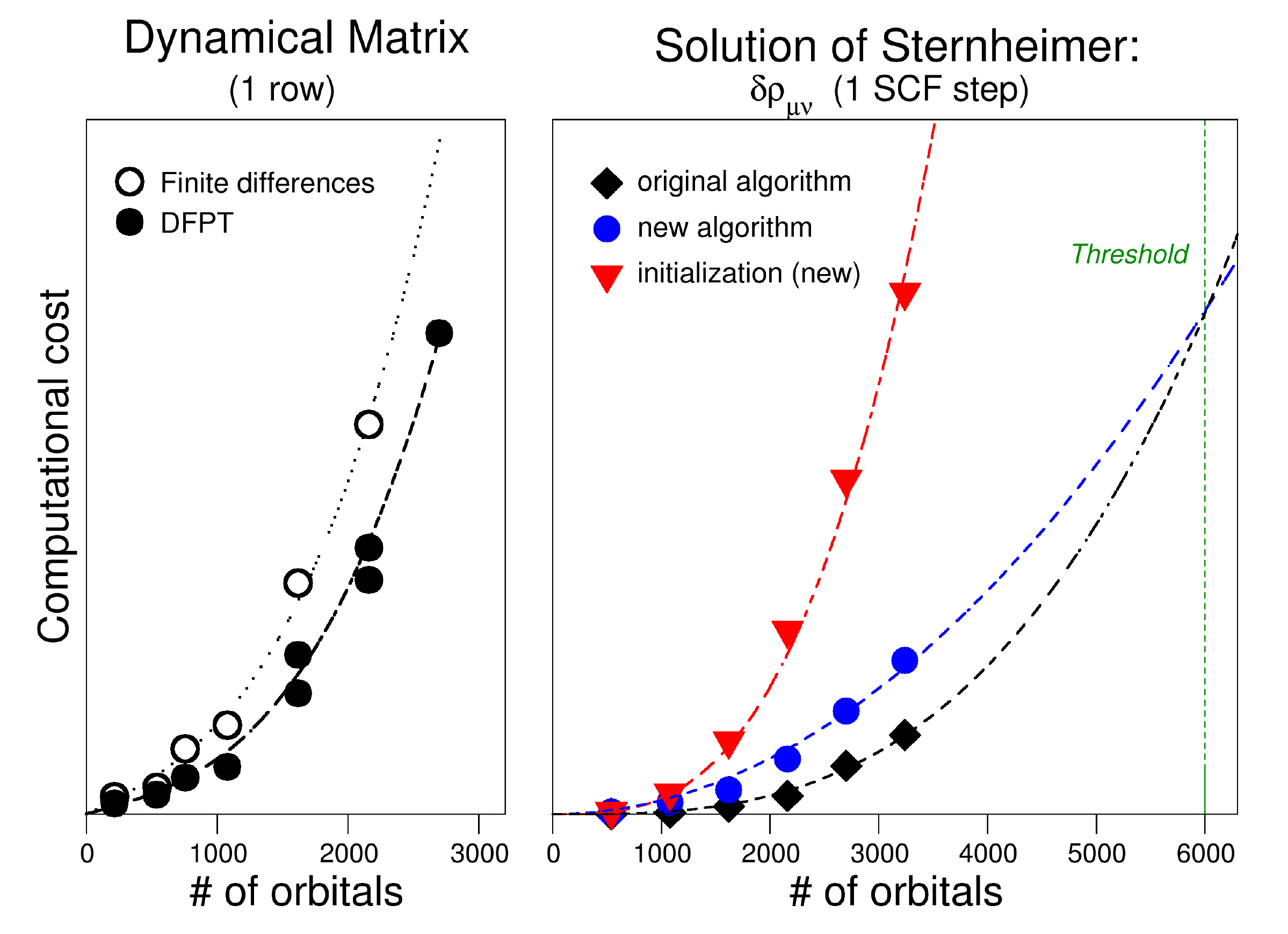}
\caption{\label{fig:dfpt} (Left panel): Comparison of the performance of the new DFPT approach with the conventional finite-differences method in \siesta. The time required to compute a whole row of the dynamical matrix (derivative of the forces on all atoms when one atom is displaced in $x,y,z$ directions) is plotted as a function of the number of orbitals in carbon fullerenes of different sizes. (Right panel) Performance of the alternative algorithm described in the text (based on Eq.~\eqref{derC2}, blue circles), as compared to the original implementation based on Eq.~\eqref{derC} (black diamonds). The initialization (computing $\Xi$ and $\Omega$) is the most time-consuming step, although it has to be performed only once, and can be used for all perturbations (each atomic displacement).  The new algorithm becomes more efficient for system sizes larger than the threshold value (green dashed line).}. 
\end{figure}

The original implementation of Density Functional Perturbation Theory, as a post-processing and independent code ({\sc linres}\cite{PrunedaPRB2002}) has been recently merged into \siesta. It allows to compute the phonon dispersions using a supercell approach ($\Gamma$-point phonons). Both LDA and GGA functionals can be used (through LibXC). Calculation of the perturbed Hamiltonian and overlap matrix elements follows the same methodology as for ground-state calculations, with similar computational costs, which are comparable to those obtained with a finite difference approach. Figure \ref{fig:dfpt} shows a comparison between both methods for model fullerene-type systems of different sizes.

The solution of the Sternheimer equation, and calculation of the perturbed density matrix is the most demanding step. It requires the perturbed coefficients of the electronic wavefunctions to be obtained,

\begin{equation}
\label{derC}
\partial c_{i\mu} = \sum_j \sum_{\alpha \beta} c_{j\alpha}^* \left[ \frac{  \partial H_{\alpha \beta} - \varepsilon_i \partial S_{\alpha \beta}}{\varepsilon_i -\varepsilon_j}   \right] c_{i\beta} c_{j\mu} = 
\sum_j A_{ij} c_{j\mu}
\end{equation}  

\noindent where $A_{ij}=\sum_{\alpha \beta}
\frac{c_{j\alpha}^*\Delta_{\alpha\beta}^ic_{i\beta}}{\varepsilon_i
  -\varepsilon_j}$ and  $\Delta_{\alpha\beta}^{i}=\left[\partial
  H_{\alpha \beta} - \varepsilon_i \partial S_{\alpha \beta}
  \right]$. The change in the density matrix is then given by

\begin{eqnarray}
\label{drho}
\nonumber
\partial\rho_{\mu\nu} &=& \sum\limits_{i}^{all}\left[ n_i \partial c_{i\mu}^*c_{i\nu} + n_i c_{i\mu}^*\partial c_{i\nu} + \partial n_i c_{i\mu}^*c_{i\nu}  \right] = \\
\nonumber
&=& \sum\limits_{i}^{all} n_i \sum\limits_{j}^{all} \left[ A_{ij}^* c_{j\mu}^*c_{i\nu}+c_{i\mu}^*A_{ij}c_{j\nu} \right] + \sum\limits_{i}^{all}\partial n_ic_{i\mu}^*c_{i\nu} = \\
\nonumber
&=& \sum\limits_{i}^{all} n_i \sum\limits_{j}^{unocc} \left[ A_{ij}^* c_{j\mu}^*c_{i\nu}+c_{i\mu}^*A_{ij}c_{j\nu} \right] 
\\
& -& 
\sum\limits_{ij}^{occ} n_j c_{i\mu}^*c_{j\nu} \sum_{\alpha\beta}c_{j\alpha}^*\partial S_{\alpha\beta} c_{i\beta} 
+ \sum\limits_{i}^{all}\partial n_ic_{i\mu}^*c_{i\nu}
\end{eqnarray}

\noindent and a similar expression applies to the change in the energy-density matrix.

The change in the occupation of the electronic state can be computed from the change in its eigenenergy $\varepsilon_i=\sum_{\alpha\beta}c_{i\alpha}\Delta_{\alpha\beta}^ic_{i\beta}$ and it is relevant in metals, for states close to the Fermi level. The Fermi level can also be shifted by the perturbation, and it can be determined through conservation of the number of electrons in the system, $N_e$.

Obtaining $\partial\rho_{\mu\nu}$ is the most computationally expensive part of the code. While the computation of $\partial H_{\mu\nu}$ basically has linear scaling with the system size, the matrix $A_{ij}$ scales as $N_b^2\cdot M$, where $N_b$ is the number of basis functions, and $M$ is the maximum number of neighbour orbitals for any orbital in the system. Equation \eqref{derC} then requires $N_b^3$ operations for each atomic perturbation, and the change in the density matrix requires $N_b^2\cdot M$ loops.  An alternative approach that offers a better computational scaling for systems with a gap has also been tested. If we define
 $\Xi_{\alpha\beta}^{i}=\sum_j\frac{c_{j\alpha}^*c_{j\beta}}{\varepsilon_i-\varepsilon_j}$, we obtain:
 
\begin{equation}
\label{derC2}
\partial c_{i\mu} = \sum_\beta \left[\sum_{\alpha} \Xi_{\alpha\mu}^i \Delta_{\alpha\beta}^i\right] c_{i\beta}  = 
\sum_\beta \Lambda_{\mu\beta}^i c_{i\beta}
\end{equation}  
where $\Lambda_{\mu\nu}^i=\sum_\eta \Xi_{\eta\mu}^i\Delta_{\eta\nu}^i$ is a smooth function of $\varepsilon_i$ and can be described by a Chebyshev's expansion with a few selected energy points and their corresponding weights: 
\begin{equation}
\nonumber
\partial c_{i\mu} = \sum_l \tilde\omega_{l,i} 
\sum_\beta \Lambda_{\mu\beta}^{(l)} c_{i\beta}= 
\sum_l \tilde\omega_{l,i} 
\sum_{\alpha\beta} \Xi_{\alpha\mu}^{(l)} \Delta_{\alpha\beta}^{(l)} c_{i\beta} 
\end{equation}

Notice that $\Xi_{\alpha\beta}^{(l)}$ is perturbation-independent, and could be computed only once and used for all the possible atomic displacements, with a cost that scales as $N_b^2\cdot N_l$, with $N_l$ being the number of Chebyshev' polynomials. The change in the electronic density is then given by
\begin{align}
\nonumber
+\partial \rho_{\mu\nu} \sim &\sum_i  c_{i\mu}^*\partial c_{i\nu}= \sum_{l\eta}\Lambda_{\nu\eta}^{(l)}\sum_i  c_{i\mu}^* \tilde\omega_{l,i} c_{i\eta}=\\
\nonumber
&\sum_{l\eta}\Lambda_{\nu\eta}^{(l)}\Omega_{\eta\mu}^{(l)}=
\sum_{l\eta\gamma}\Xi_{\gamma\nu}^{(l)}\Delta_{\gamma\eta}^{(l)}\Omega_{\eta\mu}^{(l)}
\end{align}
where only the central term requires self-consistency, and $\Omega_{\alpha\beta}^{(i)}=\sum_jc_{j\alpha}^* \tilde\omega_{i,j} c_{j\beta}$. Although computing the change in the density scales as $\mathcal{O}(N_b^2M)$, most of the computational cost is required in an initialization step to obtain $\Xi$ and $\Omega$ that are perturbation-independent, enabling the extraction of the whole dynamical matrix with $\mathcal{O}(N_b^3)$ operations. A preliminary serial calculation for C$_n$ fullerenes shows that the threshold system size for the new algorithm to become more efficient than the original implementation lies at around 650 atoms. This value can be conveniently reduced by an efficient parallelization of the initialization step.

\subsection{\label{sec:transiesta}\tsiesta}

\begingroup
\newcommand\G{\mathbf{G}}
\newcommand\gS{\mathbf{g}_S}
\newcommand\Gr{\mathbf{G}^r}
\newcommand\Ga{\mathbf{G}^a}
\newcommand\Dyn{\mathbf{D}}
\newcommand\HH{\mathbf{H}}
\newcommand\VV{\mathbf{V}}
\newcommand\EDM{\boldsymbol{\mathcal{E}}}
\newcommand\DM{\boldsymbol{\rho}}
\newcommand\DE{\DM_\Eq}
\newcommand\DN{\DM_\Neq}
\newcommand\Spec{\mathcal{A}}
\newcommand\dev{\mathcal{D}}
\newcommand\SO{\mathbf{S}}
\newcommand\ncor{\boldsymbol{\Delta}}
\newcommand\SE{\boldsymbol{\Sigma}}
\newcommand\Scat{\boldsymbol{\Gamma}}
\newcommand\rr{\mathbf{r}}
\newcommand\RR{\mathbf{R}}
\newcommand\kk{\mathbf{k}}
\newcommand\qq{\mathbf{q}}
\newcommand\kT{k_BT}
\newcommand\sumE{\sum}
\newcommand\NE{\mathfrak{E}}
\newcommand\idxE{\mathfrak{e}}
\newcommand\sumU{\sum^{\NU}}
\newcommand\NU{N_\mu}
\newcommand\idxU{\mu}
\newcommand\cd{\!\dd}
\newcommand\dd{\mathrm{d}}
\newcommand\E{\epsilon}
\newcommand\eig{\varepsilon}
\newcommand\TT{\mathcal{T}}
\newcommand\RE{\mathcal{R}}
\newcommand\wX{\widetilde{\mathbf{X}}}
\newcommand\wY{\widetilde{\mathbf{Y}}}

\newcommand\ID{\mathbf{I}}

\newcommand\BZ{\mathrm{BZ}}
\newcommand\Eq{\mathrm{eq}}
\newcommand\Neq{\mathrm{neq}}
\newcommand\varneq{\boldsymbol{\theta}}
\newcommand\bk[2]{\langle#1|#2\rangle}

\newcommand\Csix{\ensuremath{\mathrm{C}_{60}}}
\newcommand\HOMO{\mathrm{H}}
\newcommand\LUMO{\mathrm{L}}

\newcommand{\drho}{\delta\rho}
\newcommand{\dV}{\delta V}

\newcommand\dEBZ{\!\!\!}

The transport code \tsiesta, initially developed by Brandbyge and co-workers\cite{Brandbyge2002}, enables open-boundary condition calculations by extending periodic regions with bulk electrodes. It is based on the non-equilibrium Green function formalism which allows biased calculations.
\tsiesta\ has been completely re-written and now uses advanced inversion algorithms, enables $N_\idxE\ge1$ electrodes, allows thermo-electric calculations, performing real-space calculations (without $\kk$-points) and adds phonon transport calculations using the Hessian\cite{Papior-17,Papior2019},

The non-equilibrium Green function formalism can be summarized in the following equations which are generalized for $N_\idxE\ge1$ electrodes.
\begin{align}
  \label{eq:neq:elec}
  \DM &= \DE^\idxE+\sumE_{\idxE'\neq\idxE}\ncor_{\idxE'}^\idxE\equiv \DN^\idxE ,\\
  \label{eq:neq:elec1}
  \DE^\idxE &\equiv  \frac i{2\pi}\iint_\BZ\dEBZ\cd \kk\dd\E%
  \left[\G_\kk(\varepsilon)   - \G^\dagger_\kk(\varepsilon)  \right]n_{F,\idxE}(\E)e^{-i\kk\cdot\RR}, \\
  \label{eq:neq:elec2}
  \ncor_{\idxE'}^\idxE &\equiv
  \frac1{2\pi}\iint_\BZ\dEBZ\cd \kk\dd\E\, 
  \Spec_{\idxE',\kk} (\varepsilon) e^{-i\kk\cdot\RR}\big[n_{F,\idxE'}(\E)-n_{F,\idxE}(\E)\big],
\end{align}
where $\DE^\idxE$ is the equilibrium density matrix for electrode $\idxE$,
$G$ is the retarded Green's function matrix, and $\ncor^\idxE_{\idxE'}$ is the correction to the equilibrium part. The spectral function $\Spec_{\idxE}=\G\Scat_\idxE\G^\dagger$ and carries electrons from the electrode $\idxE$. Finally, $n_{F,\idxE}$ is the Fermi function with chemical potential denoted by the electrode $\idxE$. It is evident that the Fermi functions depend on the chemical potential \emph{and} the electronic temperature in the associated electrodes. By using different temperatures for each electrode one can calculate thermoelectric effects due to different reservoirs having separate electronic temperatures self-consistently.

We note that \tsiesta\ uses a multiple complex energy-contour algorithm to more accurately describe the total density $\DM$. It does this by weighing each $\DN^\idxE$ using a simple scheme\cite[Sec.~3.2]{Papior-17}. So far, few multi-electrode calculations have been performed so the importance of the multiple contour algorithm is currently unknown\cite{Jacobsen2016,Kolmer2019,Brandimarte2017}. However, for the well-known 2-electrode problem it allows smoother convergence properties\cite{Brandbyge2002}.

In the latest \tsiesta\ we implement 3 different inversion algorithms; \emph{i}) a block-tri-diagonal algorithm (BTD), \emph{ii}) MUMPS sparse algorithm and \emph{iii}) a dense algorithm (LAPACK). The performance of these (speedup compared to \tsiesta\ 3.1) is summarized in Fig.~\ref{fig:ts-perf}. Since the BTD algorithm is linear scaling for constant width it can easily outperform the dense algorithm by a factor $100$. 
This performance gain is also important for the memory footprint which enables even larger systems. The BTD algorithm favors long and narrow systems, but uses less memory for all types of systems.

\begin{figure}[htb]
  \centering
  \includegraphics[width=.98\linewidth]{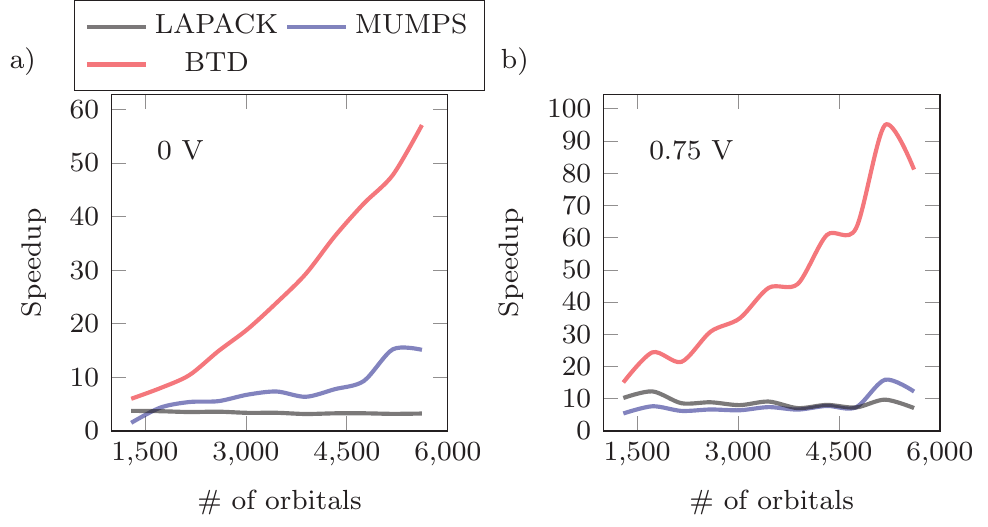}
  \caption{Performance characterization of \tsiesta\ using a pristine graphene
      cell (24 atoms wide). Speedup for (a) EGF and (b) NEGF calculations of pristine graphene 
      compared against the dense implementation. The BTD method exhibits more than
      40 times the speed of the LAPACK implementation for the largest size. MUMPS gains
      speed after $5,000$ orbitals.
      \label{fig:ts-perf}
  }
\end{figure}

A recent addition to the \tsiesta\ package is the use of real space self-energy terms\cite{zerothi_sisl,Papior2019}. These self-energies are semi-infinite in more than 1 direction and can thus be used as surrounding electrodes, for e.g. single defects in 2D or 3D structures or line defects. Real space self-energies are superior to $\BZ$ integrated quantities since they correctly describe the infinite bulk by leaving out image couplings and also removes the need for $\kk$-point sampling. When taking into account the complete procedure for a \tsiesta\ calculation the real space self-energies provide an increased throughput since the SCF $\kk$-point sampling and the subsequent $\kk$-point sampled transport calculation are completely removed\cite{Papior2019}.

Additionally, \tbtrans\ enables calculations of user defined tight-binding models and also interfaces to phonon transport using the Hessian matrix (program named \phtrans). The phonon Green function is similar to the electron
\begin{equation}
    \G_\qq(\omega)=\big[(\omega^2+i\eta^2)\mathbf I -
  \Dyn_\qq-\SE_\qq(\omega)\big]^{-1},
\end{equation}
where $\Dyn$ is the Hessian and $\omega$ the phonon frequency.
Finally, inelastic transport involving phonon-excitation can be treated with perturbation theory in a postprocessing step with the {\sc Inelastica} package\cite{inelastica,Frederiksen-07}.

\endgroup

\subsection{\label{sec:wannier}Wannierization}

The interface between \siesta\ and {\sc wannier90}~\cite{wannier90-site,Pizzi-20} (version 3.0.0) has been 
implemented, so the latter code can be called directly from \siesta\ as a library,
or used as a post-processing tool.
{\sc wannier90} is an open-source code for generating maximally-localized Wannier
functions~\cite{Marzari-97,Marzari-12} and using them to compute advanced electronic
properties of materials with high efficiency and accuracy.

\begin{figure}[htb]
\centering
  \includegraphics[width=0.7\linewidth]{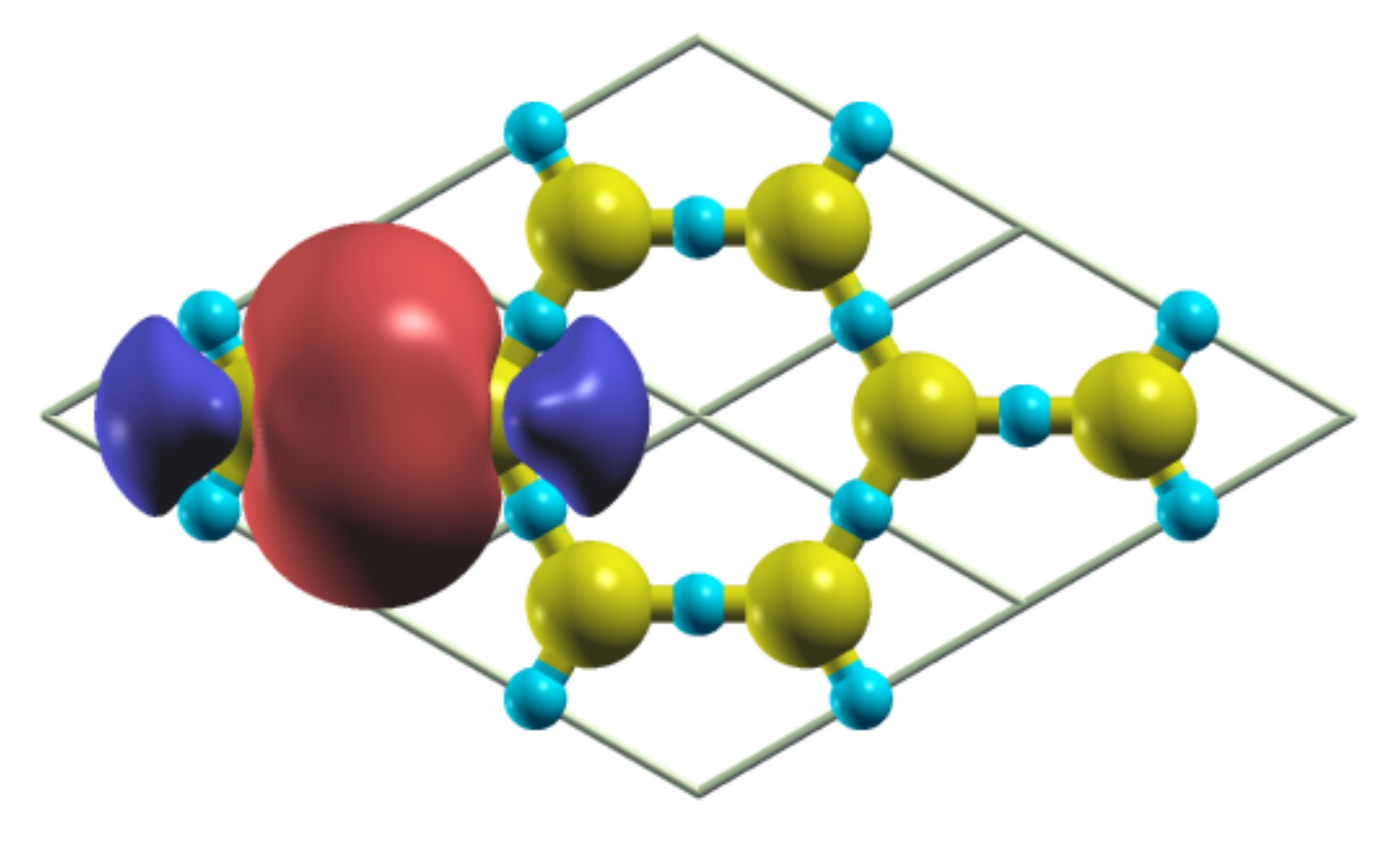} \\
  \includegraphics[width=0.7\linewidth]{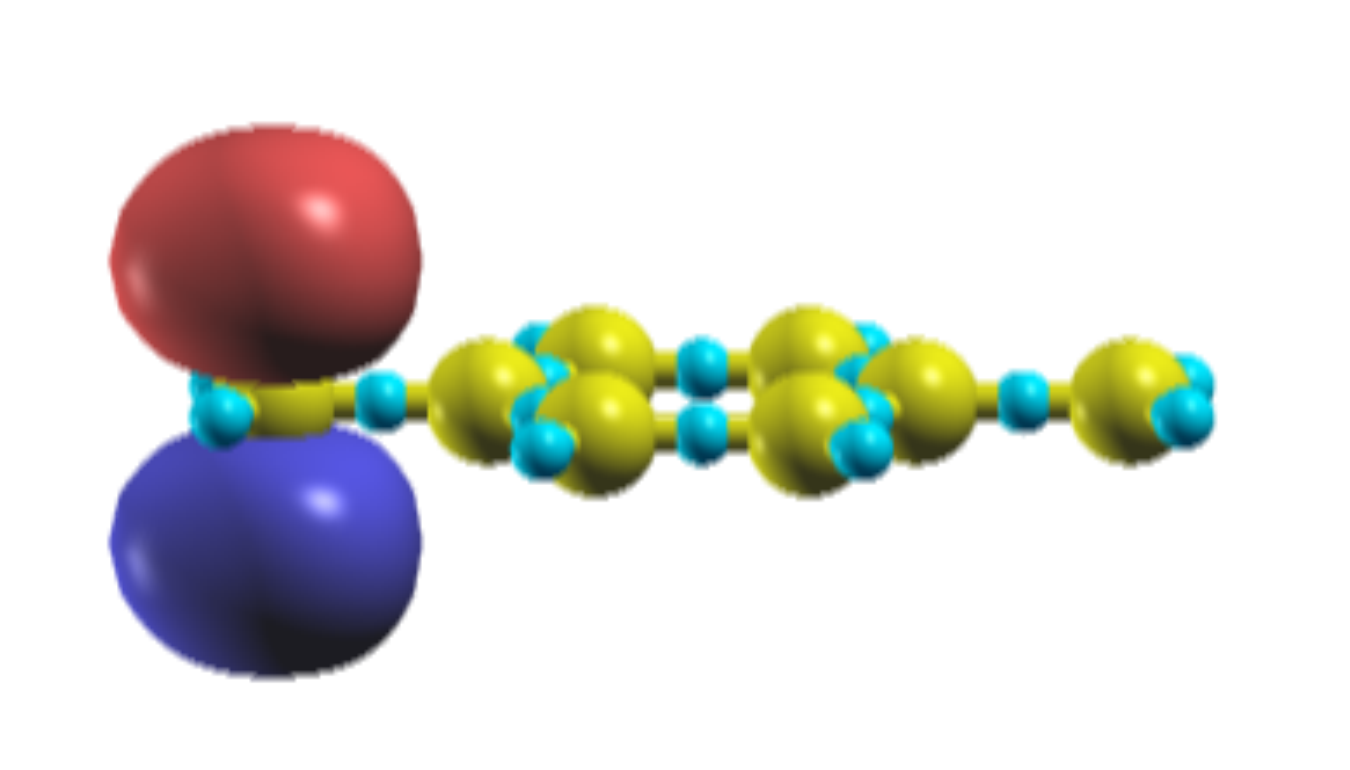}
\caption{\label{fig:wannier} Maximally localized Wannier functions (MLWFs) for graphene. 
         Top panel displays the character of
         $\sigma$-bonded combinations of $sp^{2}$ hybrids.
         Bottom panel displays the $\pi$ character of the
         bands with weight on the $p_{z}$  orbitals. 
         Isosurfaces of different colors correspond to two
         opposite values for the amplitudes of the 
         real-valued MLWFs. Yellow spheres represent the position of the 
         C atoms, while smaller blue spheres mark the center of the bonding.
        }
\end{figure}

The Wannier functions can be considered as a unitary transformation (more precisely,
a Fourier transformation) of a set of Bloch functions associated with a given
manifold of bands. 
We can view the Bloch and Wannier functions as providing two different
basis sets describing the same manifold of states associated with the electron
band manifold in question.
The Wannier functions display a number of very interesting
properties.~\cite{Vanderbilt-book}
Among them, we can enumerate:
$(i)$ they are localized in real space, each of them concentrated around a given unit cell (see Fig.~\ref{fig:wannier}); 
$(ii)$ Wannier functions centered on different cells are translational images
of one another;
$(iii)$ they form an orthonormal basis set;
$(iv)$ they span the same subspace of the Hilbert space as is spanned by the
Bloch functions from which they are constructed.
Because of the gauge freedom in the definition of the
phases of the Bloch functions, the Wannier functions are not unique. However, the location of their centers in the home unit cell
is unique to within a lattice vector, i.e. they are gauge invariant.~\cite{Vanderbilt-book}
The high degree of arbitrariness in the definition
of the phases can be exploited to produce unitary transformation matrices between Bloch and Wannier functions in such a way that a localization functional
that measures the sum of the quadratic spreads of the Wannier functions in the home unit cell around their centers is minimized.~\cite{Marzari-97}
In a practical procedure to construct Wannier functions, a set of localized functions is used to generate an initial guess for the unitary transformations.
These localized functions should be roughly located on sites where Wannier functions are expected to be centered and have appropriate angular character.
In our implementation, we can directly use the localized atomic orbitals of the basis, or the hydrogenoid localized functions (including hybrid orbitals), as suggested in {\sc wannier90}.

The Wannier functions provide an exact tight-binding representation of
the dispersion of the Bloch bands.
This property will be exploited to
extract in an automatic and user blind way the parameters required to
run multiscale simulations as described in Sec.~\ref{sec:multiscale}.

Currently, {\sc wannier90} can be used as a post-processing tool,
or it can be directly called from \siesta\ in a library mode.
Within this last approach, the unitary matrices that transform
the Bloch states into Wannier functions are directly accessible 
in \siesta,
allowing a clear and straightforward interconnection between the two 
alternatives to span the Hilbert space.
Besides, the use of Wannier functions opens the door to a wide range of
potential applications.
Already implemented in \siesta\ is the possibility of performing SCF convergence under the constraint of a rigid shift on the energy associated with a given Wannier function to be used to calculate electron-electron interactions for multiscale simulations as detailed in Sec.~\ref{sec:multiscale}.
The interface with the self-consistent dynamical mean field
theory {\sc DMFTwDFT} code~\cite{DMFTwDFT} using MLWF has been already implemented~\cite{Singh-20}.
Also, alternative approaches to compute the exact Hartree-Fock exchange
in extended insulating systems with a linear scaling computational cost using
MLWFs have been proposed, being another interesting research line for the
future.~\cite{Wu-09}

\subsection{\label{sec:multiscale}Multiscale methods}

Density Functional Theory can be used as the basis for parameterized multiscale 
 methods, that can be used to carry out simulations including 
tens or even hundreds of thousands of atoms.~\cite{Garcia-Fernandez-16} 
First-principles methods are used to produce detailed models that are subsequently used
to predict properties that require large-scale simulations. The models are created for
specific materials and their accuracy can be systematically improved to converge towards
DFT precision. 
Given the dependence on first-principles, we refer to these methods as \emph{second-principles DFT} (SPDFT)
and are run on an independent code called \textsc{Scale-Up}.~\cite{Garcia-Fernandez-16}

SPDFT is based on a division of the total electronic density, $n(\vec{r})$, into a reference ($n_0(\vec{r})$)
and a deformation ($\delta n(\vec{r})$) contributions,
   \begin{equation}
      n(\vec{r})=n_0(\vec{r})+\delta n(\vec{r}),
      \label{eq:dens_div}
   \end{equation}
where $\delta n(\vec{r})$ is considered as a small perturbation with respect to $n_0(\vec{r})$ that,
in non-magnetic cases, represents the ground state of the system\cite{Garcia-Fernandez-16}.
   This division is then used\cite{Garcia-Fernandez-16} to expand the DFT energy with $\delta n$ finding that 
   the zeroth order term, $E^{(0)}$, corresponds with the full DFT energy for the reference density.
   The corrections to this reference energy only depend on $\delta n$ (and parametrically
   on $n_0$) which, given its smallness, can be efficiently calculated leading to a fast and accurate
   approximation of the full DFT energy. 
   The expansion is usually taken to second-order,
   \begin{equation}
     E\approx E^{(0)}+E^{(1)}+E^{(2)}+...,
   \end{equation}
   resulting in a stationary problem that is equivalent to 
   Hartree-Fock with the important distinction that the interactions are \emph{screened} by the 
   exchange-correlation potential.
   In order to keep $\delta n$ small the application of the method is restricted to 
   problems where atomic bonds are not created or destroyed, i.e. to processes that display
   an invariant bond topology.

The $E^{(0)}$ term represents the exact DFT energy for the reference density. We represent it for a 
variety of geometries with an accurate force-field\cite{Wojdel_2013} that allows for fast evaluation. 
The $E^{(1)}$ and $E^{(2)}$ terms account for the changes in the electronic structure that are represented
by geometry-dependent Wannier functions. Under this basis $E^{(1)}$ becomes a tight-binding model while
$E^{(2)}$ represents electron-electron interactions. 

The interconnection between the first (\siesta) and the second (\textsc{Scale-Up}) principles
simulations is carried out through a python script, \textsc{Modelmaker}. Taking a few
cutoff distances \textsc{Modelmaker} is able to produce a model's terms and automatically 
carry out DFT simulations with \siesta\ to determine the force field, a Wannier Hamiltonian
to represent the bands, electron-lattice terms to account how the bands change with geometry and 
electron-electron interactions to describe, for example, magnetism.

While, so far, few publications with SPDFT methods include explicit treatment of electronic degrees of freedom, the lattice part has successfully been used in several applications. One of the main fields of research has been thermal conductivity in perovskites. In particular it was employed to study the electrophononic coupling in SrTiO$_3$\cite{Torres-19} and PbTiO$_3$\cite{Seijas-19} and the proposal of a thermal switch in PbTiO$_3$.~\cite{Seijas-18} It has also been used to study the competition between various ferroelectric domain structures in PbTiO$_3$/SrTiO$_3$ superlattices as a function of strain.~\cite{Das-19} As a result it was found that tensile  strains lead to the appearance of chiral ferroelectric vortices while ferroelectric skyrmions were predicted and experimentally observed for more compressive strain values.~\cite{Das-19} The calculated dielectric properties of these superlattices\cite{Yadav-19} are in very good agreement with measured values and show very large electric susceptibility consistent with regions of negative, static electric permittivity situated at the core of the vortices and the PbTiO$_3$/SrTiO$_3$ interfaces.

\subsection{\label{sec:scripting}Scripting and integration in external
  frameworks}
An ongoing trend in many areas of computational science is to move away
from rigid and monolithic codes, favoring instead a more flexible
approach in which the internal functionality of a program is somehow exposed to
the outside world. If done in a proper and well-documented way, this
can serve to enhance the interoperability of codes with different
functionalities, playing to the relative strengths of each, and/or to
implement new functionalities by combining the available basic
blocks. In \siesta\ we have followed two different but complementary
routes to these ends: the development of an internal scripting framework based on the Lua
language, which enables new functionality without code recompilation,
and the implementation of a formal interface to the AiiDA platform.

\subsubsection{\label{sec:lua}Lua interface}

Lua\cite{Ierusalimschyn} is an easy-to-learn and fast scripting language built for
embedding. It is very lightweight (its memory footprint is less than
$300\,\mathrm{kB}$), and provides very simple ways to interface to the
data structures and routines of a host program. A Lua script,
interpreted by the Lua interpreter embedded in the program, can then
control the flow of execution and the data. Different user-level
scripts can implement new functionalities, \emph{without
  recompilation} of the host code.
The strategy we have followed in \siesta\ is based on handling control
to the Lua interpreter at specific relevant points in the program flow
(e.g. at the beginning of a geometry step, at the end of a scf step,
etc). Lua scripts implement handlers appropriate to the point they
want to hook into, and can request access to specific data
structures. For example, a script intended to implement a better
scf mixing algorithm would be executed after every scf step, inspecting
the convergence data, and changing mixing parameters or schemes, as
appropriate. As another example, convergence checks over mesh-cutoffs and $\mathbf
k$-point sampling can be performed automatically.

The above mixing scenario exemplifies an important area of usefulness
of the approach: the prototyping in Lua, (followed eventually by a
full implementation), of new ideas and algorithms. We have implemented
a number of custom molecular dynamics modes, geometry relaxation
algorithms, and advanced optimization schemes, in a pure Lua library
\textsc{flos}\cite{flos}. The code in the library can be re-used, or
taken as starting point for other implementations by users. These
user-level scripts can in turn be shared, opening the way to the
development of new functionality with faster turnaround that the
traditional approach that needs a careful integration into the
program's code base.

As a specific showcase of the power of the Lua embedding, we have
developed a number of variations of the nudged-elastic band method
(NEB)~\cite{Smidstrup2014,Sheppard2008} for transition-state
search. Previously proposed implementations in \siesta\ involved
significant, hard to maintain code changes, and did not make into the
mainstream version. With Lua, we have been able to implement,
non-intrusively, not only the standard algorithm, but a Double Nudged
Elastic Band (DNEB)~\cite{doi:10.1063/1.1636455} variation, and also
another version which treats atomic coordinates and lattice variables
on an equal footing (the variable-cell NEB, or VC-NEB,
method~\cite{Qian2013}).

The integration of Lua functionality in \siesta\ has been made
possible by the development of an intermediate layer, 
\textsc{flook}\cite{flook}, (for ``fortran-Lua-hook''), which provides
wrappers for access to Fortran data structures and subroutines.

\subsubsection{\label{sec:aiida}AiiDA plugins and workflows}

The AiiDA framework~\cite{Pizzi-2016,aiida-site,aiida-core} provides
support for high-throughput computations in materials science, keeping
full provenance of the calculations and facilitating data handling and
sharing. The framework is open-source, written in Python, and designed
to support arbitrary codes via a plugin interface. A plugin for \siesta\
has been implemented and is distributed as the open-source
package \texttt{aiida-siesta}~\cite{siesta-aiida}. The plugin provides
the basic operations of preparing the input files for a calculation
using AiiDA-specific input objects, and parsing the results and
generating AiiDA output objects. The AiiDA data are stored in a graph
database that keeps a permanent record of the inputs and outputs of
the calculation, and is fully searchable for, e.g. data analytics purposes. 
    
\begin{figure}[htb]
  \centering
  \includegraphics[width=0.98\linewidth]{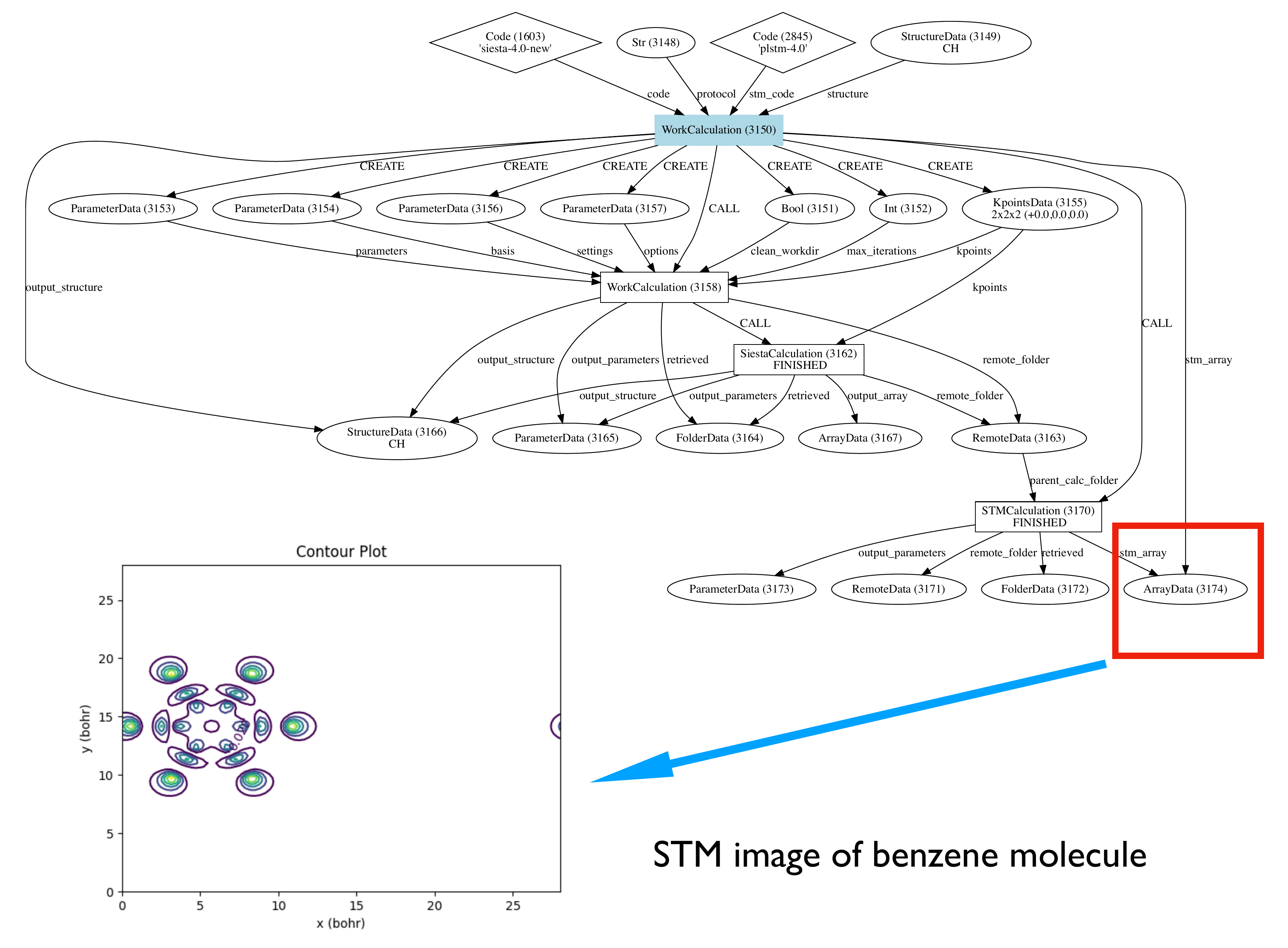}
  \caption{Automatically generated graph for the execution of an AiiDA
    workflow for simulation of STM images.
  \label{fig:aiida-graph}
  }
\end{figure}

AiiDA also provides robust support for the creation of workflows that
incorporate all the necessary steps in the calculation of potentially
complex properties, together with the proper heuristics and fail-safe
features. The \texttt{aiida-siesta} package provides a base workflow and a few
workflows for standard materials properties, such as band
structures. Fig.~\ref{fig:aiida-graph} shows the execution graph of a
workflow designed to generate a synthetic STM image from a given
structure. Work is ongoing to implement more complex ones.

In addition to an interface to the computational capabilities of the
\siesta\ code via the plugin and workflows, the \texttt{aiida-siesta} package
also provides an implementation of basic objects representing
pseudopotential files, notably one for PSML. Families of
pseudopotentials can be uploaded to an AiiDA database and shared via
the provided mechanisms for data export and import, facilitating the
interoperability of different codes.

\subsection{\label{sec:utilities}Utilities for post-processing and
  supplementary features}

\siesta\ offers several features beyond the core
functionality of solving the electronic structure problem and
performing optional geometry relaxations and molecular dynamics
runs. It is worth noting in particular that the atomic character of the basis set
enables the use of a very intuitive suite of analysis tools, which
take advantage of the fact that most of the concepts relating to
chemical bonding use the language of atomic orbitals.

The (partial) density of states, atomic and orbital
populations, and other useful output can be obtained directly from the
program. The \siesta\ distribution includes also several tools in the
\texttt{Util} directory for band-structure and wavefunction plotting,
bonding analysis, etc.
Beyond these, special tool packages that
implement a specific feature that extends the functionality of the
main program, or that provide extra options for visualization or
post-processing in general, are available in alternate distribution
points. We describe in what follows the most relevant developments.

\subsubsection{\label{core-utils}Updates to core utilities}

A number of improvements, enhancements, and additions have been made to the
core utilities shipped with the \siesta\ distribution.

There is now a  ``fat-bands'' feature, by which bands can be decorated with
information about the relative weight of given orbitals in each state.
The  wave-function-related analysis tools have been extended to the non-collinear and
spin-orbit case. This includes the COOP/COHP bonding analysis, band-structures, and a
new tool spin-texture calculation.
There have been also improvements to band-structure plotting utilities
and to the visualization of charge densities, potentials, and
other magnitudes represented in a real-space grid.

A band unfolding utility has been added. Based on the Fourier
decomposition of the Bloch wave functions, it allows to perform a
``full unfolding'' even for non-periodic systems (e. g. liquids)
calculated with a large simulation cell. By refolding the fully
unfolded bands, from the reciprocal supercell of a perturbed or
defective crystal, into the reciprocal unit cell of the primitive
crystal, one recovers the conventional ``unfolded''
bands~\cite{Mayo-20}.

\subsubsection{\label{sec:sisl}sisl}

\sisl\ is a Python toolbox that was initially conceived to handle and manipulate
\siesta/\tsiesta\ output\cite{zerothi_sisl}. It has since been extended to support other DFT
codes, with the aim of offering equivalent operations for them.

By reading the LCAO outputs from \siesta\ one can post-process the
Hamiltonian and calculate e.g. Brillouin zone integrated DOS, wave
functions expanded on grids, eigenvalues, band velocities and many
more. \sisl\ can process nearly all the \siesta\ output files. In
particular, it is also able to post-process data on the real-space grid.
Its command line interface allows data format changes,
e.g. conversion of \siesta\ \texttt{XV} files to \texttt{xyz}/\texttt{xsf}
files or \siesta\ binary grid data (\texttt{VH}, \texttt{VT}, \dots)
to \texttt{cube}/\texttt{xsf} files.

As it can process density matrices from \siesta, one can also use \sisl\
to prepare an input electronic-structure for new calculations, which
may be helpful to reduce initial SCF steps.

\sisl\ also allows creation of custom tight-binding models (both
orthogonal and non-orthogonal), and since it extracts the DFT
Hamiltonian matrix one can manipulate the Hamiltonian to retain
certain band-structure features and thus perform large-scale
simulations\cite{36268dbff41747b3ba5b1f038919154d}. This allows
calculating far-field currents using reduced basis-sets with very
little loss of accuracy.

The Atomic Simulation Environment (ASE)\cite{ase-paper} and \sisl\ have a
certain degree of overlap in terms of geometry handling functionality.
One can easily convert to and from ASE objects in \sisl, thus
allowing seamless interaction.

\subsubsection{\label{sec:postnikov}Other post-processing and visualization utilities}

The body of utilities contributed by non-core developers and other
\siesta\ users has continued to expand.  In particular, we feature in
this section two suites of utilities, one dealing with alternate
visualization tools for some {\siesta} results, and another one
specifically dealing with lattice dynamics.

For structures, the {\tt xv2xsf} and {\tt xv2vesta} converters process
data from the \siesta\ {\tt .XV} file into the native formats of
XCrySDen~\cite{xcrysden} and VESTA~\cite{vesta}, respectively. Each of these two codes offers many
options of graphical representation of structures, adding
translations, clipping fragments etc.  Three-dimensional spatial
functions (e.g., charge density, local density of states integrated
throughout the chosen energy range), computed by {\siesta} on a
real-space grid. Tools are provided for interpolating the data from
the {\siesta} output grid (fixed by the unit cell dimensions and the
{\tt MeshCutoff} parameter) onto an arbitrarily cut (and possibly
rotated or resampled) parallelepipedic box.  XCrysDen provides a
number of display options, including contour lines over grid planes,
or isosurfaces.
A special feature available in XCrySDen is plotting the Fermi
surfaces. A special script, {\tt eig2bxsf}, serves to analyze the list
of $\mathbf{k}$-points handled by \siesta, expanding it onto a regular
sequence, and writing the respective band energies in the necessary
format.

The tools concerning the lattice dynamics have been developed having
in mind the $\Gamma$ phonons calculated for a large enough supercell,
that is a typical case in a simulation of molecular crystals or
disordered substitutional alloys. For visualization, {\tt vib2xsf} and {\tt vib2vesta}
place arrows at the atoms according to the vibration pattern stored in
the eigenvectors file ({\tt .vectors}), produced by the core {\tt
Vibra} utility, and can also be used to make animations (sequences of
snapshots) of selected vibration modes. Both {\tt vib2xsf} and {\tt
vib2vesta} tools allow the selection of a part of the system to be exposed.

The {\tt phdos} tool is designed for analyzing zone-center vibration
results.  As the system is supposed to be large (e.g., a supercell
chosen for a periodic crystal), the (artificially broadened, for
convenience) discrete spectrum may serve as a fair approximation to the
total density of modes, and if weighted with (squared) components of
eigenvectors at different atoms -- provide a decomposition into
contributions of different atoms in the total density of vibration
modes.

\begin{figure}[ht]
    \centering
    \includegraphics[width=0.95\linewidth]{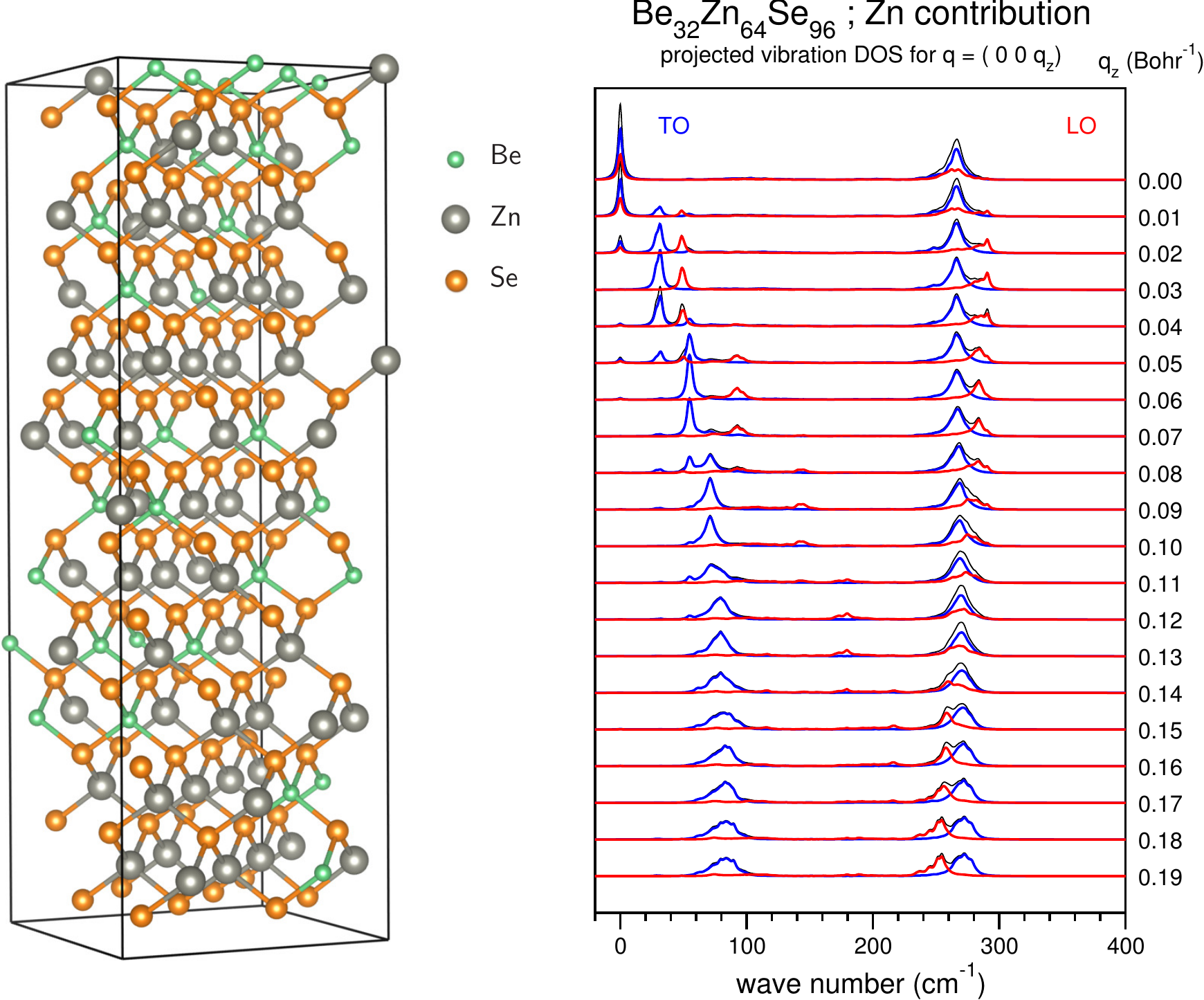}
    \caption{\label{fig:BeZnSe_phdos}
    Left panel: a 192-at. quasirandom supercell representative for 
    the Be$_{1/3}$Zn$_{2/3}$Se solid solution; 
    right panel: density of modes within the frequency range of Zn-Se
    vibrations, extracted with {\tt phdos} and projected onto different values of $q_z$ and different
    polarisations, parallel (labelled LO) and perpendicular (TO) to $\mathbf{q}$. 
    These results were partially shown in Fig.~4 of Ref.~\onlinecite{PRB89-155201}
    and discussed in that work.
    }
\end{figure}

A more sophisticated option is the \emph{projection} of different
eigenvectors according to various criteria. The typical system under
study is a supercell in which e.g. an alloying, or some kind of
deformation, breaks the underlying perfect periodicity. Still, some
trends related to the latter can be revealed by appropriate
projections. The two obvious cases are the projections onto (1)
$\mathbf{q}$-vectors of the underlying lattice and (2) irreducible
representations of the space group of the underlying lattice; the
corresponding formulas and some results can be found in
Ref.~\onlinecite{PSSA210-1332}.  The first type of projection, if done
for a sequence of $\mathbf{q}$ values, helps to reveal ``phonon
dispersions'', obviously blurred by the broken periodicity, also
making distinction between transversal and longitudinal modes -- see
Ref.~\onlinecite{PRB89-155201} for an example of use.  To make the
trends more pronounced, the supercell needs to be sufficiently long in
the direction concerned -- see, e.g., Fig.~\ref{fig:BeZnSe_phdos}.
The simplest case, a projection onto a single
$\mathbf{q}$=0 value, may also be of interest, since it enhances the modes
which are expected to dominate the infrared or Raman spectra, and thus
facilitates their comparison with experiment.

The symmetry projection may help to isolate in a possibly complex
spectrum those modes which are expected to dominate according to a
given selection rule, again in view of their verification against the
experiments. The group-symmetry information needed for the projections is
available e.g. from the Bilbao Crystallographic Server,\cite{Bilbao-SAM} and the
technical details are explained in the documentation included in the
tools.

\begin{figure}[htb]
    \centering
    \includegraphics[width=0.95\linewidth]{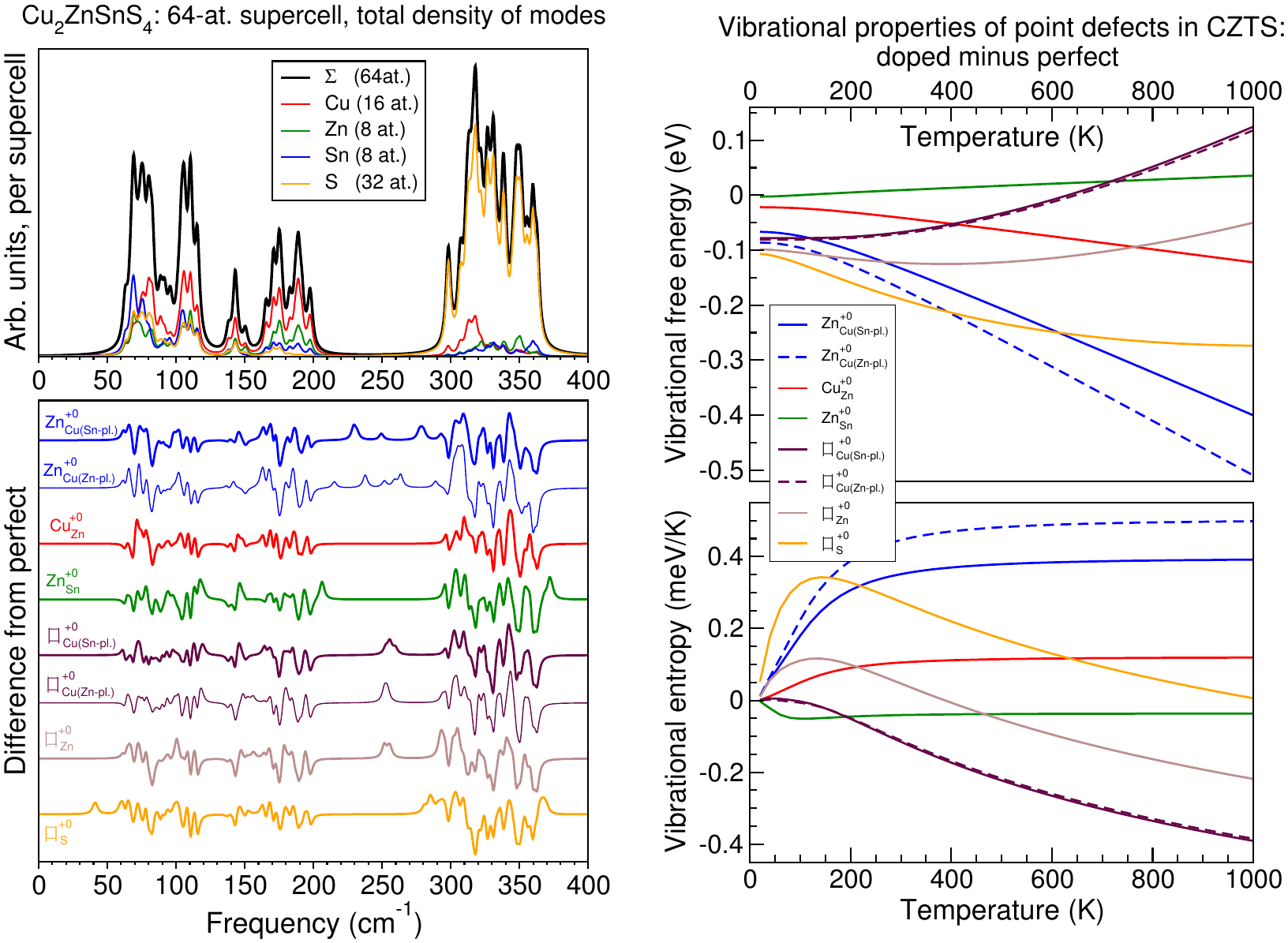}
    \caption{\label{fig:CZTS_entro}
    Vibration properties of Cu$_2$ZnSnS$_4$ (CZTS) with substitutional impurities,
    used in Ref.~\onlinecite{JAP114-124501}. Left panel: densities of modes
    (extracted with {\tt phdos}); right panel: vibration contributions to
    the free energy and entropy (calculated with {\tt vibent}).
    Adapted from Fig.~5.3 and 5.4 of Ref.~\onlinecite{Narjes-thesis}.}
\end{figure}

\begin{figure}[htb]
    \centering
    \includegraphics[width=0.95\linewidth]{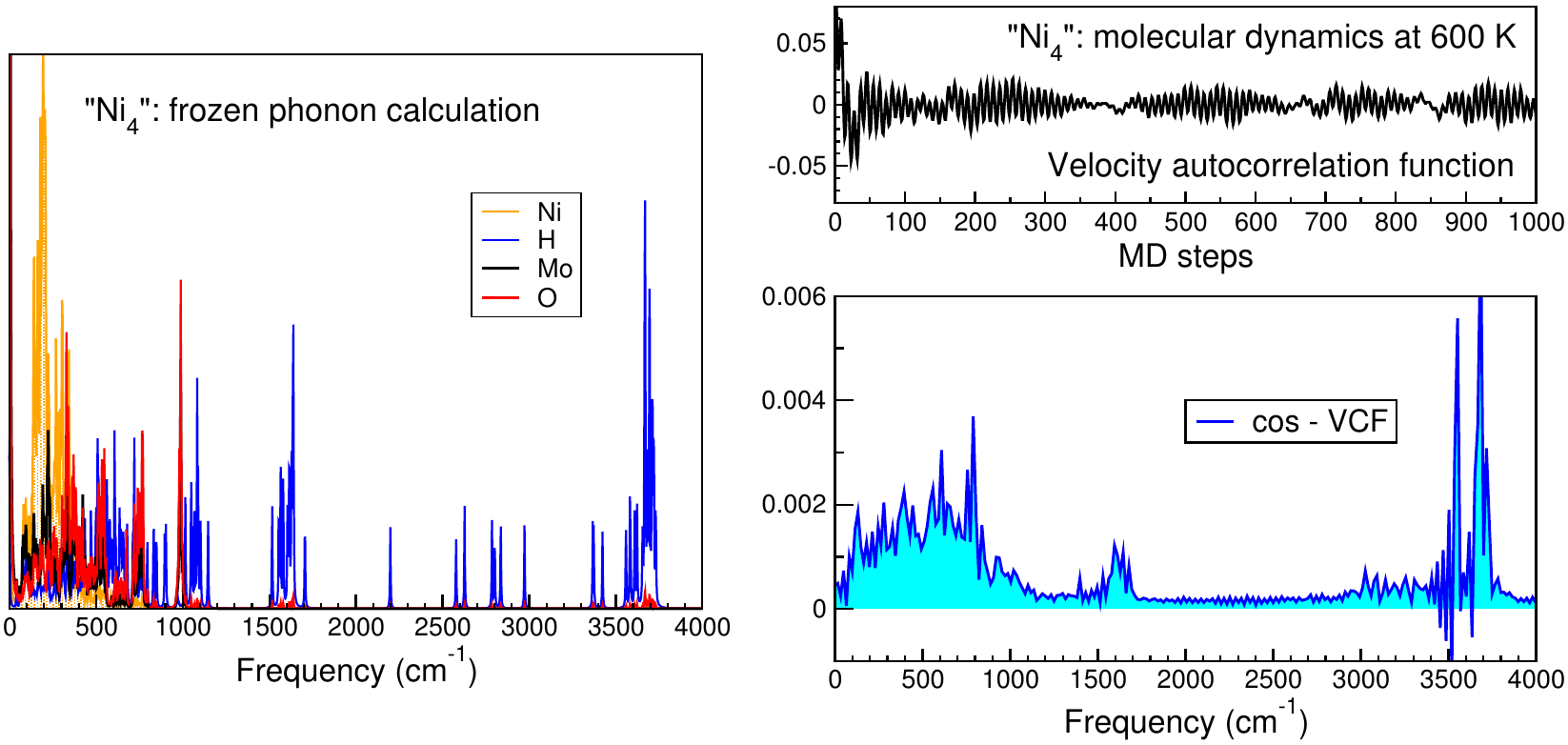}
    \caption{\label{fig:Ni4-MD}
    Vibration properties of ``Ni$_4$'' molecular magnet, 
    [Mo$_{12}$O$_{30}$($\mu_2$-OH)$_{10}$H$_2$\{Ni(H$_2$O)$_3$\}$_4$]${\cdot}$14H$_2$O.
    Left panel: density of modes from frozen phonon calculation; right panel: 
    velocity autocorrelation function, its Fourier transform and hence resulting 
    density of vibration modes.}
\end{figure}

The {\tt vibent} tool performs a straightforward calculation (see,
e.g., Sec.~II.C in Ref.~\onlinecite{JAP114-124501}, or Sec.~5.3 in
Ref.~\onlinecite{Narjes-thesis}) of temperature-dependent vibration
contributions to the free energy and entropy -- see
Fig.~\ref{fig:CZTS_entro} as an example.  The necessary input
information is the vibration spectrum, originating from the {\tt
  Vibra} frozen phonon calculation on a sufficiently large system.

The {\tt velcf} tool calculates the velocity autocorrelation function and its
Fourier transform from a (presumably sufficiently long) molecular dynamics (MD)
history, recorded in the {\tt .MD} or {\tt .ANI} file. This
technique~\cite{Allen+Tildesley} can be used to obtain phonon
frequencies, and was applied along with a \siesta\ calculation in Ref. \onlinecite{NATO01-vibra}.
An example of such simulation (1000 MD steps at 600 K) is shown in Fig.~\ref{fig:Ni4-MD}
in comparison with frozen phonon results, revealing similarities of the spectra obtained.

\subsubsection{\label{sec:optical properties}
Optical properties of finite systems: linear response TDDFT starting from \siesta\ 
orbitals}

The \siesta\ package offers at least two ways of obtaining optical properties of finite systems.
The first way uses real-time TD-DFT propagation by applying an external electric field with a simple time dependence (e.g., a Heaviside step-function)~\cite{Tsolakidis2002}.
The second way is by computing the non-interacting dielectric function~\cite{Soler-02,Artacho-08}. 
Both methods are implemented in \siesta\ and can be employed without any external tools. However,
they are limited in different aspects. The non-interacting dielectric function often 
underestimates the HOMO-LUMO gaps and calls for the use of the phenomenological scissor-shift
operator. Real-time propagation makes cumbersome the analysis of the optical response properties
in the frequency domain. Furthermore, the frequency resolution scales with the duration of the real-time simulation. Thus, accurate spectra require long simulations.

Fortunately, there are two efficient implementations of \textit{linear-response} TDDFT that use the Kohn-Sham orbitals from \siesta\ as a starting point and are available for the open-source
community \cite{Coulaud2013,Koval2019}. In both packages, the linear density response
$\delta n(\bm{r}, \omega)$ is obtained directly in the frequency domain which makes 
straightforward the analysis of derived properties. However, there are differences between both implementations on the construction of the auxiliary basis necessary to expand the orbital products. These differences can severely affect the computational cost of the calculation.   

The linear-response TDDFT is built on the concept of the induced electronic density
$\delta n(\bm{r}, \omega)$ in response to a small perturbation of the external potential
$\delta V_{\mathrm{ext}}(\bm{r}, \omega)$. The integral operator connecting $\delta n(\bm{r}, \omega)$ to
$\delta V_{\mathrm{ext}}(\bm{r}, \omega)$ is the interacting density response
function $\chi(\bm{r}, \bm{r}', \omega)$. By virtue of the KS equations,
$\chi(\bm{r}, \bm{r}', \omega)$ can be connected to the  
non-interacting density response function $\chi_0(\bm{r}, \bm{r}', \omega)$ \cite{Petersilka1996}
with a Dyson equation

\begin{equation}
\chi(\omega) = \chi_0(\omega) + \chi_0(\omega) K \chi(\omega),
\label{dyson-eq-response}
\end{equation}
where the interaction kernel $K(\bm{r}, \bm{r}')$ contains the bare Coulomb interaction and the so-called exchange and correlation kernel $K_{xc}$, which is a known operator for simple functionals like LDA and GGA.
The non-interacting response function
$\chi_0(\bm{r}, \bm{r}', \omega)$ can be expressed as a sum over electron-hole excitations
within the basis formed by the KS orbitals $\Psi_n(\bm{r})$ \cite{Petersilka1996,Koval2016,Koval2019}

\begin{equation}
\chi_0(\bm{r}, \bm{r}', \omega) = \sum_{nm}(f_n-f_m)
\frac{\Psi_n(\bm{r})\Psi_m(\bm{r})\Psi_m(\bm{r}')\Psi_n(\bm{r}')}{\omega-E_m+E_n},
\label{non-interacting-response-sos}
\end{equation}
where $f_n$ are occupations of the KS orbitals and $E_n$ are their energies.

The optical polarizability tensor $\alpha(\omega)$ is related to the induced density by
$\alpha(\omega) = \int \bm{r} \delta n(\bm{r}, \omega) d\bm{r}$ or alternatively 
\begin{equation}
\alpha(\omega) = \iint \bm{r} \chi_0(\bm{r}, \bm{r}', \omega) \delta V_{s}(\bm{r}', \omega) \,
d\bm{r} d\bm{r}'\label{optical-polarizability-vscr},
\end{equation}
where due to Eq.~\eqref{dyson-eq-response} and using the dipole approximation for the 
electron-photon coupling, the screened effective perturbation $\delta V_{s}(\bm{r}', \omega)$
satisfies the linear integral equation
\begin{equation}
(\mathbb{I} - K \chi_0(\omega))\delta V_s(\omega) = \bm{r}.
\label{iteration}
\end{equation}

\begin{figure}[htb]
  \centering
  \begin{subfigure}
         \centering
         \includegraphics[width=0.98\linewidth]{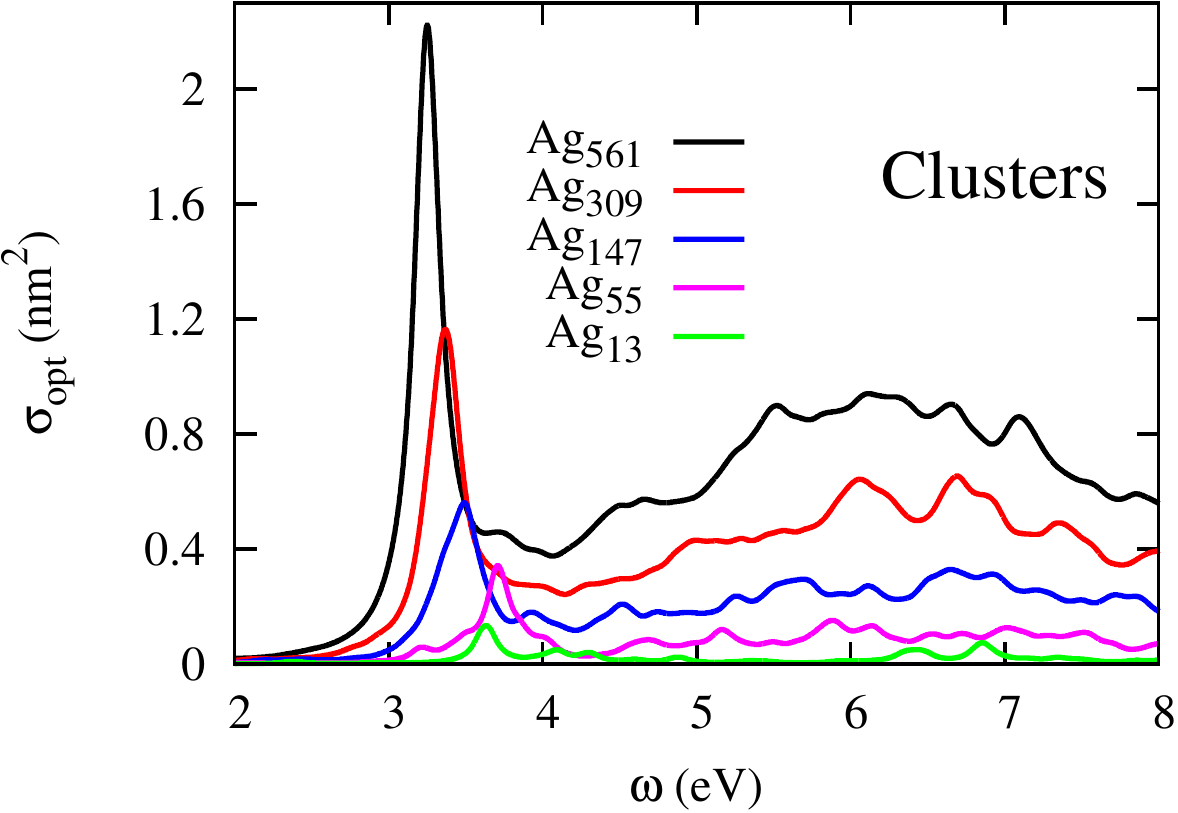}
         \caption{The absorption cross sections of silver clusters of icosahedral
  shape. One can recognize sharp surface-plasmon resonances around 3--4 eV and a broad resonance at 6--7 eV.}
         \label{fig:abs-ag-icosahedral}
     \end{subfigure}

  \begin{subfigure}
         \centering
         \includegraphics[width=0.6\linewidth]{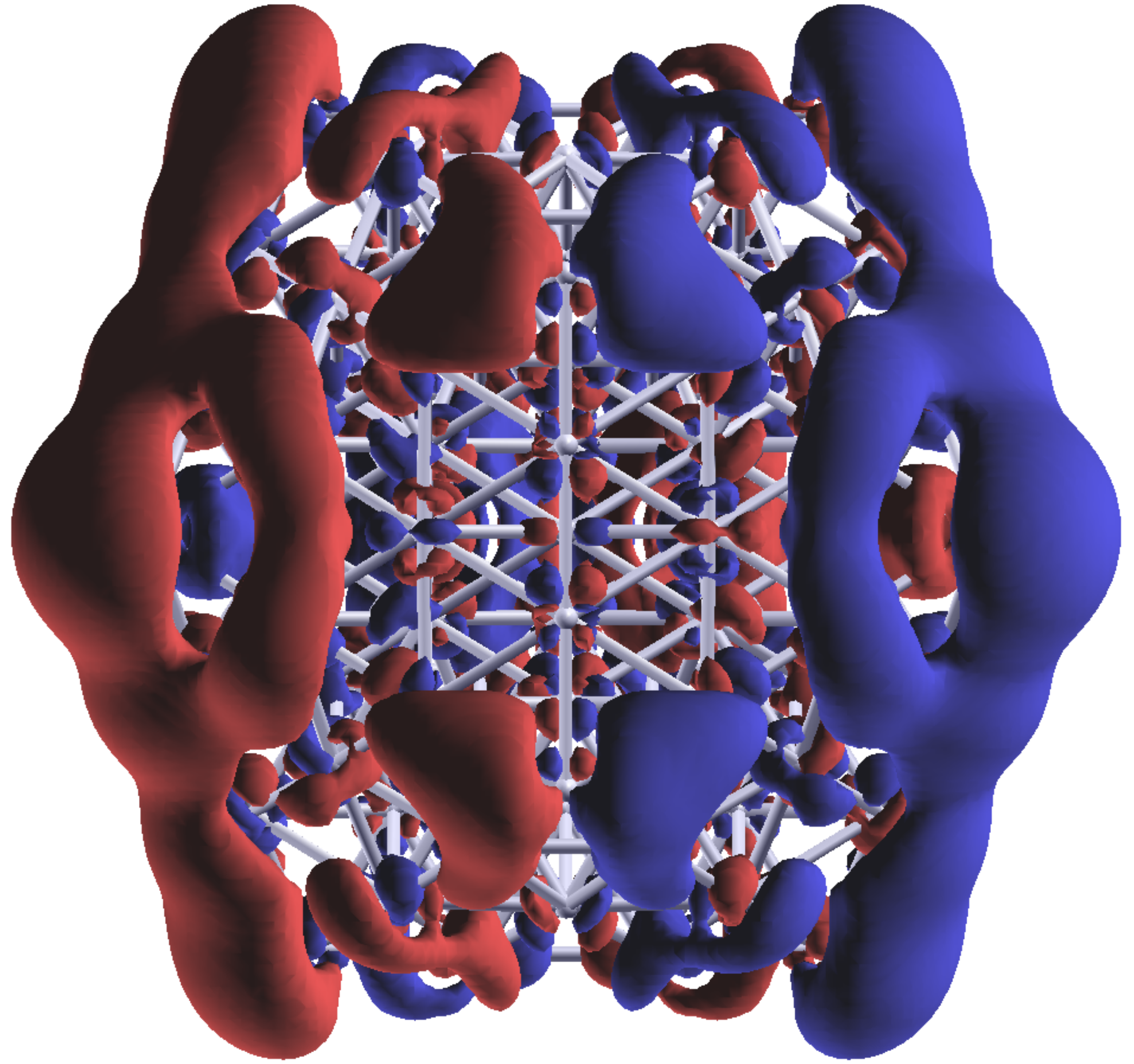}
         \caption{The isosurfaces of density change $\mathrm{Re}(\delta n(\bm{r},\omega))$ of the Ag$_{147}$ cluster 
close to the frequency of the surface-plasmon resonance of the cluster (3.4 eV). }
         \label{fig:dn-ag147-icosahedral}
     \end{subfigure}
\end{figure}

The efficiency of the methods presented in References  \onlinecite{Coulaud2013,Koval2019} comes from solving iteratively Eq.~\eqref{iteration}  for $\delta V_s(\omega)$ instead of using standard matrix inversion to obtain $\chi(\omega)$ from Eq.~\eqref{dyson-eq-response}. 
Once $\delta V_s(\omega)$ is known, 
Eqs.~\eqref{non-interacting-response-sos} and \eqref{optical-polarizability-vscr}
allow the computation of  the optical properties of the system. Furthermore, it is also possible to perform different types of analysis. For example, it is 
easy to partition the polarizability tensor $\alpha(\omega)$ in terms of electron-hole 
contributions \cite{Ullrich:2015, Koval2016} due to existence of the sum over
the electron-hole pairs in Eq.~\eqref{non-interacting-response-sos}. Similarly, one can achieve
other types of Mulliken-like analysis~\cite{Dronskowski1993-COOP, Koval2016, Koval2019} of 
the optical polarizability tensor $\alpha(\omega)$ or the induced density
$\delta n(\bm{r}, \omega)$.

The Python implementation of linear response TDDFT in the PySCF-NAO package as described in Ref.~\onlinecite{Koval2019}
is convenient to use and rather potent.  
It is capable of computing the optical properties of compact metallic
objects containing up to several hundreds of atoms \cite{Barbry2015, Koval2016, Marchesin2016}.
For example, we were able to track down the different size-dependence of the plasmon resonance in sodium and 
silver clusters due to the screening effect of silver $d$-orbitals in the latter case \cite{Barbry2018-thesis}.
In those calculations, using an optimized version that incorporates some additional memory-saving features not present in the currently distributed version of PySCF-NAO, icosahedral silver and sodium clusters containing up to 5043 atoms
were studied. 

In Figures \ref{fig:abs-ag-icosahedral} and \ref{fig:dn-ag147-icosahedral},
we show the photo-absorption cross sections 
of a series of compact silver clusters~\cite{Koval2016} and the real part of 
induced density change in the cluster Ag$_{147}$ close to its surface-plasmon 
frequency (3.4 eV), respectively.

\subsubsection{\label{sec:thermaltransport} Thermal transport by the AEMD
 method} 


The approach to equilibrium molecular dynamics (AEMD)
method\cite{AEMD} has been implemented to obtain the thermal
conductivity. In the first stage of the method, the system is decomposed
in two different regions, each one equilibrated to a different initial
temperature (canonical run with Bose, or Anneal MD). Then, a
microcanonical run (Verlet) is carried out for the whole system, and the
average temperature of each subsystem is monitored.  This temperature
transient regime is then used to extract the thermal conductivity from
the exact solution of the heat transport equation.\cite{SIllera}

\subsubsection{\label{sec:core-levels} Core level shifts}

Core-level shifts can serve to analyze changes in the local and
chemical environment of atoms of a given
species. Density-functional-theory calculations have proved to be
quite useful in complementing the experimental information, which is
sometimes hard to interpret. Two schemes have been implemented in
\siesta\ for the calculation of core-level shifts within a
pseudopotential approach~\cite{Garcia-Gil2012}.

In the so-called \textit{initial-state} approximation the electronic
relaxation in the presence of the core hole is neglected, and the
photo-electron's binding energy is directly related to the eigenvalue
of the core level. A pseudopotential calculation obviously cannot
compute the latter, but differences in core eigenvalues in different
environments can be estimated by the changes in the expectation value
of the crystal potential using the core state's atomic wavefunctions
$\psi_{n}^{lm}$ at different sites. These can be extracted from the
  matrix elements

\begin {equation}
 V^{m m'} = \int d^3 r ~ ({\psi_{n}^{lm}} ( \vec{r} - \vec{\tau}))^* V
 ( \vec{r}) \psi_{n}^{lm'} ( \vec{r} - \vec{\tau})
\label{eq:corelevel}
\end{equation}
with a further step of averaging to remove the splittings stemming
from the loss of spherical symmetry.

In the \textit{final-state} approximation, the relaxation is
explicitly taken into account, and the experimental shifts (measured
via the kinetic energy of an exiting electron) are correlated with the
differences in the energy of the crystal with a ``core-hole'' in
different sites. For this, a special
pseudopotential with a missing core electron has to be generated, and a
full \siesta\ calculation is needed for each different site.

The implemented methodology has been used to study, for example, the
shifts induced by hydrogen bonding in organic
molecules~\cite{GarciaGil-2013}.

\subsection{\label{sec:software-eng}Software-engineering advances and partnerships}

The traditional development model for scientific codes in academic
settings has been typically based on multiple contributions with
various levels of programming competence, and with very
little time to plan ahead in the face of pressing scientific demands.
\siesta\ has been no exception, and has grown in features and complexity
over the years. It is very important to keep complexity under control,
or else a project becomes un-maintainable and cannot survive. It is
not simple, however, to balance the need of incorporation of new
features, and the need to increase the computing performance in a
landscape of constantly evolving hardware and programming models.
One essential route is modularization, which allows the separation of
concerns at various levels. In the context of a code like \siesta\ , this
means that the scientific ideas and algorithms should be
handled at a high level, calling on lower-level modules for specific
functionality (domain-specific libraries, mathematical libraries,
communication protocols, etc). These lower-level modules can hopefully
be re-used by different codes and, most importantly, can be focused on
by highly-skilled programmers for optimization on relevant
architectures.

Another important method of taming complexity involves the
streamlining of the data structures of the code. This is an ongoing
process (see Sect.~\ref{sec:future}), but has already taken a very significant
step by the introduction of reference-counted data structures. They
build on a well-known and not particularly advanced technique of
memory-handling~\cite{ref-counting}, but in \siesta\ they have enabled a much simpler
bookkeeping of the data structures needed for a richer control of
molecular-mechanics and scf iterations.

Regarding performance-oriented developments, in the recent past we
have implemented a mixed MPI/OpenMP programming model, which allows,
for suitable systems, to better balance arithmetic intensity and
communications needs. The deployment of this model is more
advanced in the TranSiesta module, and significant speedups have been
obtained for large systems.

Some of the above software-engineering developments have been enabled
and strengthened by the participation of \siesta\ in a number of
international partnerships, notably the MaX (Materials at the
eXascale) EU center of excellence~\cite{MaX} and the Electronic
Structure Library initiative~\cite{ESL}. The ``separation of concerns''
described above in the context of modularization is an example of the
so-called ``open-innovation'' paradigm, at the foundation of the ESL
strategy for code reusability, and is also a cornerstone of MaX's efforts to achieve
exascale-readiness for its flagship materials science codes
(with \siesta\ among them): performance-enhancement efforts are to be
focused on relevant domain-specific modules.

A number of modules from \siesta\ have been turned into stand-alone
libraries which now feature in the ESL: \texttt{libGridXC} for
exchange and correlation calculations, \texttt{libPSML} as a handler
of PSML files, \texttt{xmlf90} for general purpose handling of XML
files, etc. Conversely, \siesta\ uses some of the libraries offered by
the ESL, notably the ELSI library of electronic-structure solvers
mentioned in Sect.~\ref{sec:elsi}, whose development, including its
API design and internal data organization, has been in turn influenced
by contributions and feedback from the \siesta\ project, among
others. There are also plans to incorporate the PSolver
library~\cite{psolver} for the solution of the Poisson problem, a
contribution to the ESL from the BigDFT project.

We should mention that the renewed dynamism of
\siesta\ development and the advances made possible by the
interaction with community initiatives are both a blessing and a
challenge. It is non-trivial, for example, to handle the building process of a code
that relies on a number of different external libraries, programming models, and
special features such as the embedded Lua interpreter.  Luckily, as
will be discussed in Sect.~\ref{sec:future}, these are issues that are
being addressed in wider contexts, and \siesta\ is well placed to take
advantage of it.

\section{Applications}
\label{sec:applications}

We present here a few showcase applications that illustrate the
capabilities of \siesta, in breadth, efficiency, and accuracy.

\subsection{4 terminal NEGF on germanium surface}

\begin{figure}[htbp]
  \centering
  \includegraphics[width=0.99\linewidth]{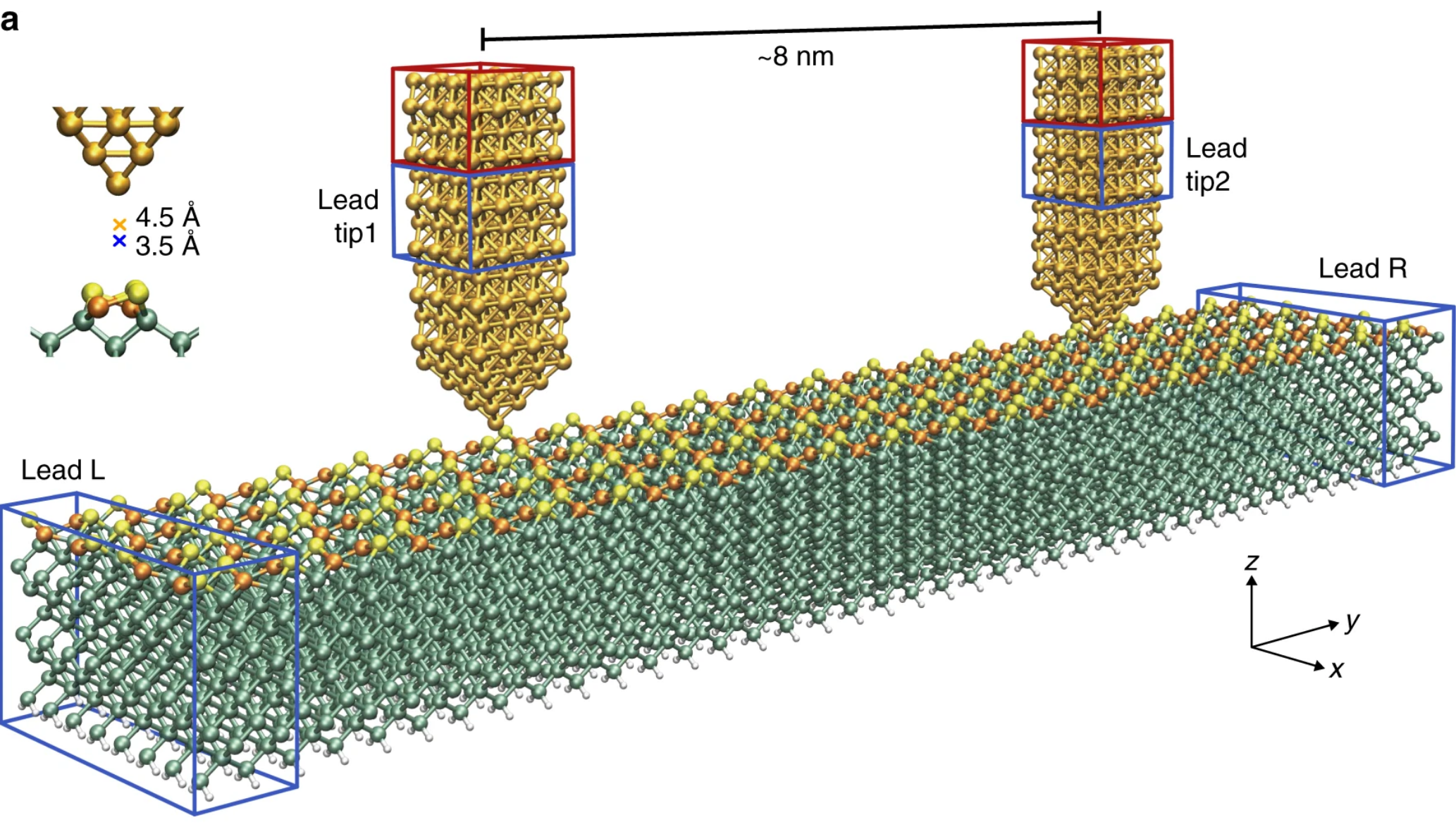}
  \caption{First-principles transport simulations for the two-probe experiments. a) Representation of the four-terminal setup. The electrode regions are highlighted by blue boxes, two of them located at each Ge(001)-c($4\times2$) slab terminations (leads left and right) and the other two at each Au model tip (leads tip1 and tip2). The 50 Ge atoms closest to each tip were allowed to fully relax, adapted from Ref.~\citenum{Kolmer2019}.
  \label{fig:4t}
  }
\end{figure}

Breakthrough simulations using the new multi-terminal implementation on \tsiesta\ were fundamental to elucidate the electronic transport mechanism on a novel and complex experiment.\cite{Kolmer2019}
For the first time a two-probe scanning tunneling microscopy/spectroscopy (STM/STS) with probes operating in tunneling conditions over the same atomic-scale system was used to extract detailed information of in-plane electronic transport.
The addressed system was the reconstructed (001) surface of germanium, where electrons injected from one STM tip at a position determined with atomic precision were collected at the same Ge dimer row at a distance as short as 30 nm.
The experiment was theoretically modeled by a system composed of a twelve-layer Ge(001)-c(4$\times$2) slab contacted by Au tips oriented along the (100) direction (Fig.~\ref{fig:4t}).
On this self-consistent 4-terminal treatment, two Ge electrodes were connected at each slab termination and other two at the Au model tips.
The whole system was defined by 4924 atoms (36442 atomic orbitals), in a super-cell of dimensions $\sim32\times160\times80\,\textup{\AA}^3$, and where 5 different tip-to-sample distances were considered.
Besides the large dimensions of the system, another important challenge of such simulation was the level alignment between the metallic and semiconducting leads and the scattering region, for which a method had to be devised.
A remarkable agreement was found between the calculated transmission function and the experimental transconductance spectra, allowing the identification and assignment of the observed resonances to transport channels existing along the surface Ge dimer rows.
Moreover, the simulations elucidated the transport directionality of the injected hot electrons, revealing a transition from 2D to quasi-1D coherent transport regime as a function of the carrier's energy. 
This work shows that complex experiment setups combined with advanced calculations can provide new insights into transport properties at the nanoscale.

\subsection{Novel topological phases in ferroelectric materials}
\label{sec:topoferro}

\begin{figure}[htbp]
  \centering
  \includegraphics[width=0.8\linewidth]{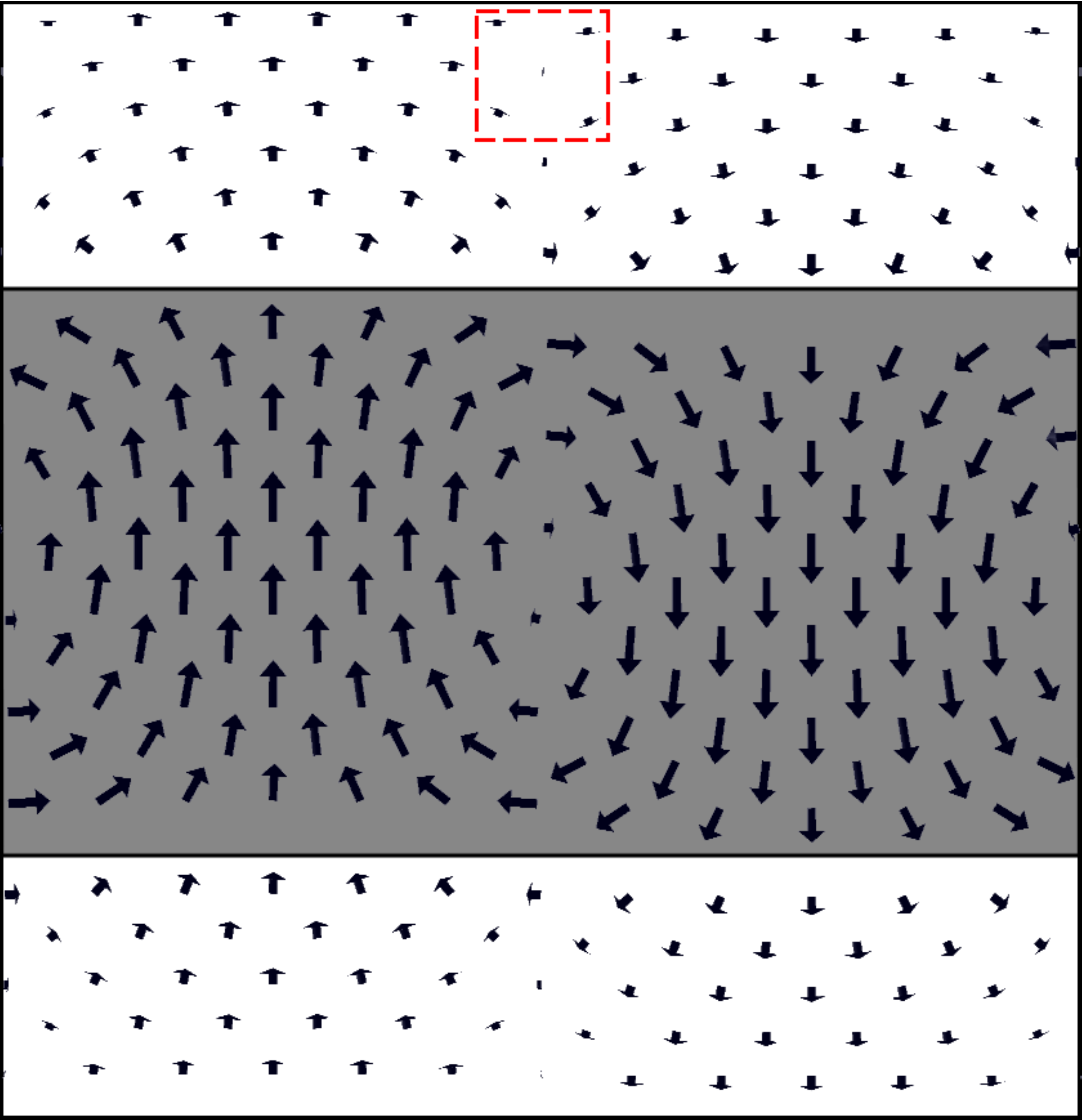}\\
  \includegraphics[width=0.8\linewidth]{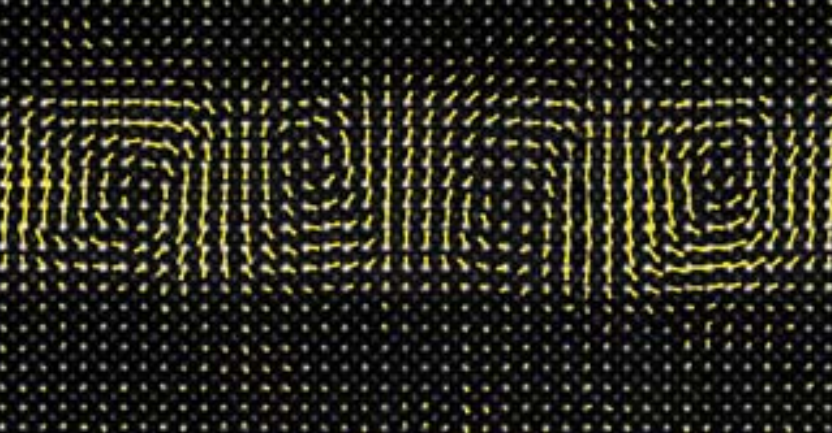}
  \caption{Top panel: local polarization profile of polydomain structures in
   (PbTiO$_{3}$)$_{n}$/(SrTiO$_{3}$)$_{n}$ with $n$=6 obtained
   from an atomic relaxation with \siesta. The PbTiO$_{3}$ and
   SrTiO$_{3}$ are depicted as grey and white regions respectively. 
   Clockwise and counterclockwise vortices within the PbTiO$_{3}$ are clearly visible.
   Red dashed square in the SrTiO$_{3}$ layers mark the position where antivortices are formed.
   Reprinted with permission from \citet{Aguado-Puente-12}
   Phys. Rev. B {\bf 85}, 184105 (2012).
   Bottom panel: experimental observation of vortex–antivortex structures
   in a cross-sectional high-resolution scanning transmission electron microscopy
   image with an overlay of the polar displacement vectors for a 
   (SrTiO$_3$)$_{10}$/(PbTiO$_3$)$_{10}$ superlattice, 
   showing that an array of vortex–antivortex pairs is present in each PbTiO$_{3}$ layer.
   Courtesy of R. Ramesh, adapted from Ref.~\citenum{Yadav-16}.
  \label{fig:vortices}
  }
\end{figure}

In material systems with several interacting degrees of freedom (such as spin, charge and
lattice distortions), the complex interplay between these factors can give rise to exotic
phases. A prototypical example are the superlattices 
of alternating lead titanate and strontium titanate layers. 
Simulations on such PbTiO$_{3}$/SrTiO$_{3}$ heterostructures,
consisting on $n$ unit cells of PbTiO$_{3}$ 
and $n$ unit cells of SrTiO$_{3}$ stacked along the [001] direction were carried out 
with \siesta.
As a function of the periodicity, the superlattices undergo a phase transition
from a monodomain configuration (small periodicity, $n \lesssim 3-4$)
with a normal component of the polarization that is preserved throughout the structure,
to a multidomain configuration (large periodicity, $n \gtrsim 3-4$) with alternating
up and down domains.\cite{Zubko-12}
In order to further reduce the electrostatic energy costs, the local dipoles
within the PbTiO$_{3}$ layer continuously rotate forming a sequence of
clock-wise/counter-clockwise array of vortices along the [100] direction.~\cite{}
The theoretical predictions, done with \siesta~\cite{Aguado-Puente-12}
after the relaxation of supercells of up to 1000 atoms, were experimentally confirmed 
five years later by atomic-scale mapping of the polar atomic displacements by scanning
transmission electron microscopy\cite{Yadav-16} (Fig.~\ref{fig:vortices})
Moreover, the appearance of an axial component of the polarization pointing in the direction
of the vortices make the systems chiral and optically active, as lately  
confirmed by circular dichroism experiments\cite{Shafer-18}.

\subsection{1D and 2D systems}
\siesta\ is particularly well suited to study low dimensional nanostructures, such as 1D and 2D systems where a large vacuum region is needed within the simulation cell. When, in addition, a large number of atoms is required to study particular physical effects is where \siesta\ could excel with respect to other methods. There is extensive literature on simulations of graphene and other exfoliated materials, where the properties of point defects, edges or grain boundaries are of much relevance. To list a few examples, the magnetic properties of impurities,~\cite{Boukhvalov2008,Yazyev2007} and edges~\cite{Slota2018}, but also electronic properties, including transport characteristics, in grain boundaries~\cite{Yazyev2010a,Yazyev2010b}, ribbons~\cite{Kim2008}, nanoporous graphene~\cite{Moreno2018}, large graphene flakes~\cite{Hu-2014,Hu2019}, or the effect of substrates~\cite{Kim2008b}. Other materials, such as mono- and multi-layered dichalcogenides\cite{Matte2010TMDC,Popov2012TMDC} or phosphorene\cite{Liu2014Phosphorene,Guan2014}, are also being widely studied, including optical properties in nanoflakes with up to a few thousand atoms.~\cite{Hu2016Phosphorene}.

\subsubsection{CDWs}
\label{sec:cdw}
A number of recent studies on charge density waves (CDW) in low dimensional materials illustrates the impressive accuracy that can be obtained with \siesta\ for systems with very subtle electronic structures.~\cite{GusterThesis} For example, in 2H-NbSe$_2$ \siesta\ calculations were able to predict the existence of six different atomic structures within a narrow energy range of a few meV, all of them compatible with the experimental 3$\times$3 CDW modulation. Careful analysis of theoretical and experimental STM images for different bias potentials allowed to identify two of these structures that can coexist in the same image.~\cite{GusterNbSe2}
In a different work,~\cite{GusterBB} the temperature dependency of the electronic Lindhard response function in blue bronze K$_{0.3}$MoO$_3$ was studied. This system has a rather complex monoclinic structure, with twenty formula units per unit cell where MoO$_6$ octahedra form chains along one direction (b-axis).
The Lindhard function shows well decoupled sharp responses that correspond to intra- and interband Fermi surface nesting. By fitting these peaks one can obtain the coherence length of the fluctuating 1D electron-hole pair (that determines the length scale of the experimental intrachain CDW correlations), and the intrachain modulation of the response (that determines the shape of the Kohn anomaly measured in experiments), providing, for the first time, a quantitative evidence of the weak electron-phonon coupling scenario for the Peierls transition.

\subsection{\siesta\ in biology: pilin proteins as conductors}

  {\siesta}'s efficiency and the clear band gaps of biomolecules in general
have made molecular biology a very suitable field for \siesta\ since the
beginning,\cite{DNA-2000} and have stimulated targeted developments of 
the code for the field, such as QM/MM.\cite{QMMM1,Sanz_2011}
  An interesting illustration of its suitability in an all-quantum 
biological problem is the study of the electrostatics around the pilin 
protein in aqueous solution.\cite{Feliciano_2012}
  The pilin considered here is the main protein in the pili (external filaments) 
of the {\it geobacter sulfurreducens bacterium}, which have been shown to be
able to transmit electronic current, allowing the microbe to feed 
by remote redox reactions on ferrous mineral particles in the soil.
  As a nanowire designed by natural evolution, understanding the 
mechanism for charge transport is of obvious interest. 
  
  Peculiar to this protein is the fact that its main alpha helix, the
main feature of this elongated protein, is singly oriented, that is, 
there is no back alpha helix (as in a common hairpin configuration)
that would counter the polarization of the single alpha helix:
  In an alpha helix all peptide-bond dipoles point in the same direction
along the axis of the helix, which, in solid-state parlance, represents
a polarization, with clear electrostatic implications. 
  Indeed, a DFT calculation of the molecule in vacuum shows a 
well defined electrostatic potential ramp along the protein, which
tends to close the effective band gap.
  The question is then, how does an aqueous environment affect this
depolarizing field. 

\begin{figure}[htbp]
  \centering
  \includegraphics[width=0.99\linewidth]{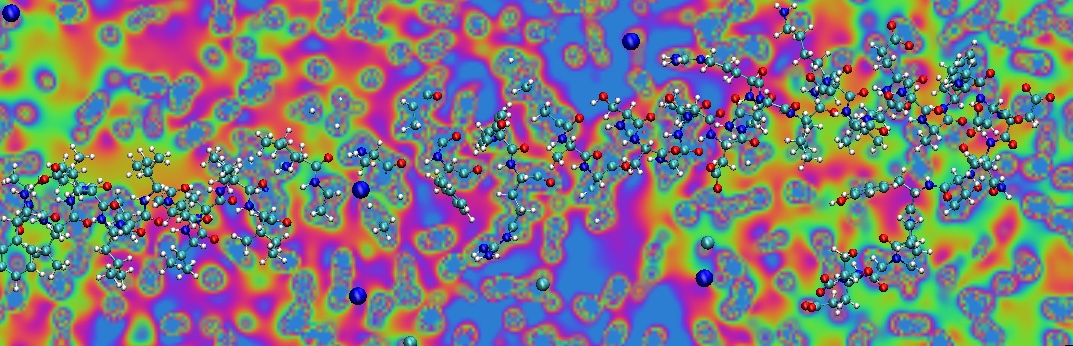}
  \caption{Colour coded electrostatic potential on a plane cutting along 
  the main axis of the geobacter sulfurreducens pilin molecule in wet conditions.
  A perspective ball rendering of the atomic strucuture of the protein is
  superposed. For the meaning and details on this Figure see
  Ref.~\onlinecite{Feliciano_2012} (Figure courtesy of Gustavo T. Feliciano).}
  \label{fig:pilin}
\end{figure}

  Long molecular mechanics (MM) simulations were performed for the 
protein in a suitable solution of NaCl at a concentration of 0.1 M.
  The protein's residues had charge states corresponding to $pH=7$,
and the MM field was validated with \siesta calculations in vacuum
(944-atom dynamic relaxation in a 104.4$^3$ \AA$^3$ box).
  The wet system contained 4580 atoms, and the statistical average
of the electrostatic potential around the molecule (see Fig.~\ref{fig:pilin})
was obtained from a sample of full \siesta\ calculations of statistically 
independent snapshots, taken every 50 ps during the last 0.5 ns of
the simulation.

  Fig.~\ref{fig:pilin} shows how the aqueous environment kills the
quite homogeneous potential ramp along the protein axis that appears 
in vacuum and replaces it with long-wave-length slow, but quite significant 
fluctuations.
 The gap remains sizeable, and coherent transport is not likely. 
 However, the frontier orbitals evolve in a very suggestive way for 
enhanced diffusive electron transport.\cite{Feliciano_2012}

\subsection{Use of \siesta\ in other fields}

Although an exhaustive summary of all the recent results
obtained with \siesta\ is out of the scope of this work, 
we would like to point the attention of the reader
to a sample of recent reviews in various fields in which the program
is featured. These cover
biological sciences\cite{Cole-16} (including interaction between organic and inorganic materials\cite{Darvish-16,Wenxuan-16}),
geology and materials under high-pressure\cite{Hermann-17}, 
isotopic fractionation predictions for Martian geochemistry\cite{TaoLiu-19},
the engineering of typical core structural materials used in nuclear reactors,\cite{Mayoral-17}
or even in astrophysical and atmospheric systems\cite{Escribano-18}.
The reactivity of metallic nanoparticles for catalysis was treated by \citet{Vines-14}, and the role of \siesta\ in the 
computation of the kinetic and dynamics of catalytic reaction at surfaces 
(including adsorption and desorption of reactants or products) was explored in 
Chapter 8 of Ref.~\citenum{Tao-15} by Catapan and coworkers.

\section{\label{sec:future}Future evolution}

Work on enhancing \siesta's capabilities, performance, and robustness
is continuing, driven by a good number of developers and
collaborators. A mature and flexible development platform and
practices are essential to keep them productive. Our recent platform
changes have forced developers to shift workflows twice in the past
four years. Through the changes we have learned a lot but also spent a
significant amount of time on ensuring \siesta's continuous
development. At the current state we believe we have stabilized the
development platform on GitLab while we will add more integrated
development features in the coming years, e.g. continuous integration
(CI) and source code checks. Using CI will also enable easier
code-style checks to conform to coding standards. We hope that our
open-platform initiative will keep external contributions coming into
the program.

Our basic-development plans include also refactoring, apparently
unexciting but essential to streamline the code base to enable further
implementations. Also, we foresee a change in the release model,
moving away from coexisting long-lived release branches whose
maintenance takes up a lot of time, and offering instead more frequent
and short-maintenance releases.

We plan to exploit the idea of modularization, continuing the
abstraction of relevant reusable pieces, but also dealing with a
higher-level, exposing the core electronic-structure capabilities of
\siesta\ to other programs.  It will be necessary to redesign some of
the internal data structures to remove global variables and
encapsulate them into objects or derived types associated to
particular configurations and stages of the calculations. This
encapsulation will be matched by a streamlining of the input/output
operations. This work will open the door to the creation of complex
workflows leveraging the strengths of various codes.
    
Accelerated hybrid architectures (including, for example, GPUs) are
very likely going to feature prominently in the upcoming exascale
machines. In the case of \siesta, the data indirection associated to
the handling of sparse matrices limits the acceleration possibilities
of the section of the code that builds the Hamiltonian and overlap
matrices, but the solver stage is more amenable to porting, and in
fact several solver libraries used by \siesta\ are being enhanced to
offer GPU support, as mentioned in Sec.~\ref{sec:elsi}.

Modularization and the use of new programming models cause an increase
in the complexity of the building and deployment of the code. We will
leverage the ESL bundle, created to
facilitate the use of the modules in the ESL collection, to
streamline \siesta's building process, and explore containerization
as an option for deployment of the code.
     
The ``pseudopotential barrier to entry'' has been lowered by the
availability of curated databases supporting the PSML format. Basis
sets are a perennial challenge, but new tools and ideas are being
explored to provide users with appropriate basis sets: High-throughput
workflows for optimization; "tiers" of quality/cost,
but perhaps not just of a simple ``periodic table'' form, as offered by other
codes (e.g., FHI-aims~\cite{fhiaims_blum_2009}), but with a possible dependence on
an approximate characterization of the chemical environment in which
a given atom finds itself.
    
Complementary to the underlying basis-set optimization that focuses on providing
an adequate variational freedom, an on-the-fly contraction of the
basis set, which results in a set of lower-cardinality adapted to the
description of the occupied subspace can be exploited for increased
efficiency. This is particularly relevant for FOE methods (see
Sect.~\ref{sec:stand-alone-solvers}, in which the number of
polynomial terms depends on the extent of the spectrum.
   
The original claim to fame of \siesta\ was based on its linear-scaling
solver. We are in the process of a re-design of the $\mathcal O(N)$
code with a new, more efficient backend, based on the DBCSR library
for handling distributed block-sparse matrices~\cite{DBCSR,
  Sivkov2019} with the MatrixSwitch library \cite{MatrixSwitch} acting
as an intermediary interface between it and high-level physical ideas
and algorithms. A connection between the internal
\siesta\ formats and MatrixSwitch itself has been recently
provided, using initially the cubic-scaling libOMM library~\cite{libOMM} as a test bed, hence
still using a dense coefficient matrix, as it corresponds to the case
without localization constraints in the solution of the
electronic-structure problem. The implementation 
of a sparse coefficient matrix will make it possible to perform
efficient $\mathcal O(N)$ calculations. The computational effort can
be further reduced through the analysis of sparsity of the
Hamiltonian and overlap matrices and their re-organization in the
block-compressed sparse form.

Other developments in the pipeline are linear-response calculations
for arbitrary distortions, electronic transport calculations with spin-orbit
coupling, thermal transport with the Green-Kubo formalism, as
described in Ref.~\onlinecite{baroni-gk}, a redesign of the molecular
dynamics subsystem, and the development of workflows for
the generation of data for \textsc{Scale-Up}.

\begin{acknowledgments}
  \siesta\ development has been historically supported by different Spanish National Plan projects:
  MEC-DGES-PB95-0202, MCyT-BFM2000-1312, MEC-BFM2003-03372, FIS2006-12117, FIS2009-12721, FIS2012-37549, FIS2015-64886-P,
  and RTC-2016-5681-7, the latter one together with Simune Atomistics Ltd.
  Currently, we thank financial support from the Spanish Ministry of Science, Innovation and Universities
  through the grant No. PGC2018-096955-B.

  We acknowledge the Severo Ochoa Centers of Excellence Program under
  Grants No. SEV-2015-0496 (ICMAB), and SEV-2017-0706 (ICN2), the
  GenCat Grant No. 2017SGR1506, and the European Union MaX Center of Excellence (EU-H2020 Grant No. 824143).

  P.G.-F. acknowledges support from Ram\'on y Cajal Grant No. RyC-2013-12515.
  J.I.C acknowledges RTI2018-097895-B-C41.

  R.C. acknowledges to the European Union’s Horizon 2020 research and innovation program under the 
  Marie Sk\l odoswka--Curie grant agreement no. 665919.
  
  D.S.P, P.K, and P.B acknowledge MAT2016-78293-C6, FET-Open
  No. 863098, and UPV-EHU Grant IT1246-19.
  
  V. Yu was supported by a MolSSI fellowship (U.S. NSF award 1547580), and
  the ELSI development (V.B.,V.Yu) by NSF award 1450280.
  We also acknowledge Honghui Shang and Xinming Qin for giving us access to the
  {\sc Honpas} code, where a preliminary version of the hybrid functionals support described here 
  was implemented.

  We are indebted to other contributors to the \siesta\ project, whose names can
  be seen in the  file in the \texttt{Docs/Contributors.txt}
  file of the \siesta\ distribution, and we thank those, too many to list, contributing fixes, comments,
  clarifications, and documentation for the code.

\end{acknowledgments}

The data that support the findings of this study are available from the corresponding author upon reasonable request.



\nocite{*}
\providecommand{\noopsort}[1]{}\providecommand{\singleletter}[1]{#1}%

\end{document}